%% file: hyper_master.tex
\documentclass[10pt,a4paper]{article}
\usepackage{graphicx}
\usepackage{amsmath}
\usepackage{dsfont}
\usepackage{amsfonts}
\usepackage{amsthm}
\usepackage{amssymb}
\usepackage{rotating}
\usepackage{verbatim}

\input definitions

\newcommand{\new}[1]{#1}

\newcommand{\todo}[1]{}

\title{Classical and Quantum Transport Through Entropic Barriers Modelled by Hardwall Hyperboloidal Constrictions}

\author{R. Hales$^1$ and H. Waalkens$^{1,2}$\\
\noindent
{\small $^1$University of Bristol, UK, $^2$University of Groningen, The Netherlands}}

\begin{document}
\maketitle
\date

\begin{abstract}
We study the quantum transport through entropic barriers induced by hardwall constrictions of hyperboloidal shape in two and three spatial dimensions. Using the separability of the Schr\"odinger equation and the classical equations of motion for these geometries we study in detail the quantum transmission probabilities and the associated quantum resonances, and relate them to the classical phase structures which govern the transport through the constrictions. These classical phase structures are compared to the analogous structures which, as has been shown only recently, govern reaction type dynamics in smooth systems. 
Although the systems studied in this paper are special due their separability they can be taken as a guide to study entropic barriers resulting from constriction geometries that lead to non-separable dynamics.
\end{abstract}


\input{sec1_intro}

\input{sec2_transprob}

\input{sec3_sepschrod}
\input{sec4_classical}
\input{sec5_transtate}

\input{sec6_resonances}

\input{sec7_conclusions}

\section*{Acknowledgements}

The authors gratefully acknowledge financial support (doctoral training account) from the EPSRC.

\end{document}

%% file: definitions.tex
\newcommand{\sn}{{\,\mbox{sn}}}
\newcommand{\cn}{{\,\mbox{cn}}}
\newcommand{\dn}{{\,\mbox{dn}}}
\newcommand{\tn}{{\,\mbox{tn}}}

\newcommand{\ud}{\mathrm{d}}
\newcommand{\ui}{\mathrm{i}}
\newcommand{\ue}{\mathrm{e}}

\newcommand{\T}{\mathbb{T}}
\newcommand{\R}{\mathds{R}}
\newcommand{\Z}{\mathds{Z}}
\newcommand{\N}{\mathds{N}}
\newcommand{\C}{\mathds{C}}

\newcommand{\sphere}{\mathds{S}}

\renewcommand{\Re}{\operatorname{Re}}

%% file: sec1_intro.tex
\section{Introduction}
\label{sec:intro}

A system displays reaction type dynamics if its phase space possesses bottleneck type structures. Such a system spends a long time in one phase space region (the region of `reactants') and occasionally finds its way through a bottleneck to another phase space region (the region of `products') or vice versa.  This type of dynamics does not only characterize chemical reactions but is of great significance in many different fields of physics and biology. Examples include ballistic electron transport problems \cite{Eckhardt95}, surface migration of atoms in solid state physics \cite{Jacucci}, ionisation of Rydberg atoms in electromagnetic fields \cite{wwju,ujpyw}, and on a macroscopic scale, the capture of moons near giant planets and asteroid motion \cite{WaalkensBurbanksWiggins05b,Jaffe}.

In systems where the dynamics is smooth and Hamiltonian, the phase space bottlenecks eluded to above are induced by saddle-centre-\ldots-center type equilibrium points, i.e. equilibrium points at which the matrix associated with the linearization of Hamilton's equations has one pair of real eigenvalues, $\pm \lambda$, and otherwise purely imaginary eigenvalues $\pm \ui \omega_k$, $k=2,\ldots,f$, where $f$ is the number of degrees of freedom.
In chemistry terms there is a `transition state' associated with the bottleneck, i.e. a state  the system has to pass `through' on its way from reactants to products. The most efficient and commonly used approach to compute reaction rates is \emph{transition state theory},
where the main idea is to place a dividing surface in the transition state region and compute the reaction rate from the flux through the dividing surface (for recent references, see the perspective paper \cite{PollakTalkner}). 
This approach has major computational benefits over other methods to compute the reaction rate because the latter typically require the integration of trajectories in order to decide whether they are reactive (i.e. extend from reactants to products, or vice versa) or nonreactive (i.e. stay in the regions of products or reactants). Rather than this \emph{global} information about trajectories, which to obtain is computationally expensive, transition state theory requires only \emph{local} information about the phase space structures near the saddle-centre-\ldots-center equilibrium point -- namely the construction of the dividing surface.
However, in order to be useful and not to overestimate the reaction rate the dividing surface needs to have the property that it divides the phase space into a reactants and a products region in such a way that it is crossed exactly once by reactive trajectories
and not crossed at all by nonreactive trajectories. 
The question  how to construct such a dividing surface for systems with an arbitrary number of degrees of freedom 
has posed a major problem for many years, and has been solved only recently based on ideas from dynamical systems theory 
(see \cite{ujpyw} and the recent review paper \cite{WaalkensSchubertWiggins08} with the references therein). 
The main building block in this construction is formed by a so called \emph{normally hyperbolic invariant manifold} (NHIM)  which is a manifold that is invariant under the dynamics (i.e. trajectories with initial conditions in the manifold stay in the manifold for all time) and is unstable in the sense that the expansion and contraction rates associated with the directions tangent to the manifold are dominated by those expansion and contraction rates associated with the directions transverse to the manifold \cite{Wiggins94}. The NHIM is the mathematical manifestation of  the transition state. In fact, the NHIM which is a sphere of dimension $2f-3$ (with $f$ again denoting the number of degrees of freedom) can be viewed to form  the equator of the dividing surface  which itself is a sphere of dimension $2f-2$   located in a $(2f-1)$-dimensional energy surface  if it has an energy slightly above the energy of the equilibrium point. The NHIM separates the dividing surface into two hemispheres. All forward reactive trajectories (trajectories evolving from reactants to products) cross one of these hemispheres; all backward reactive trajectories (trajectories evolving from products to reactants) intersect the other of these hemispheres.  Moreover, the NHIM has stable and unstable manifolds. These have the structure of spherical cylinders $\R\times S^{2f-3}$.  Since they are of one dimension less than the energy surface they have sufficient dimensionality to serve as impenetrable barriers in phase space \cite{wwju}.  They enclose the regions in the energy surface which contain the reactive trajectories and this way form the phase space conduits for reactions. 

Due to the spatial confinement, quantum effects are particularly strong for the passage through a phase space bottleneck, and accordingly, there is a strong interest in the quantum mechanical manifestation of the transition state.   In molecular collision experiments, for example, high resolution spectroscopic techniques have been developed to directly or indirectly probe the transition state (see, e.g., \cite{sy}).  Two quantum mechanical imprints of the transition state are given by the quantization of the so-called cumulative reaction probability which is the quantum analogue of the classical flux, and the quantum resonances associated with the transition state. The quantization of the cumulative reaction probability concerns the stepwise increase of the cumulative reaction probability each time a new transition channel opens as energy is increased.  While this is quite difficult to observe in chemical reactions (see, e.g.,  the controversial experiment on the isomerization of ketene \cite{LovejoyMoore93}) this effect can be seen almost routinely as a quantization of the conductance in the ballistic electron transmission through point contacts in semiconductor hetero-structures \cite{vanWees88,Wharam88}, metal nano-wires \cite{Olesen94,Krans95}  and even liquid metals. 
The quantum resonances on the other hand, describe how wavepackets initialised on the transition state decay in time (see  \cite{WaalkensSchubertWiggins08} for a detailed study). 

Moreover, there is a strong interest in the development of a quantum version of transition state theory, i.e. in a method to compute quantum reaction rates in such a way that it has similar computational benefits as (classical) transition state theory.  Though much effort has been devoted to this problem it is still considered an open problem in the recent perspective paper \cite{PollakTalkner}. One major problem here seemed to be the lacking geometric insight which ultimately led to the realization of classical transition state theory.  In \cite{SchubertWaalkensWiggins06,WaalkensSchubertWiggins08} a quantum version of transition state theory has been developed which incorporates the classical phase space structures mentioned above in a natural way. It has been demonstrated to yield quite efficient procedures to compute cumulative reaction probabilities as well as resonances.

In this paper we are concerned with phase space bottlenecks which are \emph{not} induced by equilibrium points. In the chemistry literature such bottlenecks are referred to as \emph{entropic barriers}: in the microcanonical pictures this means that despite of the absence of a potential barrier, there is a minimum in
the number (or to be more precise phase space volume) of possible configurations transverse to a reaction path. More concretely, we will consider potentialless systems with two and three degrees of freedom where the entropic barriers result from hard wall constrictions with the shape of an hyperbola and an (asymmetric) hyperboloid, respectively.
We will be particularly interested in the phase space structures which govern the reaction dynamics in these systems and thus play an analogous role as in the case of a smooth Hamiltonian system with reaction type dynamics as mentioned above, and their quantum mechanical manifestations.

The motivation for studying hyperboloidal geometries is that the resulting classical and quantum mechanical dynamics in such geometries are separable and in this sense completely solvable for such systems. This leads, as we will see, to a very transparent study of the influence of the phase space structures on the quantum transmission, and this way can serve as a first guide to study also non-separable dynamics in other constriction geometries.

This paper is organized as follows. In Sec.~\ref{sec:transprob} we introduce  in detail the systems studied in this paper and the associated  transmission problems. In Sec.~\ref{sec:separation} we show how the Schr\"odinger equation of the transmission problem can be separated. The corresponding separations of the classical equations of motion are studied in Sec.~\ref{sec:classical}.
The quantum and classical transmission probabilities are computed in Sec.~\ref{sec:cum_reac_prob}. Finally we compute and discuss quantum resonances in Sec.~\ref{sec:resonances}, and give a summary of the results and an outlook in Sec.~\ref{sec:conclusions}.

%% file: sec2_transprob.tex
\section{The transmission problem}
\label{sec:transprob}

\begin{figure}
\centerline{
\includegraphics[angle=0,width=4cm]{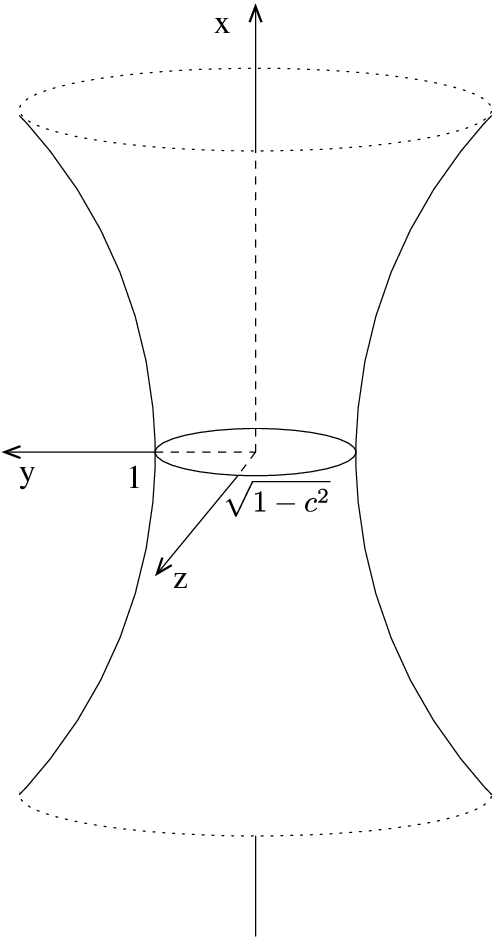}
}
\caption{\label{fig:fig1}
Accessible region confined by the boundary hyperboloid (\ref{eq:hyper}). 
The region has a ``bottleneck'' in the $y-z$ plane with the shape of an ellipse with semimajor axis 1 and semiminor axis $\tilde{c}=\sqrt{1-c^2}$ (in scaled coordinates). 
For the 2D case ($\tilde{c}=0$, or equivalently $c=1$),  the accessible region is the area between the two branches of the hyperbola (\ref{eq:hyper_2D}) in the $x-y$ plane. 
}
\end{figure}

In the 2D case we consider a point particle moving freely in a region of  the plane   defined by 
\begin{equation}\label{eq:2Dregion}
-\frac{x^2}{\tilde{a}^2}  + \frac{y^2}{\tilde{b}^2} \le 1 \,,
\end{equation}
where $(x,y)$ are  Cartesian coordinates in the plane, and $\tilde{a}$ and $ \tilde{b}$ are positive constants.
We assume that, classically, the particle is specularly reflected when it hits either of the branches of the boundary hyperbola 
\begin{equation}
\label{eq:hyper_2D}
-\frac{x^2}{\tilde{a}^2}  + \frac{y^2}{\tilde{b}^2} = 1
\end{equation}
(see Fig.~\ref{fig:fig1}).
Quantum mechanically, this leads to the boundary condition that the wavefunction which describes the position of the point particle has to vanish on the boundary hyperbola \eqref{eq:hyper_2D}. In the wide-narrow-wide geometry of the region \eqref{eq:2Dregion} we can associate the part which has $x\ll-1$ with the region representing the `reactants' and the part which has  $x\gg1$ as the `products', and that a `reaction' has taken place when the particle has moved from reactants to products. This interpretation directly applies to the ballistic transmission of electrons through a point contact formed by a lead of the shape   \eqref{eq:2Dregion}, but more generally can be viewed as a model describing the collective motion of a many body problem like a molecule from one configuration (or `isomer') to another.

In the 3D case we consider an analogous region in the three-dimensional space defined by 
\begin{equation}\label{eq:3Dregion}
-\frac{x^2}{\tilde{a}^2} + \frac{y^2}{\tilde{b}^2} + \frac{z^2}{\tilde{c}^2} \le 1 \,,
\end{equation}
where $(x,y,z)$ are Cartesian coordinates, and $\tilde{a}$, $ \tilde{b}$ and $ \tilde{c}$ are positive constants for which we impose the condition $ \tilde{b}\ge  \tilde{c}$. Note that this condition is only imposed for convenience and does not restrict the generality since one can simply swap the $y$ axis with the $z$ axis. The region \eqref{eq:3Dregion} is 
bounded by the (asymmetric) hyperboloid
\begin{equation}
\label{eq:hyper}
-\frac{x^2}{\tilde{a}^2} + \frac{y^2}{\tilde{b}^2} + \frac{z^2}{\tilde{c}^2} = 1
\end{equation}
(see Fig.~\ref{fig:fig1}). We again assume that, classically, the particle is specularly reflected when it hits the boundary hyperboloid and hence also that the quantum mechanical (position) wavefunction vanishes on the boundary hyperboloid \eqref{eq:hyper}.
We note that the region   \eqref{eq:2Dregion} in 2D can be  formally obtained  from the region  \eqref{eq:3Dregion} in 3D by letting $\tilde{c}\to 0$ which implies $z\to0$. While taking this limit leads to no problems for the classical dynamics, one has to be more careful, due the Heisenberg uncertainty relation, when considering this limit in the quantum case. One can view the 2D transmission problem to be contained in the 3D transmission problem either by considering a small but finite $\tilde{c}>0$ which leads to a flat region near the $x-y$ plane where for the energies under consideration no excitations in the $z$ direction are possible, or by considering a cylindrical region in 3D where the base of the cylinder has the shape \eqref{eq:2Dregion}.

The region (\ref{eq:2Dregion}) has a ``bottleneck'' contained in the $y$ axis which is given  by the line segment with minimal and maximal $y$ values $-\tilde{b}$ and $+\tilde{b}$, respectively. Similarly, the region \eqref{eq:3Dregion} has a bottleneck in the $y-z$ plane which is bounded by the ellipse
$y^2/\tilde{b}^2+z^2/\tilde{c}^2=1$.  In order to reduce the number of (effective) parameters we use as the length scale the maximum value of $y$ in the bottleneck. So formally we have $\tilde{b}=1$ and the number of parameters specifying the accessible regions is 1 in the 2D case and 2 in the 3D case.

The transmission through the bottlenecks can be viewed as a scattering problem. To this end we assume that a beam of (noninteracting) particles is incident from $x\ll -1$ (the `reactants') and we want to compute the transmission probability to $x\gg1$ (the `products'). We will compute the transmission probability both classically and quantum mechanically in the spirit of transition state theory in Sec.~\ref{sec:cum_reac_prob}.

As mentioned in the introduction the motivation for choosing  constrictions of the types (\ref{eq:hyper_2D}) and (\ref{eq:hyper}) is that they are the most general type of hard wall constrictions for which the transmission problem can be separated and in this sense solved explicitly.  
We will discuss the separation in the following section (Sec.~\ref{sec:separation}).
In fact, in the 2D case the transmission problem is still separable if the constriction is composed of two branches of different confocal hyperbolas. However, the asymmetric case has no 3D analogue and we therefore restrict ourselves to the symmetric case \eqref{eq:hyper_2D}.
Some aspects of the quantum transmission and the associated resonances through constrictions of the types  \eqref{eq:hyper_2D} and  \eqref{eq:hyper} have been addressed already in earlier papers.  The quantum resonances for an asymmetric 2D  constriction consisting of the branches of different hyperbola have been studied by Whelan \cite{Whelan}.  The quantum transmission problem (without resonances) through a constriction of the type \eqref{eq:hyper_2D} has been studied by Yosefin and Kaveh \cite{YosefinKaveh90}. 
Similarly, the transmission problem (again without resonances) has been studied for an axially symmetric hyperboloidal constriction in 3D by  Torres, Pascual and S{\'a}enz \cite{Torresetal94}, and for the asymmetric case by Waalkens \cite{Waalkens05}. The main purpose of the present paper is to study the quantum transmission and the assoicated resonances through the 2D and 3D constrictions \eqref{eq:hyper_2D} and  \eqref{eq:hyper} in a coherent way using the perspective of transition state theory.

%% file: sec3_sepschrod.tex
\section{Separation of the Schr{\"o}dinger equation}
\label{sec:separation}
 
For the quantum transmission problem,
we have to find solutions of the free Schr\"odinger or Helmholtz equations
\begin{equation}\label{eq:Helmholtz2D}
-\frac{\hbar^2}{2m} \left( \frac{\partial^2}{\partial x^2} +  \frac{\partial^2}{\partial y^2} \right) \psi = E \psi \quad (\text{2D}) 
\end{equation}
or
\begin{equation}\label{eq:Helmholtz3D}
-\frac{\hbar^2}{2m} \left( \frac{\partial^2}{\partial x^2} +  \frac{\partial^2}{\partial y^2}  +  \frac{\partial^2}{\partial z^2} \right) \psi = E \psi \quad (\text{3D})\,,
\end{equation}
which for $x\gg1$ are waves propagating in the positive $x$ direction and fulfill Dirichlet boundary conditions, i.e. we require the restriction of $\psi$ on the boundary hyperbola \eqref{eq:hyper_2D} resp. hyperboloid \eqref{eq:hyper} to vanish.
The Helmholtz equations \eqref{eq:Helmholtz2D} and   \eqref{eq:Helmholtz3D}  together with their boundary conditions can be separated in elliptic and ellipsoidal coordinates, respectively, as we will discuss in the following two subsections which separately consider the 2D case and 3D case.

\subsection{The 2D system}
\label{subsec:sep2D}

\begin{figure}
\centerline{
\includegraphics[angle=0,width=8cm]{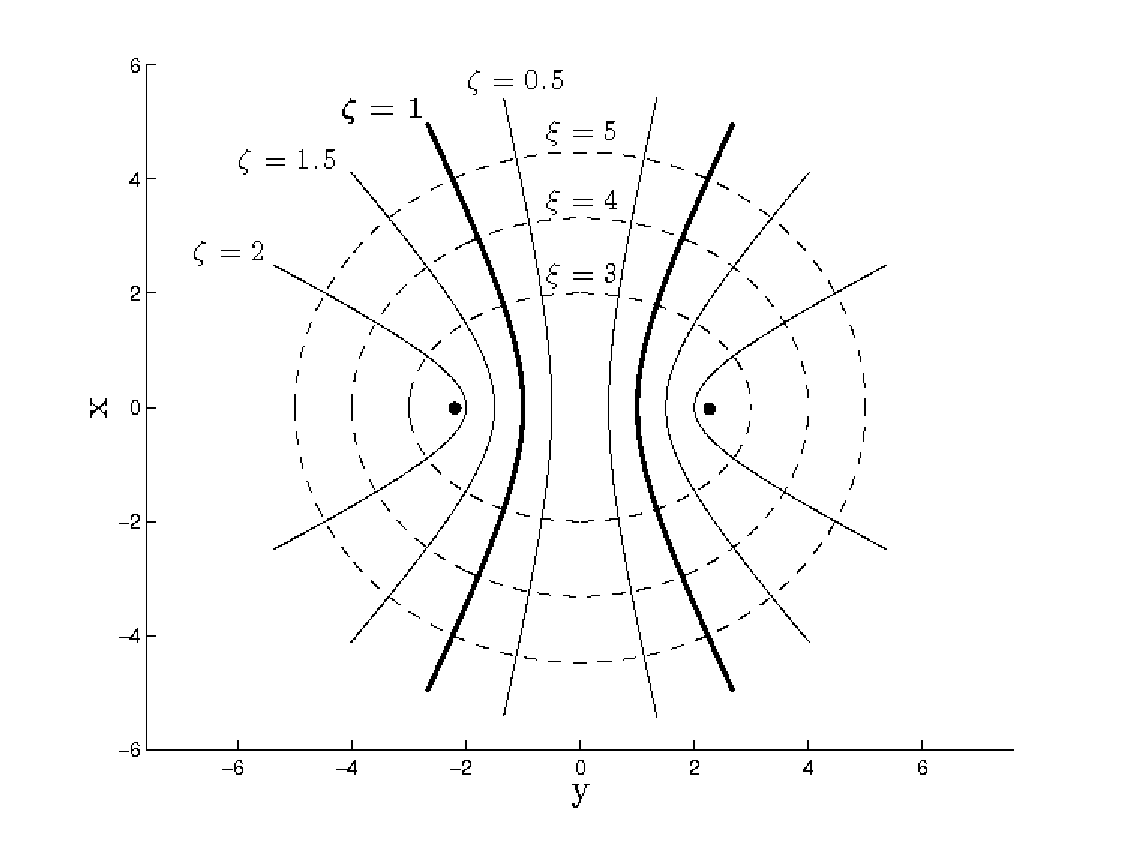}
}
\caption{\label{fig:coord_2D}
Coordinate lines $\xi=$const. (dashed ellipses) and $\zeta=$const. (solid hyperbolae) with the boundary hyperbola (\ref{eq:hyper_2D}) in bold.  The bold dots mark the focus points $(x,y)=(0,\pm a)$. ($a^2=5$.) 
}
\end{figure}

The 2D Helmholtz equation (\ref{eq:Helmholtz2D}) together with
the Dirichlet boundary conditions can be separated in elliptic coordinates $(\zeta,\xi)$ \cite{Morse53,WWD99}.
Each of them parametrizes a family of confocal quadrics

\begin{equation}
\label{eq:eqb3d_confocalsurfaces_2D}
\frac{x^2}{s^2-a^2} + \frac{y^2}{s^2} = 1,
\end{equation}
where $s\in \{\zeta,\xi\}$ and $a^2=1+\tilde{a}^2/\tilde{b}^2$.

For $s=\xi > a$,  both terms on the left hand side of Eq.~(\ref{eq:eqb3d_confocalsurfaces_2D}) are positive and the equation defines a family of confocal ellipses with foci at $(x,y)=(0,\pm a)$. Their intersections with the $x$ axis and $y$ axis are at $x=\pm \sqrt{\xi^2-a^2}$ and $y=\pm \xi$, respectively. For $a > s=\zeta > 0$, the first term on the left hand side of
Eq.~(\ref{eq:eqb3d_confocalsurfaces_2D}) is negative giving
confocal (two sheeted) hyperbolae with foci also at $(x,y)=(0,\pm a)$.
Their intersections with the $y$ axis are at $y=\pm \zeta$; they do not
intersect the $x$ axis.

\begin{figure}
\centerline{
\includegraphics[angle=0,width=12cm]{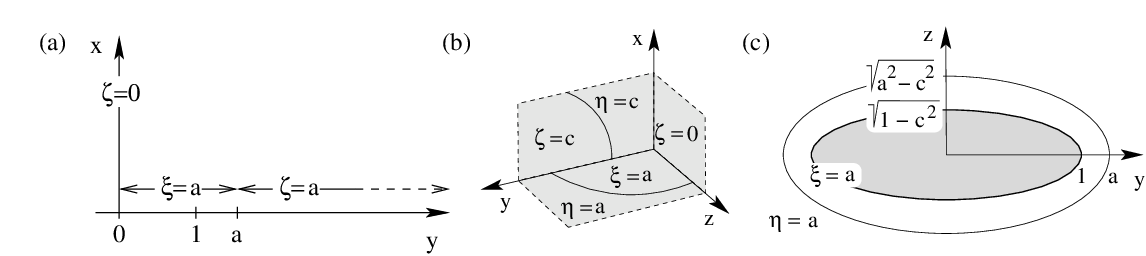}
}
\caption{\label{fig:singular}
(a) Singular elliptic coordinate surfaces, for the 2D system. The line segment $[0,1]$ is half of the ``bottleneck'' on the $y$ axis. (b) Singular ellipsoidal coordinate surfaces, for the 3D system. (c) ``Bottleneck'' (shaded region) in the $y-z$ plane bounded by the 3D hyperboloidal constriction whose intersection with the $y-z$ plane is the ellipse $y^2/1+z^2/(1-c^2)=1$, and singular coordinate patches $\xi=a$  and $\eta=a$ (inside and outside of the
ellipse $y^2/a^2+z^2/(a^2-c^2)=1$, respectively).
}
\end{figure}

The coordinate lines of $\zeta$ and $\eta$ are shown in 
Fig.~\ref{fig:coord_2D}.  
Inverting Eq.~(\ref{eq:eqb3d_confocalsurfaces_2D}) within the positive
$x-y$ quadrant gives
\begin{eqnarray}
\label{eq:eqb3d_transxeztoxyz_a_2D}
x &=& \frac{\sqrt{(\xi^2-a^2)(a^2-\zeta^2)}}{a}\,,\\
\label{eq:eqb3d_transxeztoxyz_c_2D}
y &=& \frac{\xi\zeta}{a}\,,
\end{eqnarray}
with
\begin{equation}
\label{eq:eqb3d_xezranges_2D}
0 \le \zeta \le a \le \xi.
\end{equation}
The remaining quadrants are obtained from appropriate reflections. 
However, it is also useful to reduce the discrete reflection symmetry of the system about the $x$ axis and the $y$ axis. In fact the solutions of the Helmholtz equation \eqref{eq:Helmholtz2D} fulfilling the Dirichlet boundary conditions along the boundary hyperbola \eqref{eq:hyper_2D} can be classified in terms of their parities $\pi_x$ and $\pi_y$ which correspond to the reflections about the $y$ axis and $x$ axis, respectively.  We therefore introduce the symmetry reduced system which only has the positive $x-y$ quadrant as the fundamental domain and impose Dirichlet (negative parity) or Neumann boundary conditions (positive parity) on the $x$ and $y$ axes. The Cartesian coordinate axes are obtained from the elliptic coordinates $\zeta$ and $\xi$ in terms of the equalities in 
 (\ref{eq:eqb3d_xezranges_2D}):
$\zeta=0$ gives the $x$ axis; 
$\xi=a$ gives the segment of the $y$ axis between the focus points, the rest of the $y$ axis has  $\zeta=a$ (see Fig.~\ref{fig:singular}(a)).  

The boundary hyperbola (\ref{eq:hyper_2D}) (in scaled coordinates) coincides with the coordinate line $\zeta = 1$, i.e. in the region \eqref{eq:2Dregion} $\zeta$ takes values in  $[0,1]$. 
Considering only the region enclosed by the boundary hyperbola (\ref{eq:hyper_2D}),
the coordinate lines $\xi=$const. $\ge a$ are transverse to the $x$ direction. To this end note that the singular coordinate line
$\xi=a$ contains the `bottleneck'  $(x,y)\in\{0\}\times[-1,1]$. 
The coordinate $\xi$ thus parametrizes the direction of the transmission; $\zeta$ parametrizes the direction transverse to the transmission.

The parameter $a$ determines how strong the narrowness of the constriction 
changes with $x$: for $a\rightarrow \infty$ the constriction becomes an infinitely long rectangualar strip; for $a\rightarrow 1$ the constriction  degenerates to the $y$ axis with a hole of width 2 about the origin.

With the ansatz $\psi(\zeta,\xi)=\psi_\zeta(\zeta)\psi_\xi(\xi)$ the partial differential equation~(\ref{eq:Helmholtz2D}) can be separated and turned into the set of ordinary differential equations 
\begin{equation}
\label{eq:sephelmholtz_2D}
-\frac{\hbar^2}{2m}\left(\sqrt{s^2-a^2}\frac{\mbox{d} }{\mbox{d} s}\right)^2 \psi_s(s) = 
E \left(s^2-s_2^2\right) \, \psi_s(s)\,, 
\end{equation}
where $s\in\{\zeta, \xi\}$ and $s_2^2$ denotes the separation constant.
The equations for $\zeta$ and $\xi$ are identical, but they have to be considered on the different intervals \eqref{eq:eqb3d_xezranges_2D} and for different boundary conditions. 
In fact the equations have regular singular points \cite{Morse53} at $\pm a$. These regular singular points have indices $0$ and $1/2$, i.e. 
there are solutions, which near $\pm a$ are of the form $\psi_s(s) = (s\mp a)^{{q}}\tilde{\psi}(s)$ where $\tilde{\psi}(s)$ is analytic and 
$q=0$ or $q=1/2$.   
As the elliptic coordinates $(\zeta,\xi)$ give
for the regular singular point $a$
the Cartesian  $y$ axis, the indices determine the parities  $\pi_x$ of the total wave function  $\psi(\zeta,\xi)$ \cite{WWD97}, i.e.
$q=0$ or $q=1/2$ correspond to total wave functions which have  $\pi_x=+$  or $\pi_x=-$, respectively. The value of $\psi_\zeta$ at the ordinary point $\zeta=0$ determines the parity $\pi_y$.

For the computation and interpretation of the results below, it is useful to remove the singularities
in (\ref{eq:sephelmholtz_2D}). This can be achieved by the transformation
\begin{equation}
\label{eq:regularization_2D}
(\zeta(\nu),\xi(\lambda)) = a (\cos(\nu),\cosh(\lambda))\,,
\end{equation}
which is the standard parametrization of elliptic coordinates
by triginometric functions. Inserting \eqref{eq:regularization_2D} into \eqref{eq:eqb3d_transxeztoxyz_a_2D} and \eqref{eq:eqb3d_transxeztoxyz_c_2D} 
gives
\begin{equation} \label{eq:xy_lambdanu}
x =  a \sin (\nu) \sinh(\lambda) \,, \quad 
y = a \cos(\nu) \cosh(\lambda)\,.  
\end{equation}
To cover the positive $x-y$ quadrant $(\nu,\lambda)$ have to vary in the intervals
\begin{equation}\label{eq:nu_lambda_range_2D}
0 \le \nu \le \pi/2\,, \quad
0\le \lambda < \infty\,.
\end{equation}
The boundary hyperbola (\ref{eq:hyper_2D}) has 
\begin{equation}
\nu=\nu_{\text{B}} = \arccos(1/a)\,.
\end{equation}
Extending the intervals \eqref{eq:nu_lambda_range_2D} to 
\begin{equation}
\nu_{\text{B}} \le \nu \le \pi-\nu_{\text{B}}\,, \quad
-\infty < \lambda < \infty\,.
\end{equation}
we get a full regular cover of the region \eqref{eq:2Dregion} in terms of the strip $[\nu_{\text{B}} ,\pi-\nu_{\text{B}}]\times\R$.

Transforming (\ref{eq:sephelmholtz_2D}) to the coordinates $(\nu,\lambda)$ leads to
\begin{equation}
\label{eq:helmreg_2D}
-  \frac{\hbar^2}{2m}  \frac{\mbox{d}^2}{\mbox{d} \hat{s}^2}\psi_{\hat{s}}(\hat{s}) 
= 
\sigma_{\hat{s}}
E\left(s^2(\hat{s}) - s_2^2\right) \, \psi_{\hat{s}}(\hat{s})\,,
\end{equation}
where 
$\hat{s}\in \{\nu,\lambda\}$,
$s(\hat{s}) \in \{\zeta(\nu),\xi(\lambda)\}$ are the functions from~(\ref{eq:regularization_2D}) and the $\sigma_{\hat{s}}$ are the signs
$\sigma_{\lambda}=+$ and $\sigma_{\nu}=-$. 
Each of these equations can be interpreted as a
one-dimensional Schr\"odinger equation with a Hamiltonian of the standard type $H=-\hbar^2(\mbox{d}^2/\mbox{d}x^2)/2+V$ (``kinetic plus potential energy'') with effective energy and potential
\begin{equation}
\label{eq:effenerpot_2D}
E_{\hat{s},\textrm{eff}}  = - \sigma_{\hat{s}}E s_2^2\,,\quad
V_{\hat{s}, \textrm{eff}}(\hat{s})  =
-\sigma_{\hat{s}}Es^2(\hat{s})\,.
\end{equation}
The effective energies and potentials (\ref{eq:effenerpot_2D}) are shown for ``representative'' values of the separation constant $s^2_2$ in 
Fig.~\ref{fig:pots_xfig_c_2D}(a) of Sec.~\ref{sec:classical}. Here  $\nu$ varies in an interval of length $\pi$ which is the period of the effective potential $V_{\nu, \textrm{eff}}$. What we mean by ``representative'' will be explained in Sec.~\ref{sec:classical}, where we analyze the corresponding classical system.
Since the effective potential $V_{\lambda, \textrm{eff}}$ is symmetric under the reflection $\lambda\mapsto -\lambda$ (i.e. the reflection about $\lambda=0$) there are solutions of \eqref{eq:helmreg_2D} that are symmetric or antisymmetric under this reflection. Using \eqref{eq:xy_lambdanu} we can relate  the behaviour of solutions under this reflection to the parity $\pi_x$. Similarly, since the effective potential $V_{\nu, \textrm{eff}}$ is symmetric under the reflection $\nu\mapsto -\nu + \pi$ (i.e. the reflection about $\nu=\pi/2$) there are solutions of \eqref{eq:helmreg_2D} that are symmetric or antisymmetric under this reflection. Again using \eqref{eq:xy_lambdanu} we can relate the behaviour of solutions under this reflection to the parity $\pi_y$. The parities $\pi_x$ and $\pi_y$ are marked at the top of Fig.~\ref{fig:pots_xfig_c_2D}(a).
The fact that the algebraic counterparts of \eqref{eq:helmreg_2D} in \eqref{eq:sephelmholtz_2D} are identical, is reflected in \eqref{eq:helmreg_2D} by the substitution $\nu \to \ui \nu$ which relates the equation for $\nu$ to the equation for $\lambda$.
\subsection{The 3D system}
\begin{figure}
\centerline{
\includegraphics[angle=0,width=10cm]{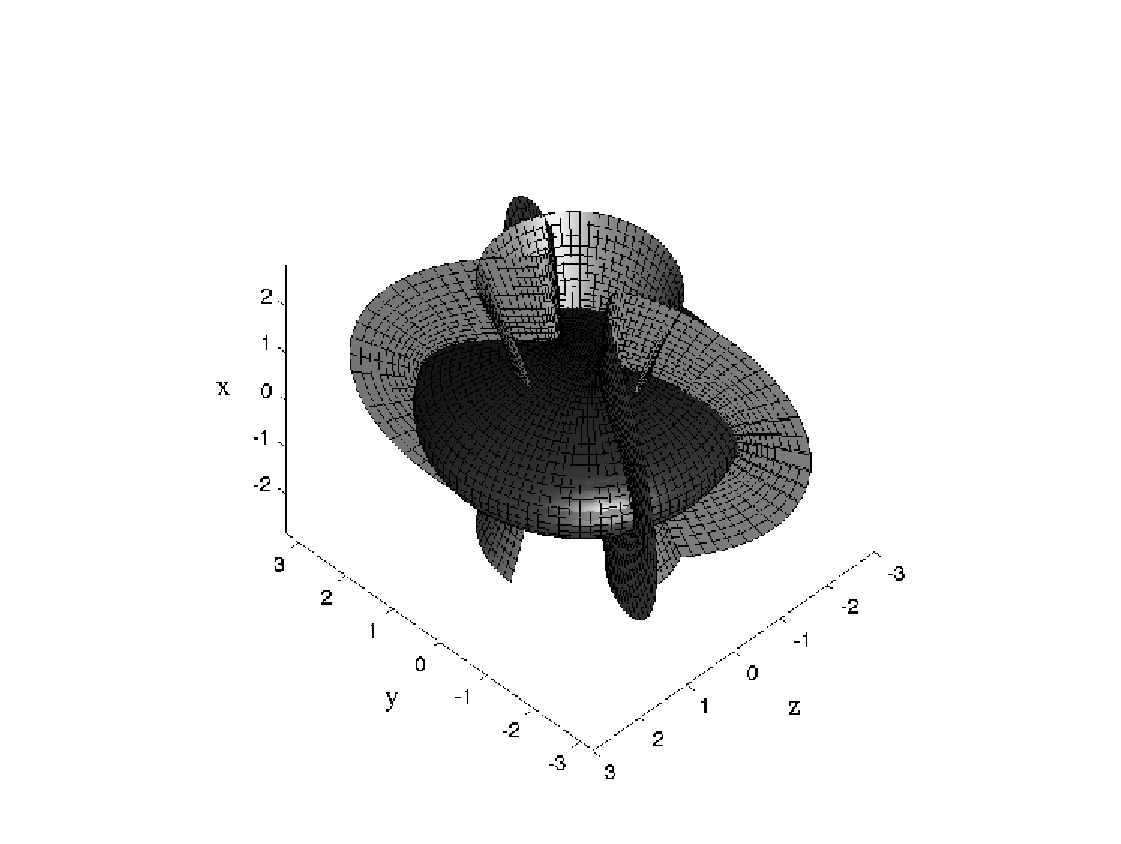}
}
\caption{\label{fig:coord}
Ellipsoidal coordinates surfaces $\xi=5$ (ellipsoid in red), $\eta=1$ (the one sheeted boundary hyperboloid (\ref{eq:hyper}) in blue) and $\zeta=1/2$ (two sheeted hyperboloid in green), for $(a^2,c^2)=(5,0.2)$.
}
\end{figure}

Similarly to the 2D case, the 3D Helmholtz equation \eqref{eq:Helmholtz3D} together with
the Dirichlet boundary conditions can be separated in ellipsoidal coordinates $(\zeta,\eta,\xi)$ \cite{Morse53,WWD99}. Each of them parametrizes a family of confocal quadrics
\begin{equation}
\label{eq:eqb3d_confocalsurfaces}
\frac{x^2}{s^2-a^2} + \frac{y^2}{s^2} + \frac{z^2}{s^2-c^2} = 1\, ,
\end{equation}
where $s\in \{\zeta,\eta,\xi\}$, $c^2=1-\tilde{c}^2/\tilde{b}^2$ and $a^2=1+\tilde{a}^2/\tilde{b}^2$.

For $s=\xi > a$,  all terms on the left hand side
of Eq.~(\ref{eq:eqb3d_confocalsurfaces}) are positive and the equation
defines
a family of confocal
ellipsoids.  Their
intersections with the $y-z$ plane, the $x-y$ plane and the
$x-z$ plane   are planar ellipses with  foci at
$(y,z)=(\pm c,0)$, $(x,y)=(0,\pm a)$ and $(x,z)=(0,\pm (a^2-c^2)^{1/2})$,
respectively. 
For $a> s=\eta > c$, the first term on the left hand side of
Eq.~(\ref{eq:eqb3d_confocalsurfaces}) becomes
negative. Eq.~(\ref{eq:eqb3d_confocalsurfaces}) thus gives confocal one
sheeted hyperboloids.
Their intersections with the $y-z$ plane are planar ellipses with foci at
$(y,z)=(\pm c,0)$;
the intersections with the $x-y$ plane and the
$x-z$ plane   are planar hyperbolas with  foci at
$(x,y)=(0,\pm a)$ and $(x,z)=(0,\pm (a^2-c^2)^{1/2})$, respectively.
For $c > s=\zeta > 0$,
the first and third
terms on the left hand side of Eq.~(\ref{eq:eqb3d_confocalsurfaces}) are negative giving
confocal two sheeted hyperboloids.
Their intersections with the $y-z$ plane and the $x-y$ plane  are
planar  hyperbolas
with  foci at
$(y,z)=(\pm c,0)$ and $(x,y)=(0,\pm a)$, respectively; they do not
intersect the $x-z$ plane.

The coordinate surfaces of $\zeta$, $\eta$ and $\xi$ are shown in Fig.~\ref{fig:coord}. 
Inverting Eq.~(\ref{eq:eqb3d_confocalsurfaces}) within the positive
$x-y-z$ octant gives 
\begin{eqnarray}
\label{eq:eqb3d_transxeztoxyz_a}
x &=& \frac{\sqrt{(\xi^2-a^2)(a^2-\eta^2)(a^2-\zeta^2)}}{a\sqrt{a^2-c^2}}\,, \\
\label{eq:eqb3d_transxeztoxyz_b}
y &=& \frac{\xi\eta\zeta}{ac}\,, \\
\label{eq:eqb3d_transxeztoxyz_c}
z &=& \frac{\sqrt{(\xi^2-c^2)(\eta^2-c^2)(c^2-\zeta^2)}}{c\sqrt{a^2-c^2}}\,,
\end{eqnarray}
with
\begin{equation}
\label{eq:eqb3d_xezranges}
0\le \zeta \le c \le \eta \le a \le \xi.
\end{equation}
The remaining octants are obtained from appropriate reflections. 
Again, we also introduce a symmetry reduced system which has the positive $x-y-z$ octant as the fundamental domain. The solutions of the Helmholtz equation \eqref{eq:Helmholtz3D} fulfilling Dirichlet boundary conditions along the boundary hyperboloid \eqref{eq:hyper} with parities $\pi_x$,   $\pi_y$ and  $\pi_z$ are then obtained from the symmetry reduced system by imposing Dirichlet or  Neumann boundary conditions along the Cartesian coordinate planes which in terms of the elliptic coordinates $(\zeta,\eta,\lambda)$ are given by the 
equalities in  one of the  equations in (\ref{eq:eqb3d_xezranges}): 
$\zeta=0$ gives the $x-z$ plane; 
$\zeta=c$ and $\eta=c$ give two surface patches which together cover the $x-y$ plane; 
$\eta=a$ and $\xi=a$ give two surface patches which together cover the
$y-z$ plane (see Fig.~\ref{fig:singular}(b)).  

The boundary hyperboloid (\ref{eq:hyper}) (in scaled coordinates) coincides with the coordinate surface $\eta = 1$, i.e. within the region \eqref{eq:3Dregion}  $\eta$ is restricted to  $[c,1]$. 
Considering only the region enclosed by the boundary hyperboloid (\ref{eq:hyper}),
the coordinate planes $\xi=$const.$\ge a$ are transverse to the $x$ direction. Note that the singular coordinate plane 
$\xi=a$ is a region in the $y-z$ plane which is enclosed by an ellipse which lies outside of the 
hyperboloidal constriction (see Fig.~\ref{fig:singular}(c)). 
The coordinate $\xi$ thus parametrizes the direction of transmission; 
$\eta$ and $\zeta$ parametrize the two directions transverse to transmission.

The parameter $c$ determines the asymmetry of the cross-section of the constriction with $c=0$ leading to an axially symmetric constriction and $c=1$ leading to the 2D case.
The parameter $a$ determines how strong the narrowness 
changes with $x$: for $a\rightarrow \infty$ the constriction becomes cylindrical with an elliptical cross-section; for $a\rightarrow 1$ the constriction 
degenerates to the $y-z$ plane with a hole having the shape of an ellipse.

With the ansatz $\psi(\zeta,\eta,\xi)=\psi_\zeta(\zeta)\psi_\eta(\eta)\psi_\xi(\xi)$ 
the Helmholtz equation~(\ref{eq:Helmholtz3D}) can be separated and turned into the set of ordinary differential equations 
\begin{equation}
\label{eq:sephelmholtz}
-\frac{\hbar^2}{2m}\left(\sqrt{(s^2-a^2)(s^2-c^2)}\frac{\mbox{d} }{\mbox{d} s}\right)^2 \psi_s(s) =
 E \left(s^4 - 2 k s^2 + l\right) \, \psi_s(s)\,,
\end{equation}
where $s\in \{\zeta,\eta,\xi\}$ and  $k$ and $l$ denote the separation constants. 
The equations for $\zeta$, $\eta$ and $\xi$ are identical, but they have to be considered on the different intervals (\ref{eq:eqb3d_xezranges}) and for different boundary conditions. For later purposes it is useful to rewrite \eqref{eq:sephelmholtz}
in the form
\begin{equation}
\label{eq:sephelmholtz_s1_s2}
-\frac{\hbar^2}{2m}\left(\sqrt{(s^2-a^2)(s^2-c^2)}\frac{\mbox{d} }{\mbox{d} s}\right)^2 \psi_s(s) =
 E \big(s^2-s_1^2\big)\big(s^2-s_2^2\big)  \, \psi_s(s)\,,
\end{equation}
where 
\begin{equation}\label{eq:def_s_1_s_2}
s_1^2 = k-(k^2-l)^{1/2} \,, \quad
s_2^2 = k+(k^2-l)^{1/2} \,,
\end{equation}
and conversely
\begin{equation}
k=\frac12 \big( s_1^2 + s_2^2  \big) \,, \quad
l = s_1^2 s_2^2\,. 
\end{equation}

\todo{Compare the equations to the 2D case; are these Lame equations?}
Similarly to equations~\eqref{eq:sephelmholtz_2D} in the 2D case
 the equations \eqref{eq:sephelmholtz} have regular singular points \cite{Morse53} at $\pm a$ and $\pm c$. All these regular singular points again have indices $0$ and $1/2$ like in the 2D case. 
Thus there are solutions, which near $\sigma=\pm a$ or $\sigma=\pm c$ are of the form $\psi_s(s) = (s-\sigma)^{{q_{\sigma}}}\tilde{\psi}(s)$ where $\tilde{\psi}(s)$ is analytic and 
$q_\sigma =0$ or $q_\sigma =1/2$.   
As the \new{ellipsoidal} coordinates $(\zeta,\eta,\xi)$ give
for the regular singular points $\pm a$ and $\pm c$ 
the Cartesian  $y-z$ plane and $x-y$ plane, respectively, the indices determine the parities  $\pi_x$ and $\pi_z$ of the total wave function  $\psi(\zeta,\eta,\xi)$ \cite{WWD99}. 
More precisely, $q_a=0$ or $q_a=1/2$ correspond to total wave functions which have  $\pi_x=+$  or $\pi_x=-$, respectively, and
$q_c=0$ or $q_c=1/2$ correspond to total wave functions which have  $\pi_z=+$  or $\pi_z=-$, respectively. As in the 2D case, the value of $\psi_\zeta$ at the ordinary point $\zeta=0$ determines the parity $\pi_y$.

For the computation and interpretation of the results below it is useful to remove the singularities
in (\ref{eq:sephelmholtz}). This can be achieved by the transformation
\begin{equation}
\label{eq:regularization}
(\zeta(\nu),\eta(\mu),\xi(\lambda)) = a (q\sn(\nu,q),\dn(\mu,q'),\frac{\dn(\lambda,q)}{\cn(\lambda,q)}  )\,,
\end{equation}
where $\sn(\phi,q)$, $\cn(\phi,q)$ and $\dn(\phi,q)$ are Jacobi's
elliptic functions with ``angle'' $\phi$ and modulus $q$ \cite{GradRyzh65}. Here the modulus
is given by $q=c/a$ and $q'=(1-q^2)^{1/2}$ denotes the conjugate
modulus. This is the standard parametrization of ellipsoidal coordinates
by elliptic functions \cite{Morse53}. 

\begin{figure}
\centerline{
\includegraphics[angle=0,width=8cm]{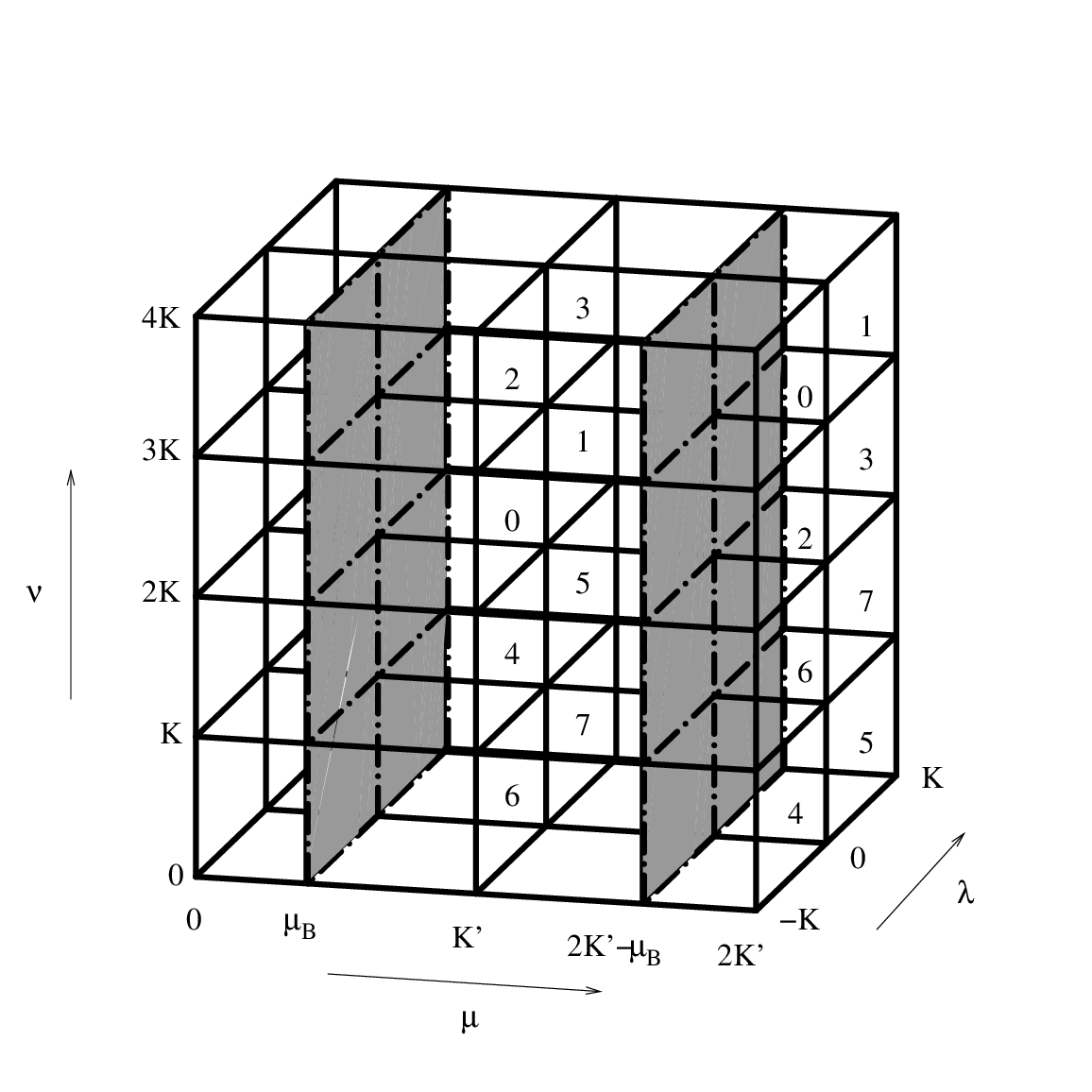}
}
\caption{\label{fig:lambda_mu_nu_grid}
Representation of the solid torus \eqref{eq:double_cover_3D_cube} as a cube with periodic boundary conditions in $\nu$. Each small cube represents one Cartesian $x-y-z$ octant. The octants corresponding to the smaller cubes  are indicated  by  a `binary' labeling with respect to the signs of $x$, $y$ and $z$ (e.g., (--,--,--) corresponds to 0, (--,--,+) corresponds to 1, etc.). The shaded planes mark the boundary hyperboloid which on the double cover is given by  $\mu=\mu_{\text{B}}$ and $\mu=2K'(q)-\mu_{\text{B}}$, where $K'(q)=K(q')$.
}
\end{figure}

Expressing the Cartesian coordinates in terms of $(\nu,\mu,\lambda)$ gives
\begin{equation}\label{eq:xzy_numulambda}
\begin{split}
x &= q' a \frac{\sn(\lambda,q)\sn(\mu,q')\dn(\nu,q)}{\cn(\lambda,q)}\,, \\
y &= a \frac{\dn(\lambda,q) \dn(\mu,q') \sn(\nu,q)}{\cn(\lambda,q)}\,, \\
z &= q' a \frac{\cn(\mu,q') \cn(\nu,q)}{\cn(\lambda,q)}\,.  
\end{split}
\end{equation}
To cover the positive $x-y-z$ octant $(\nu,\mu,\lambda)$ have to vary in the intervals
\begin{equation}\label{eq:nu_mu_lambda_ranges_3D}
0 \le \nu \le K(q)\,, \quad
0 \le \mu \le K(q')\,, \quad
0\le \lambda \le K(q)\,,
\end{equation}
where $K(q)$ and $K(q')$ are Legendre's complete elliptic integral of first kind with modulus $q$ and $q'$, respectively. The boundary hyperboloid (\ref{eq:hyper}) has 
\begin{equation}
\label{eq:boundary}
\mu=\mu_{\text{B}} = F(((a^2-1)/(a^2-c^2))^{1/2},q')\,,
\end{equation}
where $F$ is Legendre's incomplete
elliptic integral of first kind which in (\ref{eq:boundary}) has argument $((a^2-1)/(a^2-c^2))^{1/2}$ and modulus $q'$.
Extending the intervals \eqref{eq:nu_mu_lambda_ranges_3D} to
\begin{equation} \label{eq:double_cover_3D_cube}
0 \le \nu \le 4K(q)\,, \quad
\mu_{\text{B}} \le \mu \le 2 K(q')-\mu_{\text{B}}\,, \quad
-K(q) \le \lambda \le K(q)
\end{equation}
we get a double cover of the region \eqref{eq:3Dregion}  in terms of the `solid torus' $\R/(4K(q)\Z) \times[\mu_{\text{B}} , 2K(q')-\mu_{\text{B}}]\times[-K(q),K(q)]$, where $\R/(4K(q)\Z)$ denotes the topological circle resulting from identifying points in $\R$ differing by integer multiples of the period in $\nu$ which is $4K(q)$. 
In Fig.~\ref{fig:lambda_mu_nu_grid} we present the solid torus  as the cube \eqref{eq:double_cover_3D_cube}, where the opposite sides $\nu=0$ and $\nu=4K(q)$ have to be identified. Each of the smaller cubes
\begin{equation}\label{eq:smaller_cubes}
\begin{split}
&[n_\nu K(q),(n_\nu+1) K(q)] \times [0, K(q')] \times [0,\pm K(q)] \quad \text{and}\\ 
&[n_\nu K(q),(n_\nu+1) K(q)] \times [K(q'),2K(q')] \times [0,\pm K(q)] 
\end{split}
\end{equation} 
in Fig.~\ref{fig:lambda_mu_nu_grid}, with $n_\nu\in\Z$ represents one Cartesian $x-y-z$ octant of the region \eqref{eq:3Dregion}.
Note that each of the smaller cubes \eqref{eq:smaller_cubes} has four neighbours. 
This property can be understood from the fact that in order to regularise the coordinates $(\zeta,\eta,\xi)$
in terms of the coordinates $(\nu,\mu,\lambda)$ we have to regularise each of the four singular transition between two $x-y-z$ octants shown in Fig.~\ref{fig:singular} (note that the singular patch $\eta=a$ in Fig.~\ref{fig:singular} is not accessible 
in \eqref{eq:3Dregion}). 
The two covers of the double cover \eqref{eq:double_cover_3D_cube} are related by the involution
\begin{equation}\label{eq:involution_S}
S(\nu,\mu,\lambda) = (2K(q)-\nu,-\mu-2K(q'),\lambda)\,,
\end{equation}
which leaves the Cartesian coordinates \eqref{eq:xzy_numulambda} fixed (see also Fig.~\ref{fig:lambda_mu_nu_grid}).

Transforming (\ref{eq:sephelmholtz}) to the coordinates $(\nu,\mu,\lambda)$ leads to
\begin{equation}
\label{eq:helmreg}
-  \frac{\hbar^2}{2m}  \frac{\mbox{d}^2}{\mbox{d} \hat{s}^2}\psi_{\hat{s}}(\hat{s}) 
= 
\sigma_{\hat{s}}
\frac{E}{a^2}\left(s^4(\hat{s}) - 2k s^2(\hat{s}) + l\right) \, \psi_{\hat{s}}(\hat{s})\,,
\end{equation}
where 
$\hat{s}\in \{\nu,\mu,\lambda\}$,
$s(\hat{s}) \in \{\zeta(\nu),\eta(\mu),\xi(\lambda)\}$ are the functions from~(\ref{eq:regularization}) and the $\sigma_{\hat{s}}$ are the signs
$\sigma_{\lambda}=\sigma_{\nu}=+$ and $\sigma_{\mu}=-$. 
Each of these equations can be interpreted as a
one-dimensional Schr\"odinger equation with a Hamiltonian of the standard type $H=-\hbar^2(\mbox{d}^2/\mbox{d}x^2)/2+V$ (``kinetic plus potential energy'') with effective energy and potential
\begin{equation}
\label{eq:effenerpot}
E_{\hat{s},\textrm{eff}}  = \sigma_{\hat{s}}\frac{E}{a^2} l\,,\quad
V_{\hat{s}, \textrm{eff}}(\hat{s})  =
-\sigma_{\hat{s}}\frac{E}{a^2}(s^4(\hat{s})-2ks^2(\hat{s}))\,.
\end{equation}
The effective energies and potentials (\ref{eq:effenerpot}) are shown for representative (again see Sec.~\ref{sec:classical}) values of the separation constants $k$ and $l$ in 
Fig.~\ref{fig:pots_xfig_c}(a) of Sec.~\ref{sec:classical}, where  $\mu$ and $\nu$ vary in intervals of length $2K(q')$ and $2K(q)$, which are the periods of the effective potentials 
$V_{\mu, \textrm{eff}}$ and $V_{\nu, \textrm{eff}}$, respectively.

The reflection symmetry of the effective potential $V_{\lambda, \textrm{eff}}$ about $\lambda=0$ leads to solutions of \eqref{eq:helmreg} that are symmetric or antisymmetric under this reflection. Similar to the 2D case we can use \eqref{eq:xzy_numulambda} to relate the behaviour of solutions under this reflection to the parity $\pi_x$.
The effective potential $V_{\mu, \textrm{eff}}$ is symmetric about $\mu=K(q')$, and using \eqref{eq:helmreg} the symmetry or antisymmetry of solutions of \eqref{eq:helmreg} under the corresponding reflection $\mu\mapsto 2K(q')-\mu$ can be related to the parity $\pi_z$. The effective potential $V_{\nu, \textrm{eff}}$ has reflection symmetry about $\nu=0$ and $\nu=K(q)$. Eq.~\eqref{eq:xzy_numulambda} relates the symmetry or antisymmetry of the solutions under the corresponding reflections $\nu\mapsto -\nu$ and $\nu\mapsto 2K(q)-\nu$ to parities $\pi_y$ and $\pi_z$, respectively. 
We note that, like their algebraic counterparts \eqref{eq:sephelmholtz}, the wave equations \eqref{eq:helmreg} for $\nu$, $\mu$ and $\lambda$ are identical, if one considers them on different intervals (in the complex plane). The equations for $\mu$ and $\lambda$ can, e.g., be related to the equation for $\nu$ using the identities  $\sn(u+K(q)+\ui K(q'),q)=q^{-1}\dn(u,q)/\cn(u,q)$ and $\sn(-\ui u+K(q)+\ui    K(q'),q)=q^{-1}\dn(u,q')$  in \eqref{eq:regularization}.  This is similar to the statement on the wave equations \eqref{eq:helmreg_2D} in the 2D case. 

%% file: sec4_classical.tex
\section{The classical systems}
\label{sec:classical}

We will now study the classical dynamics of the transmission problem described in Sec.~\ref{sec:transprob}.
As mentioned in Sec.~\ref{sec:transprob} 
the classical motions consist of motions along straight lines in the regions \eqref{eq:2Dregion} and \eqref{eq:3Dregion} 
with specular reflections at the boundary hyperbola  and hyperboloid, respectively.
Like the Helmholtz equations with the Dirichlet boundary conditions  imposed along the boundary hyperbola and hyperboloid the classical equations of motion can also be separated in elliptic (2D) and ellipsoidal coordinates (3D).
The separability implies that the classical dynamics is integrable, i.e. there are as many constants of the motion (the separation constants) that are independent and in involution as degrees of freedom.  
A modification of the Liouville-Arnold theorem  \cite{Arnold78} says that the  space of the classical motion is (up to singular sets of measure zero) foliated by invariant cylinders (the analogues of invariant tori in closed systems). In the following we will have a closer look at these foliations for both the 2D and 3D system.

\subsection{The 2D system}
\label{subsection:hyperbola_classical}

\subsubsection{Phase space foliation}

Separating the equations of motions for the free motion in the plane in the elliptic coordinates $(\zeta,\xi)$
introduced in Sec.~\ref{subsec:sep2D} yields that the momenta $p_s$ conjugate to $s$, $s\in\{\zeta,\xi\}$, are given by
\begin{equation}\label{eq:sep_s_2D}
p_s^2 = 2mE \frac{s^2 - s_2^2 }{s^2-a^2}
\end{equation}
(see \cite{WWD97}), where $s_2^2$ is a separation constant which acts as the square of the turning point of the respective degree of freedom $s\in\{\zeta,\xi\}$. 
These equations are the analogues of the separated Helmholtz equations in the algebraic form  \eqref{eq:sephelmholtz_2D}.
Similarly, for the coordinates $\hat{s}\in(\nu,\lambda)$ and their conjugate momenta $p_{\hat{s}}$, the analogue of the regularized separated Helmholtz equations  \eqref{eq:helmreg_2D} are given by
\begin{eqnarray}
\label{eq:sep_s_hat_2D}
p_{\hat{s}}^2 &=& \sigma_{\hat{s}}2mE (s^2(\hat{s})-s_2^2)  \\ \label{eq:sep_s_hat_2D_eff_E_V}
&=& 2m (E_{\hat{s},\textrm{eff}} -V_{\hat{s}, \textrm{eff}}(\hat{s}))\,,
\end{eqnarray}
where $\hat{s}\in\{\nu,\lambda\}$ and
 $s(\hat{s})\in\{\zeta(\nu),\xi(\lambda)\}$   in \eqref{eq:sep_s_hat_2D} are the functions defined in \eqref{eq:regularization_2D}, and 
the effective energy and potential in \eqref{eq:sep_s_hat_2D_eff_E_V} are defined as in \eqref{eq:effenerpot_2D}.

The specular reflection at the 2D  boundary hyperbola $\zeta=1$ or equivalently $\nu=\nu_{\text{B}}$ and $\nu=\pi-\nu_{\text{B}}$ is described by 
mapping the phase space coordinates right before the reflection to the phase space coordinates right after the reflection according to
\begin{equation}
(\zeta,\xi,p_\zeta,p_\xi)\mapsto (\zeta,\xi,-p_\zeta,p_\xi) 
\end{equation}
or 
\begin{equation}
(\nu,\lambda,p_\nu,p_\lambda)\mapsto (\nu,\lambda,-p_\nu,p_\lambda)\,,
\end{equation}
respectively. As opposed to the phase space coordinates $(s,p_s)$, $s\in\{\zeta,\xi\}$ the phase space coordinates $(\hat{s},p_{\hat{s}})$, $\hat{s}\in\{ \nu,\lambda \}$ lead to a smooth description of the motion (apart from the specular reflections).

\begin{figure}
\centerline{
\includegraphics[angle=0,width=8.6cm]{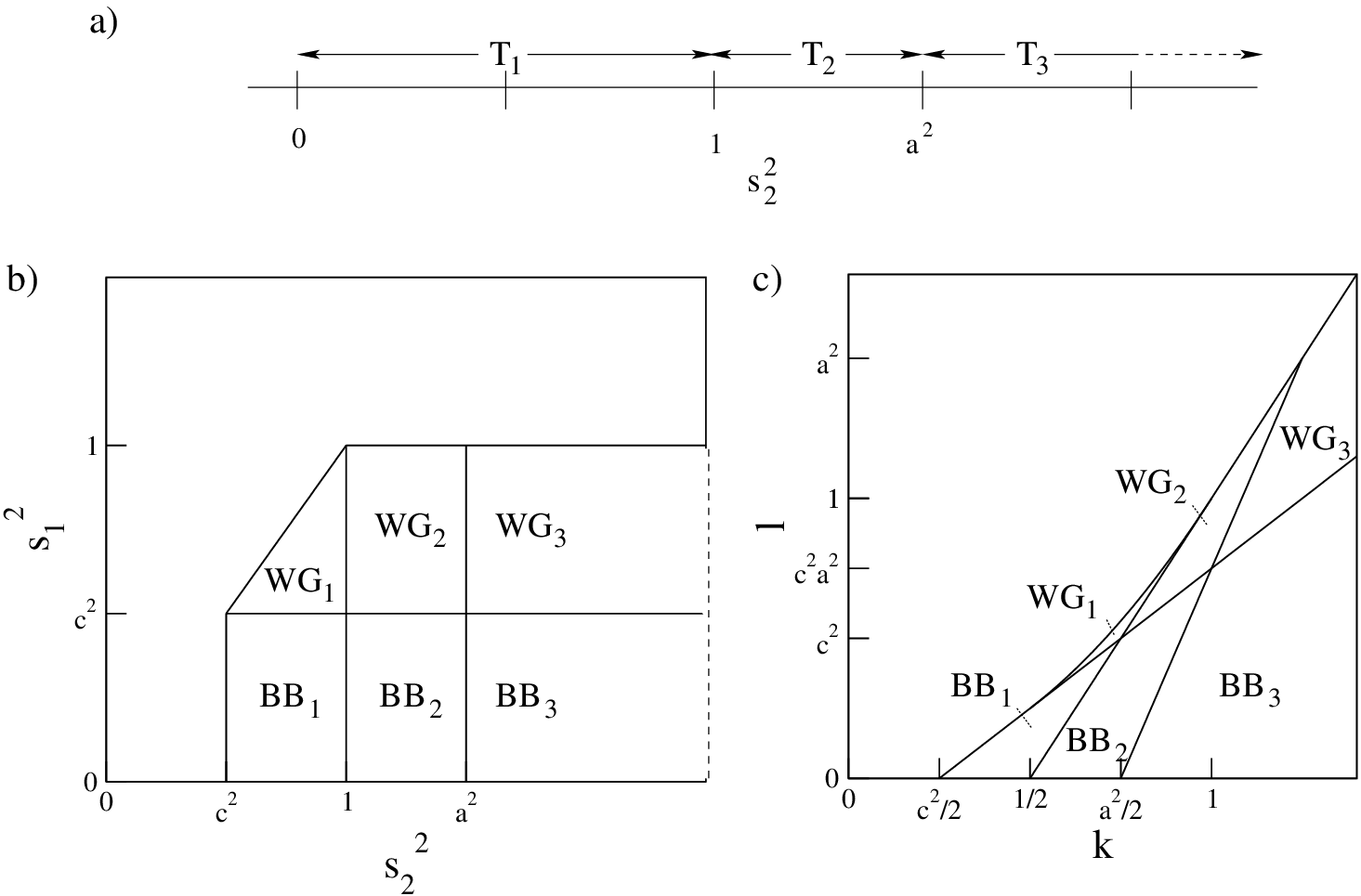}
}
\caption{\label{fig:bifdiag_a_b}
Bifurcation diagram in terms of the variable $s_2^2$ for the 2D system, with $a^2=3/2$ (a),  and in terms of the variables $(s_1^2,s_2^2)$ (b) and $(k,l)$ (c) for the 3D system, with $(a^2,c^2)=(3/2,1/2)$.
}
\end{figure}

\begin{figure}
\begin{displaymath}
\includegraphics[angle=0,width=14cm]{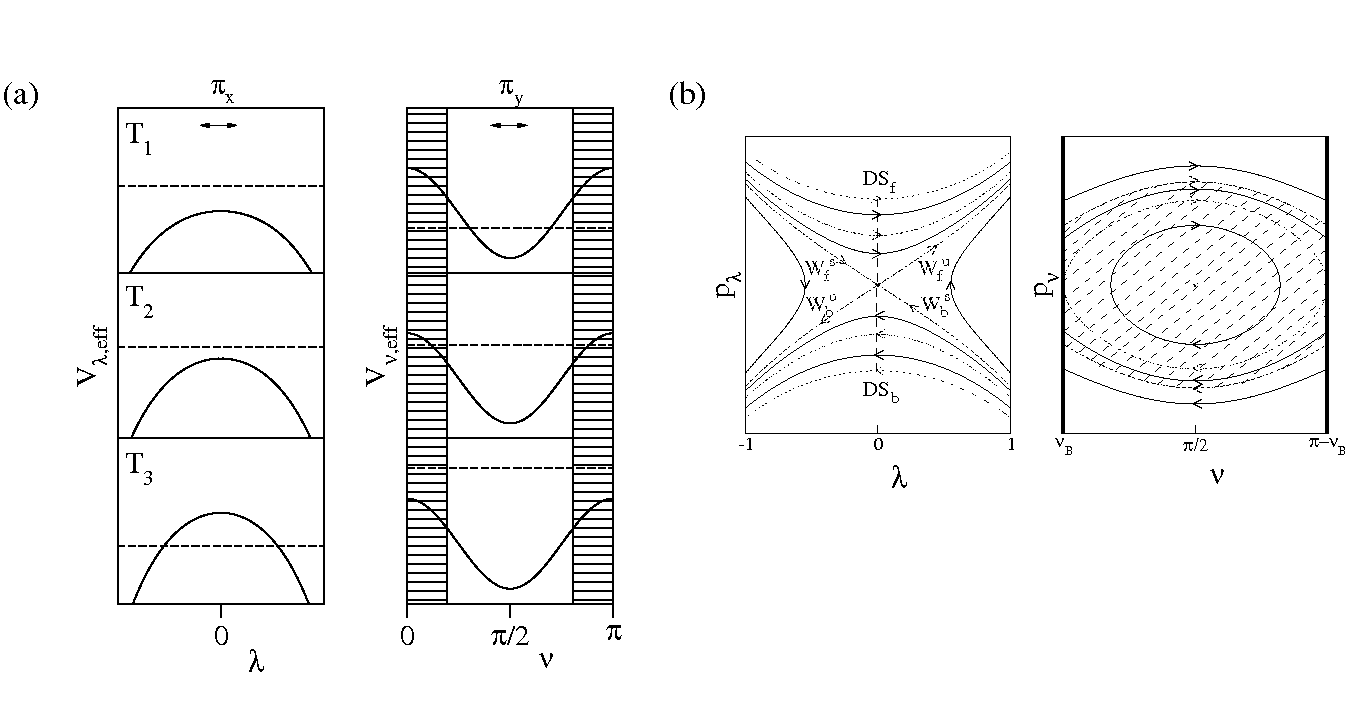}
\end{displaymath}
\caption{\label{fig:pots_xfig_c_2D}
(a) Effective potentials and energies for the types of motion 
T$_1$, T$_2$ and T$_3$  defined in
Fig.~\ref{fig:bifdiag_a_b}(a). For the $\nu$ degree of freedom the hatched regions mark the forbidden regions $[0,\nu_{\text{B}}]$ and $[\pi-\nu_{\text{B}},\pi]$
which are not contained in the region  (\ref{eq:2Dregion}). (b) Phase curves parametrized by $s_2^2$ (with $E=\text{const.}>0$).  
For $s^2_2=0$, we have the forward and backward reaction paths (the free flight motions along the $x$ axis) which correspond to the branches of the green solid hyperbola in the left panel and the green central dot in the right panel. 
For a fixed $s_2^2\in(0,1)$, we have two invariant cylinders of forward and backward reactive trajectories which do not involve specular reflections at the boundary hyperbola \eqref{eq:hyper_2D}. These appear as the two branches of the black solid `horizontal' hyperbola in the left panel  and the inner black solid circle in the right panel. 
The two cylinders which have $s_2^2=1$ and have forward and backward  trajectories that touch the boundary hyperbola 
\eqref{eq:hyper_2D}  tangentially are marked by the red dashed curves in either panel. 
A fixed $s_2^2\in (1,a^2)$ represents two cylinders which have forward and backward reactive trajectories that involve specular reflections at the boundary hyperbola.  These are marked by the two branches of the black solid `vertical' hyperbola in the left panel and the corresponding chopped circle in the right panel. 
The value $s_2^2=a^2$ represents the periodic orbit (or `transition state')  TS which corresponds to the origin in the left panel and the corresponding blue chopped circles in the right panel, and its stable and unstable manifolds with their forward and backward branches $W^{s/u}_{f,b}$ forming the blue cross in the left panel and coinciding with the blue chopped circles in the right panel.
 A fixed $s_2^2>a^2$ represents two cylinders of nonreactive trajectories on the reactants side ($\lambda<0$) and products side ($\lambda>0$), respectively. These correspond to the two branches of the black solid `vertical' hyperbola in the left panel 
 and the corresponding chopped circle in the right panel. 
 The black dashed lines in the left panel mark the forward ($p_\lambda>0$) and backward ($p_\lambda<0$) dividing surfaces 
 DS$_{\text{b/f}}$.  In the right panel these appear as the hatched chopped disk. ($a^2=5$.)
}
\end{figure}

\todo{In Fig.~\ref{fig:pots_xfig_c_2D}: Mark TS, $W^{s/u}_{f/b}$, DS$_{\text{f/b}}$.}

Like in the quantum case in Sec.~\ref{subsec:sep2D} we can also introduce a symmetry reduced system in the classical case.
For the symmetry reduced system the  motion is confined to the positive $x-y$ quadrant of the region \eqref{eq:2Dregion} with specular reflections not only at the boundary hyperbola \eqref{eq:hyper_2D} but also at the Cartesian coordinate axes.

The physical meaning of the separation constant $s^2_2$ becomes more clear from multiplying it with $(2mE)^{1/2}$ and expressing it in terms of Cartesian coordinates. A little bit of algebra then gives
\begin{equation}\label{eq:2D_sep_const}
\sqrt{2mE} s_2^2 = L_z^2 +a^2p_y^2 = \frac12 \big( L_{z-}^2 + L_{z+}^2\big)\,,
\end{equation}
where $L_z=xp_y-yp_x$ is the angular momentum about the origin, and $L_{z-}=(x+a)p_y-yp_x$ and $L_{z+}=(x-a)p_y-yp_x$ are the angular momenta about the focus points $(x,y)=(0,\pm a)$.
This is the second constant of the motion beside the energy which makes the system integrable. 
A modification of the  Liouville-Arnold theorem then implies that the four dimensional phase space is foliated by invariant cylinders (see below) which are given by the common level sets of the constants of motion $E$ and \eqref{eq:2D_sep_const}
or equivalently $E$ and $s^2_2$.
 In fact the energy plays no major role for the motions. 
It just determines the speed of the motion along the straight lines (in configuration space). As indicated by the occurrence of the energy  as a multiplicative factor in the equations for the separated momenta \eqref{eq:sep_s_2D} and \eqref{eq:sep_s_hat_2D},  energy surfaces of different positive energies only differ by the scaling of the momenta, and accordingly they all have the same type of foliation by invariant cylinders. To discuss the foliations of the energy surfaces it is  thus sufficient to consider a single energy surface of fixed energy $E>0$. 
The different types of cylinders contained in the energy surface of this energy are then parametrized by the second constant of motion, $s_2^2$, in the following way.
 
First of all, in order to simultaneously have real momenta in the physical ranges $\zeta\in [0,a] $ and $\xi \in [a,\infty)$ (see \eqref{eq:eqb3d_xezranges_2D}) the separation constant $s^2_2$ can only take nonnegative values.  We therefore will occasionally write $s_2=\sqrt{s_2^2}$.
The interval $[0,\infty)$ 
contains three subintervals which correspond to different smooth families of cylinders
which we denote by 
\begin{equation}
\begin{split}
T_1&: \quad 0<s_2^2 <1\,,\\
T_2&: \quad 1<s_2^2 <a^2\,, \\
T_3&: \quad a^2<s_2^2\,.
\end{split}
\end{equation}
At the values $s^2_2=0$, $s^2_2=1$ and $s^2_2=a^2$ the families of cylinders bifurcate, and these parameter values thus present critical motions to which we will come back below (also see the bifurcation diagram in Fig.~\ref{fig:bifdiag_a_b}).  

\begin{figure}
\centerline{
\raisebox{2cm}{T$_1$}
\includegraphics[angle=0,height=2cm]{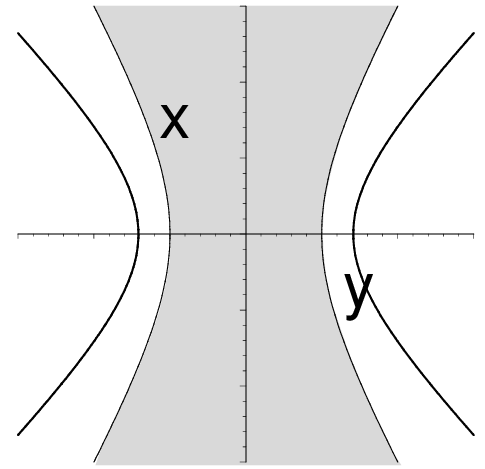}
\raisebox{2cm}{T$_2$}
\includegraphics[angle=0,height=2cm]{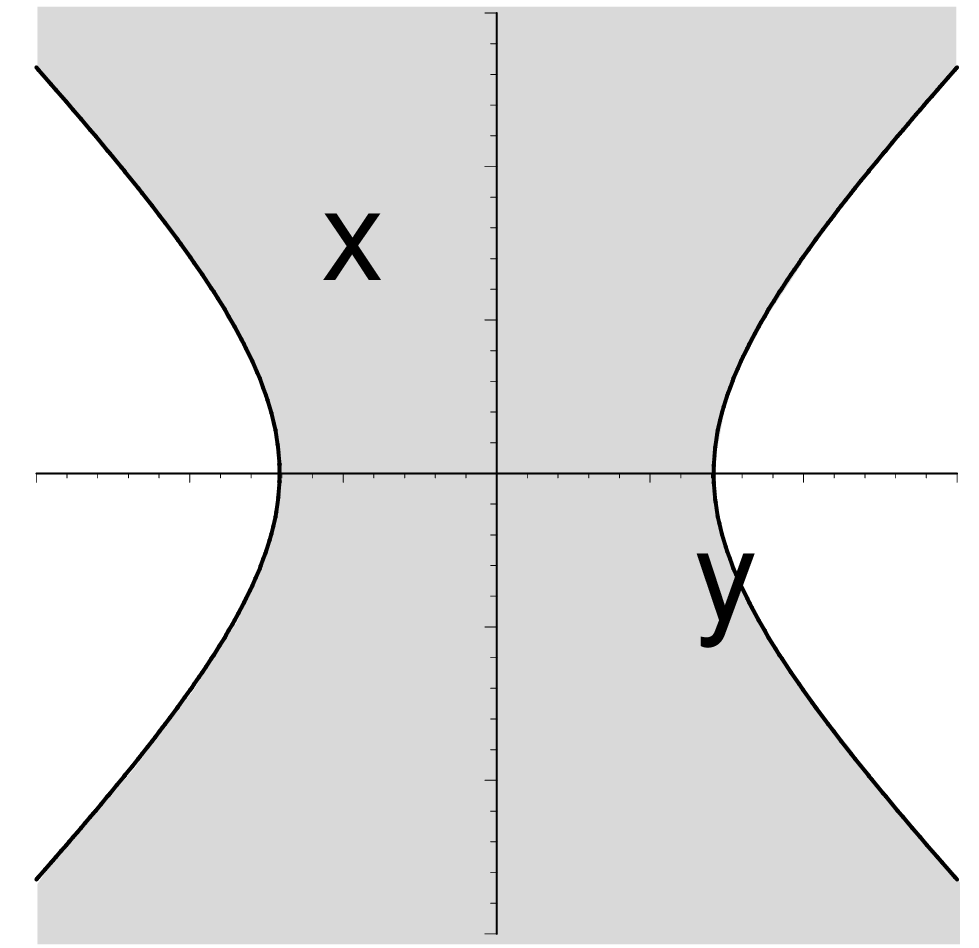}
\raisebox{2cm}{T$_3$}
\includegraphics[angle=0,height=2cm]{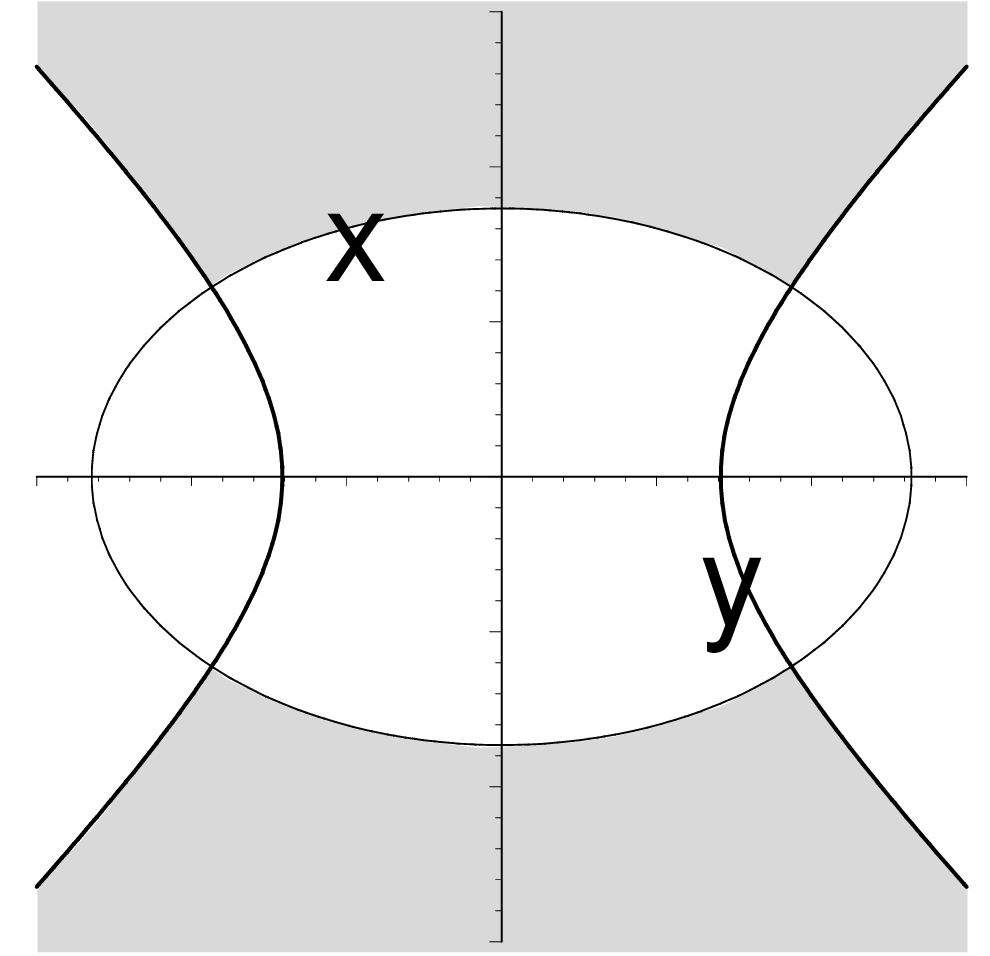}
}
\caption{
\new{\label{fig:caustics_2D}
Configuration space projections (shaded regions) of the invariant cylinders corresponding to the three types of 2D  motions T$_1$, T$_2$ and T$_3$. The bold lines mark the  boundary  hyperbola (\ref{eq:hyper_2D}). The ranges for $x$ and $y$ are both $[-3/2,3/2]$. ($a^2=3/2$.)
}
}
\end{figure}

To understand the  motions on the different types of cylinders T$_1$, T$_2$  and $T_3$ it is useful to consider the corresponding effective potentials and energies \eqref{eq:effenerpot_2D} and
phase portraits in the planes $\nu-p_\nu$ and $\lambda-p_\lambda$ in  Fig.~\ref{fig:pots_xfig_c_2D} in combination with 
the  projections of the cylinders to configuration space which are shown in Fig.~\ref{fig:caustics_2D}. 

The common level set of the constants of motion $E$ and $s_2^2$ in T$_1$ consists of two disjoint cylinders which both extend over all values of $x$. On one of these cylinders $p_\lambda$ is always greater than zero, and on the other $p_\lambda$ is always less than zero. 
These cylinders are thus foliated by forward and backward reactive trajectories, respectively.
The motion oscillates in the $\nu$ degree of freedom in such a way that the  trajectories do not hit the boundary hyperbola \eqref{eq:hyper_2D}. The topology of the cylinders, $\R\times \sphere^1$, becomes apparent from taking the Cartesian product of the lines ($\sim \R$) in the phase plane $\lambda-p_\lambda$ in the left panel of Fig.~\ref{fig:pots_xfig_c_2D}(b) with the corresponding topological circle in the phase plane $\nu-p_\nu$ in the right panel of Fig.~\ref{fig:pots_xfig_c_2D}(b).

Similarly, a common level set of the constant of motion in  T$_2$ consists of two disjoint cylinders of which one again consists of forward reactive trajectories and the other again consists of backward reactive, but 
the oscillations in the transverse degree of freedom $\nu$
now involve reflections at the boundary hyperbola. To simplify the discussion, we will glue together the two line segments in the $\nu-p_\nu$ plane in the right panel of Fig.~\ref{fig:pots_xfig_c_2D}(b) which have positive and negative $p_\nu$, respectively, at the points $\nu=\nu_{\text{B}}$ and $\nu=\pi-\nu_{\text{B}}$, i.e. at  $\nu=\nu_{\text{B}}$ and $\nu=\pi-\nu_{\text{B}}$ we identify $p_\nu$ and $-p_\nu$. Note that strictly speaking the momenta are not defined along the boundary hyperbola. However, the gluing   can also be justified from physical considerations by viewing the hard wall potential which causes the reflections as the limiting case  of a smooth potential that becomes steeper and steeper.
The resulting object can then again be viewed as a topological circle, $\sphere^1$, and taking the Cartesian products with the corresponding lines $\sim \R$ in the $\lambda-p_\lambda$ planes we again obtain topological cylinders $\R\times\sphere^1$ 
similar to those in T$_1$.

In contrast to the cylinders above, the common level set of the constants of motion  in  T$_3$ consists of two disjoint cylinders which when projected to configuration space are both bounded away from the $y$ axis by the ellipse $\xi^2=s_2^2$.
These cylinders are foliated by nonreactive trajectories which stay on the side of reactants and products, respectively.

The critical value  $s_2^2=1$ corresponds to the limiting motion between T$_1$ and T$_2$. The level set of the constants of motion  $E>0$ and $s^2_2=1$ consists of two disjoint cylinders which contain forward and backward reactive trajectories, respectively, which hit the boundary hyperbola \eqref{eq:hyper_2D} tangentially (see the dotted line in the right panel of Fig.~\ref{fig:pots_xfig_c_2D}(b)). 

At the critical value $s_2^2=0$ the two cylinders of type T$_1$ degenerate to two lines $\sim \R$ given by the Cartesian products of the dot at the centre of the $\nu-p_\nu$ plane in the right panel of Fig.~\ref{fig:pots_xfig_c_2D}(b) with the corresponding lines in the $\lambda-p_\lambda$ plane in the left panel of the same figure. One of these lines corresponds to a trajectory along the $x$ axis which has $p_x=(2mE)^{1/2}$; the other line corresponds to a trajectory along the $x$ axis which has  $p_x=-(2mE)^{1/2}$. These can be viewed as the forward and backward \emph{reaction paths}, i.e. they are the unique trajectories, which for a fixed energy, are reactive and do not involve any motion in the transverse degree of freedom \cite{WaalkensBurbanksWigginsb04,WaalkensSchubertWiggins08}. 

The critical value $s^2_2=a^2$ represents the unstable periodic orbit along the $y$ axis which, for a fixed energy $E>0$, bounces back and force between the two branches of the boundary hyperbola \eqref{eq:hyper_2D}.
The common level set of $E>0$ and $s_2^2=a^2$ consists not only of this periodic orbit but also of  the stable and unstable manifolds $W^{s}$ and $W^{u}$ of this periodic orbit.  In the $\lambda-p_\lambda$ phase plane in the left panel of Fig.~\ref{fig:pots_xfig_c_2D}(b)
the stable and unstable manifolds occur as the cross shaped structure which has the periodic orbit at the center.
With our interpretation of reflections at the boundary hyperbola to be smooth the
periodic orbit has the topology $\sphere^1$, and its stable and unstable manifolds are cylinders $\R\times \sphere^1$. The stable and unstable manifolds are of special significance for the classical transmission since they are of codimension 1 in the energy surface, i.e. they have one dimension less than the energy surface, and this way have sufficient dimensionality  to act as impenetrable barriers in the energy surface \cite{WWJU01}.   In fact, the stable and unstable manifolds form the separatrices between reactive and nonreactive trajectories. More precisely, $W^s$ and $W^u$ each have two branches: we denote the branch of $W^s$ which has $p_\lambda>0$ (resp. $p_\lambda<0$) the \emph{forward} (resp. \emph{backward}) branch, $W_f^s$ (resp. $W_b^s$), of the stable manifold. Similarly we denote the branch of $W^u$ which has $p_\lambda>0$ (resp. $p_\lambda<0$) the \emph{forward} (resp. \emph{backward}) branch, $W_f^u$ (resp. $W_b^u$), of the unstable manifold (see Fig.~\ref{fig:pots_xfig_c_2D}(b)). Moreover, we call the union of the forward banches,
\begin{equation}
W_f   := W^s_f \cup W^u_f\,,
\end{equation}
the \emph{forward reactive cylinder}, and the union of the backward branches, 
\begin{equation}
W_b  := W^s_b \cup W^u_b\,,
\end{equation}
the \emph{backward reactive cylinder}. The forward reactive cylinder encloses all trajectories in an energy surface of the respective energy  $E>0$ which are forward reactive; the backward reactive cylinder encloses all trajectories in such an energy surface  which are backward reactive. The nonreactive trajectories are contained in the complement of these regions. The forward and backward reactive cylinders thus play a crucial role for the classification of trajectories with respect to their reactivity. They can be viewed to form the phase space conduits for reaction. In particular, the forward and backward reaction paths mentioned above can be viewed to form the centerlines of the regions enclosed by these cylinders. 

The periodic orbit, or more precisely, the family of periodic orbits oscillating along the $y$ axis with different energies $E>0$ can be viewed to form the \emph{transition state} or \emph{activated complex}. Reactive trajectories of a given energy $E>0$ pass `through' the periodic orbit  at that energy (the `transition state at energy $E$') in the following sense. Setting $x=0$ on the energy surface defines  a two-dimensional surface in the energy surface which is given by 
\begin{equation} \label{eq:def_DS_2D}
\begin{split}
\text{DS} &= \{ (x,y,p_x,p_y) \in \R^4\,:\, x=0, \, y\in [-1,1], \,\\
&\phantom{=} p_x^2+p_y^2=2mE  \} \\
&=  \{ (\nu,\lambda,p_\nu,p_\lambda) \in \R^4\,:\, \lambda=0, \, \nu \in [\nu_{\text{B}},\pi-\nu_{\text{B}}], \,\\
&\phantom{=} p_\lambda^2+p_\nu^2=2mEa^2(1-a^2\cos^2 \nu)  \} \,.
\end{split}
\end{equation}
With our convention to identify the momenta $-p_\nu$ and $+p_\nu$ at $\nu=\nu_{\text{B}}$ and $\nu=\pi-\nu_{\text{B}}$, the surface DS has the topology of a  two-dimensional sphere, $\sphere^2$.
It defines a so called \emph{dividing surface} that has all the desired properties that are crucial for the transition state computation of the classical transmission probability from the flux through a dividing surface. First of all, it divides the energy surface into a reactants part ($x<0$) and a products part ($x>0$). In order to be reactive a trajectory thus has to intersect the dividing surface.
In fact the periodic orbit or transition state at energy $E$ given by
\begin{equation} \label{eq:TS2D}
\text{TS} = \{(x,y,p_x,p_y)\in \R^2\times\R^2 \,: \,   x=0, \,p_x=0,\, y\in[-1,1],  \, p_y^2=2m E  \}
\end{equation}
can be viewed to form the equator of the dividing surface \eqref{eq:def_DS_2D}. It separates the dividing surface into two hemispheres which we call the 
the \emph{forward dividing surface}
\begin{eqnarray} \label{eq:def_DS_f_2D}
\begin{split}
\text{DS}_{\text{f}} &= \{ (x,y,p_x,p_y) \in \R^4\,:\, x=0, \, y\in [-1,1], \,\\
&\phantom{=} p_x^2+p_y^2=2mE, \,p_x>0 \}\\
&=  \{ (\nu,\lambda,p_\nu,p_\lambda) \in \R^4\,:\, \lambda=0, \, \nu \in [\nu_{\text{B}},\pi-\nu_{\text{B}}], \,\\
&\phantom{=} p_\lambda^2+p_\nu^2=2mEa^2(1-a^2\cos^2 \nu) ,\,p_\lambda>0 \} \,.
\end{split}
\end{eqnarray}
and the \emph{backward dividing surface}
\begin{eqnarray} \label{eq:def_DS_b_2D}
\begin{split}
\text{DS}_{\text{b}} &= \{ (x,y,p_x,p_y) \in \R^4\,:\, x=0, \, y\in [-1,1], \,\\
&\phantom{=} p_x^2+p_y^2=2mE, \,p_x<0 \} \\
&=  \{ (\nu,\lambda,p_\nu,p_\lambda) \in \R^4\,:\, \lambda=0, \, \nu \in [\nu_{\text{B}},\pi-\nu_{\text{B}}], \,\\
&\phantom{=} p_\lambda^2+p_\nu^2=2mEa^2(1-a^2\cos^2 \nu) ,\,p_\lambda<0 \} \,.
\end{split}
\end{eqnarray}
These two hemispheres appear in the right panel of Fig.~\ref{fig:pots_xfig_c_2D}(b) as the disk enclosed by the blue curve that represents the transition state periodic orbit TS in the $\nu-p_\nu$ plane. Note that the circles contained in this disk have to be combined with the two corresponding lines in the $\lambda-p_\lambda$ plane in the right panel of Fig.~\ref{fig:pots_xfig_c_2D}(b)  which have $p_\lambda>0$ (corresponding to forward reactive trajectories) or $p_\lambda<0$ (corresponding to backward reactive trajectories).
All forward reactive trajectories have a single intersection with the forward dividing surface, and all backward reactive trajectories have a single intersection with the backward dividing surface. Nonreactive trajectories do not intersect the dividing surface at all. The dividing surface is everywhere transverse to the Hamiltonian flow apart from its equator, which is a periodic orbit and thus is  invariant under the Hamiltonian flow.

\subsubsection{Action integrals}

In the previous subsection we have seen that the phase space is (up to critical motions which form a set of measure zero) foliated by invariant cylinders where the cylinders are given by the Cartesian products of circles in $(\nu,p_\nu)$ and unbounded lines in 
$(\lambda,p_\lambda)$. For the  $(\nu,p_\nu)$ component of the motions we can directly introduce action-angle variables \cite{Arnold78}. 
As we will see below we can also associate an action type integral with the unbounded
$(\lambda,p_\lambda)$ component of the motion. Both of these actions will play a role in the semiclassical  
computation of the cumulative reaction probability  and the quantum resonances (see Sections~\ref{sec:cum_reac_prob} and \ref{sec:resonances}, respectively).

Action integrals depend on the type of motion, and typically change from one type of motion to another. For the actions associated with the $\nu$ or equivalently $\zeta$ degree of freedom we find
\begin{equation}
\label{eq:action_ints_2D}
I_\zeta=\frac{1}{2\pi}\oint p_\zeta \ud \zeta = \sqrt{2mE} \frac{4}{2\pi}\int_{\zeta_{-}}^{\zeta_{+}}  \sqrt{ \frac{\zeta^2 - s_2^2 }{\zeta^2-a^2} } \ud \zeta\,,  
\end{equation}
where we took $p_\zeta$  from (\ref{eq:sep_s_2D})
and the integration boundaries $\zeta_{-}$ and $\zeta_{+}$ 
are given by $\zeta_-=0$ and  $\zeta_+=s_2$ for motions of type T$_1$ and $\zeta_+=1$ for motions of type T$_2$ and T$_3$.
The corresponding action integral for the symmetry reduced system, which we denote by $\tilde{I}_\zeta$, is given by 
\begin{equation}
\label{eq:action_ints_red_2D}
\tilde{I}_\zeta= \frac12 I_\zeta \,.
\end{equation}

\begin{figure}
\centerline{
\includegraphics[angle=0,width=6cm]{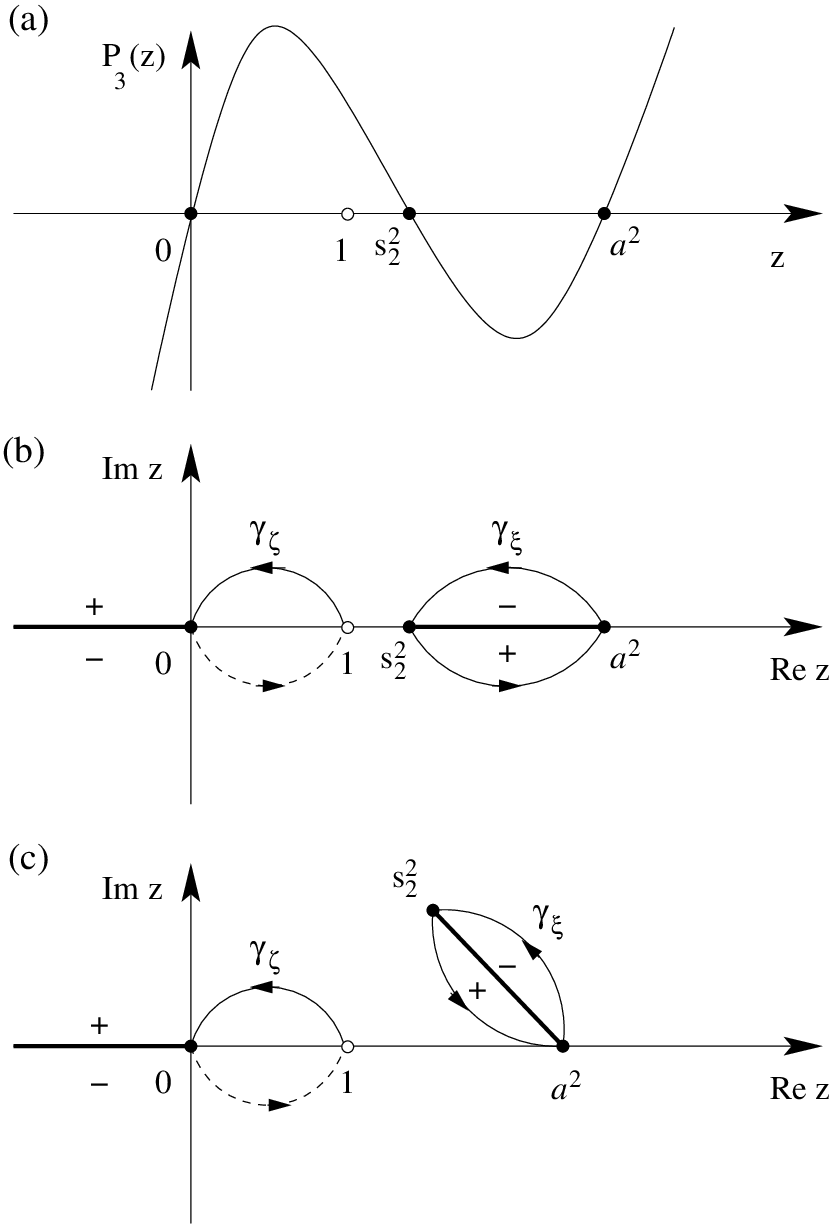}
}
\caption{\label{fig:ellcurve_res_2D}
(a) The graph of the polynomial $P_{3}$ defined in (\ref{eq:elliptic_curve}) for a real separation constant $s_2^2$ satisfying $1<s_2^2<a^2$ (motion type T$_2$). (b) Elliptic curve $\Gamma_w$ for $w^2=P_3(z)$ with $P_3$ as in (a), 
and integration paths $\gamma_\zeta$ and $\gamma_\xi$ which define the integrals in (\ref{eq:I_s_on_Gamma}) and (\ref{eq:def_I_xi_2D}). 
Only one half (Riemann sheet) of the elliptic curve is shown. 
This half is obtained from introducing the branch cuts which connect the branch points $-\infty$ and $0$, 
and $s_2^2$ and  $a^2$ (bold lines along the real axis). 
The signs below and above these branch cuts indicate the value of $w$ `right above' and `right below' the branch cut. At `$+$',  $w$ has the value $+\ui |w|$;    at `$-$',  $w$ has the value $-\ui |w|$.
The integration path $\gamma_\xi$ is a closed loop which encircles the right branch cut. The integration path $\gamma_\zeta$ consists of two parts. It starts at $z=1$ on the shown Riemann sheet of $w$ and intersects the branch cut between 0 and $-\infty$. This part is marked by a solid line. The integration path continues on the other Riemann sheet (which is a copy of the one shown and which joins this copy at the branch cuts) and ends at the point 1 on the other sheet.  This part is shown as the dashed line.
(c) The continuation of  (b) for $s_2^2$ leaving the real axis (see Sec.~\ref{sec:resonances}).
}
\end{figure}

To understand the analytic nature of the action integral $I_\zeta$ we substitute $z=\zeta^2$ in Eq.~(\ref{eq:action_ints_2D}) which gives 
\begin{equation}
\label{eq:riemann_ints_2D}
I_\zeta= \sqrt{2mE} \frac{  1}{\pi}\int_{z_-}^{z_+}\frac{(z-s_2^2)}{w(z)} \ud z\,,
\end{equation}
where
\begin{equation}
\label{eq:riemann_poly_2D}
w^2(z)=P_3(z):=\prod_{i=1}^{3}(z-z_i)\,,
\end{equation}
and $z_-,z_+$ are consecutive elements of the set $\{z_1=0,z_2=a^2,z_3=s_2^2,z_{\text{b}}=1\}$. 
Here  $z_{\text{b}}=1$ 
corresponds to the boundary hyperbola. 
The differential $\ud z/w(z)$ has the four critical points $\{z_1,z_2,z_3,\infty\}$ which means that the integral (\ref{eq:riemann_ints_2D}) is elliptic. We refrain from expressing this integral in terms of Legendre's standard integrals \cite{Byrd71}. Instead, and for later purposes (see Sections~\ref{sec:cum_reac_prob} and \ref{sec:resonances}),
we  interpret the integral $I_\zeta$ for motions of type T$_2$ and T$_3$ as an Abelian integral on the elliptic curve
\begin{equation}
\label{eq:elliptic_curve}
\Gamma_w = \{ (s,w)\in \overline{\C}^2 \, :  \, w^2 =  P_{3}(z)  \}\,.
\end{equation}
Here $\overline{\C}$ defines the compactified complex plane (i.e. the Riemann sphere).  The algebraic curve $\Gamma_w  $ is of genus 1, i.e. it has the topology of a 1-torus.
For motions of type T$_2$ or T$_3$ 
the action $I_\zeta$ in \eqref{eq:riemann_poly_2D} can then be written as
\begin{equation} \label{eq:I_s_on_Gamma}
I_\zeta =  \sqrt{2mE} \frac{1}{2\pi}  \int_{\gamma_\zeta}  (z-s_2^2) \frac{\ud z}{w}\,,
\end{equation}
where for T$_2$, the integration path $\gamma_\zeta$ is defined as illustrated in Fig.~\ref{fig:ellcurve_res_2D}(b).
For  T$_3$, the order of  $s_2^2$ and $a^2$  along the real axis in  Fig.~\ref{fig:ellcurve_res_2D}(b) is reversed. However this does not affect the definition of $\gamma_\zeta$ for T$_3$.
Due to the billiard boundary the integration path $\gamma_\zeta$ is not closed on $\Gamma_w$, i.e. the integral $I_\zeta$ is an incomplete elliptic integral.

On $\Gamma_w$ we also define the complete elliptic integral
\begin{equation} \label{eq:def_I_xi_2D}
I_\xi =  \ui \sqrt{2mE} \frac{1}{2 \pi}  \int_{\gamma_\xi}  (z-s_2^2) \frac{\ud z}{w}
\end{equation}
and its symmetry reduced partner
\begin{equation} \label{eq:def_I_xi_reduced_2D}
\tilde{I}_\xi = \frac12 I_\xi 
\end{equation}
with the closed integration path $\gamma_\xi$ in \eqref{eq:def_I_xi_2D} defined as in Fig.~\ref{fig:ellcurve_res_2D}(b). This assigns a finite, positive real valued integral also to the 
unbounded  degree of freedom $\lambda$ or equivalently $\xi$ for motion T$_2$. For motion T$_3$ the order of $s_2^2$ and $a^2$ along the real axis in Fig.~\ref{fig:ellcurve_res_2D}(b) is reversed. The integral $I_\xi$ defined according to \eqref{eq:def_I_xi_2D} is then negative real. 
Though at first not important for the classical dynamics, this integral will play an important role in  the semiclassical computations in Sections~\ref{sec:cum_reac_prob} and \ref{sec:resonances}.
\todo{are the action integrals of second kind? Should we mention this? Should we talk about periods of Abelian integrals?}

\subsection{The 3D system}
\label{subsection:hyperboloid_classical}

\subsubsection{Phase space foliation}
\begin{figure}
\begin{displaymath}
\includegraphics[angle=0,width=12cm]{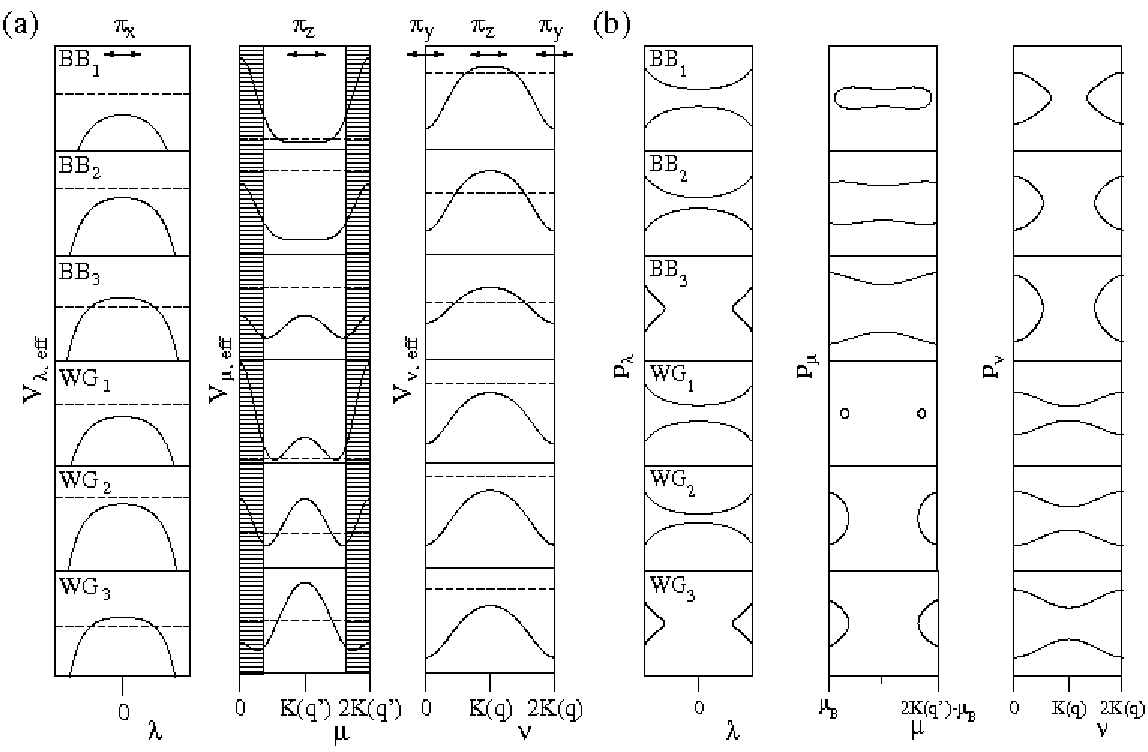}
\end{displaymath}
\caption{\label{fig:pots_xfig_c}
Effective potentials and energies (a) and phase portraits (b) for constants of motions $(s_1^2,s_2^2)$ (or equivalently $(k,l)$) in the regions 
BB$_1$, BB$_2$, BB$_3$, WG$_1$, WG$_2$ and WG$_3$  defined in
Fig.~\ref{fig:bifdiag_a_b}. For the $\mu$ degree of freedom, the hatched regions mark the forbidden regions $[0,\mu_{\text{B}}]$ and $[2K(q')-\mu_{\text{B}},2K(q')]$
which are not contained in the region (\ref{eq:3Dregion}). ($(a^2,c^2)=(3/2,1/2)$.)
}
\end{figure}
Similarly to \eqref{eq:sep_s_2D}
the separated momenta conjugate to $(\zeta,\eta,\xi)$ can be written for the 3D system as
\begin{equation}
\label{eq:sep_s}
p_s^2 =
2mE \frac{s^4-2ks^2+l}{(s^2-a^2)(s^2-c^2)}  = 2mE\frac{(s^2-s_1^2)(s^2-s_2^2)}{(s^2-a^2)(s^2-c^2)} 
\,,
\end{equation}
where $(s\in\{\zeta,\eta,\xi\})$ (see \cite{WWD99}), $k$ and $l$ are separation constants, and $s_1^2\le s_2^2$ are defined as in \eqref{eq:def_s_1_s_2}.
Since the latter are the squares of the zeroes of the numerator polynomial on the right of \eqref{eq:sep_s}, they  are the squares of  the turning points in the respective degree of freedom $s\in\{\eta,\zeta,\xi\}$. 

The corresponding equations for the coordinates $(\nu,\mu,\lambda)$ are
\begin{equation}
\label{eq:sep_s_hat}
p_{\hat{s}}^2 = \sigma_{\hat{s}} \frac{2mE}{a^2} (s^4(\hat{s})-2ks^2(\hat{s})+l)\,,
\end{equation}
where $\hat{s}\in\{\nu,\mu,\lambda\}$, and
$s(\hat{s})\in\{\zeta(\nu),\eta(\mu),\xi(\lambda)\}$ are the functions defined in (\ref{eq:regularization}).
Similarly to the 2D case,  (\ref{eq:sep_s}) and (\ref{eq:sep_s_hat}) are the classical analogues of the separated wave equations (\ref{eq:sephelmholtz}) and (\ref{eq:helmreg}), respectively.
The specular reflection at the hyperboloidal boundary $\eta=1$ or equivalently $\mu=\mu_{\text{B}}$ and $\mu=2K(q')-\mu_{\text{B}}$ becomes 
\begin{equation}
(\zeta,\eta,\xi,p_\zeta,p_\eta,p_\xi)\mapsto (\zeta,\eta,\xi,p_\zeta,-p_\eta,p_\xi)
\end{equation}
or 
\begin{equation}
(\nu,\mu,\lambda,p_\nu,p_\mu,p_\lambda)\mapsto (\nu,\mu,\lambda,p_\nu,-p_\mu,p_\lambda), 
\end{equation}
respectively. 
Note that, apart from the specular reflection, the motion described in terms of the phase space coordinates $(\hat{s},p_{\hat{s}})$, $\hat{s}\in\{ \nu,\mu,\lambda\}$ is smooth on the double cover \eqref{eq:double_cover_3D_cube}.

Expressing the separation constants $k$ and $l$, or their energy scaled counterparts $K := 2m E k$ and $L := 2 m E l$, in terms of Cartesian coordinates and
momenta gives
\begin{eqnarray}
\label{eq:def_K_L_a}
K  &=& \frac12 (|{\bf L}|^2 + (a^2+c^2)p_y^2 + a^2 p_z^2+c^2p_x^2),\\
\label{eq:def_K_L_b}
L  &=&  c^2 L_y^2 + a^2L_z^2 + a^2 c^2 p_y^2 \,,
\end{eqnarray}
where $L_x$, $L_y$ and $L_z$ denote the components of the angular momentum about the origin ${\bf L} = {\bf r}\times {\bf p}$.
The separation constants together with the total energy $E=(p_x^2+p_y^2+p_z^2)/(2m)$ 
give three constants of motion. Hence, the classical system is integrable, and a modification of the Liouville-Arnold theorem implies that the six dimensional phase space is foliated by the common level sets of $E$, $K$ and $L$, or equivalently $E$, $s_1^2$ and $s_2^2$, which are invariant  cylinders.

Like in the 2D case the energy plays no major role since it only determines the speed of the motion along the straight lines (in configuration space). So in order to describe the foliation of the energy surfaces by the invariant cylinders it is sufficient to consider the energy surface of a fixed positive energy $E>0$. Other foliations of other energy surfaces are then obtained from a suitable scaling. On such an energy surface, there are then smooth two parameter families of invariant cylinders parametrized by $s_1^2$ and $s_2^2$. The parameterization intervals of these cylinders can be obtained from 
requiring the momenta \eqref{eq:sep_s} to be real and analyzing the disposition of zeroes $s^2_1$ and $s^2_2$  relative to the poles $a^2$ and $c^2$ in \eqref{eq:sep_s}. To obtain real momenta, $s_1^2$ and hence $s_2^2$ can only take nonnegative values. Similar to the 2D case we will therefore occasionally use $s_1= \sqrt{s_1^2}$ and $s_2=\sqrt{s_2^2}$.  
It then turns out that there are six different smooth families of invariant cylinders which we denote by 
BB$_1$, BB$_2$, BB$_3$, WG$_1$, WG$_2$ and WG$_3$ 
as shown in the bifurcation diagram in Figures~\ref{fig:bifdiag_a_b}(b) and (c). 

\begin{figure}
\centerline{
\raisebox{2cm}{BB$_1$}
\includegraphics[angle=0,height=2cm]{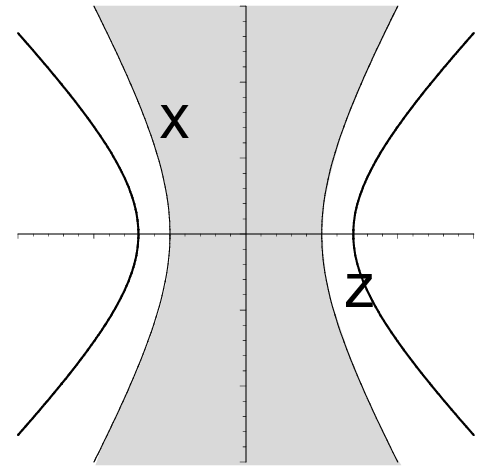}
\includegraphics[angle=0,height=2cm]{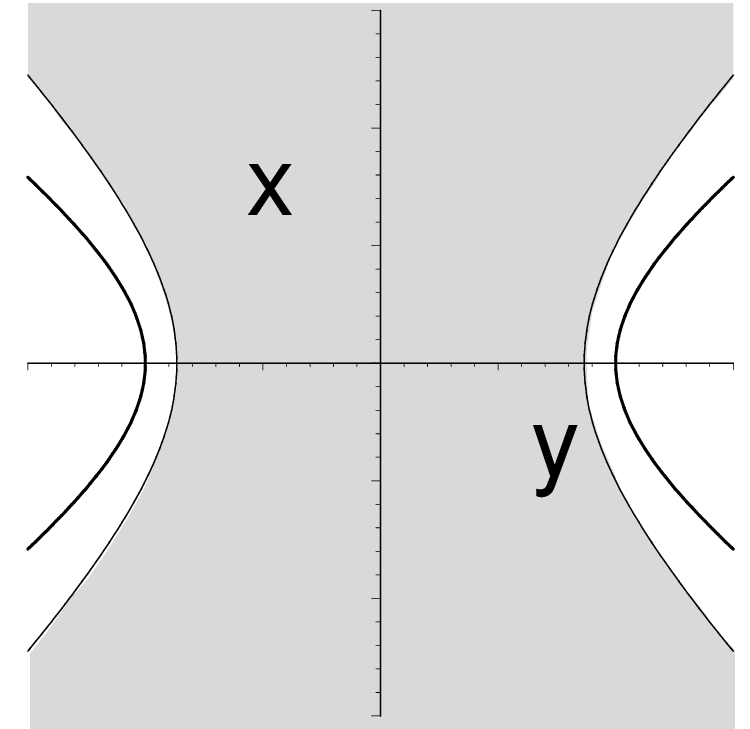}
\includegraphics[angle=0,height=2cm]{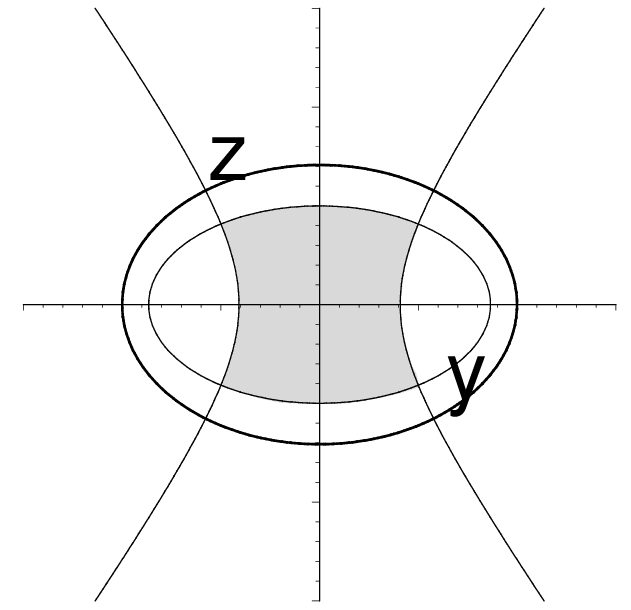}
}
\centerline{
\raisebox{2cm}{BB$_2$}
\includegraphics[angle=0,height=2cm]{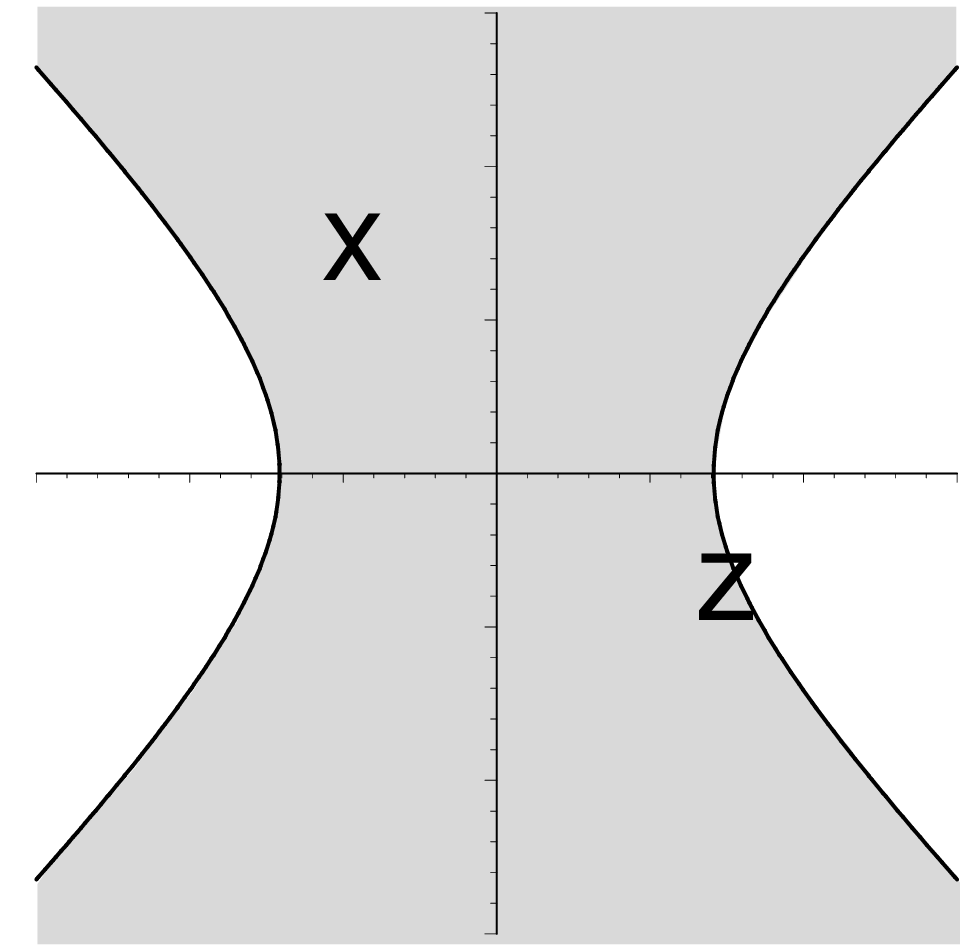}
\includegraphics[angle=0,height=2cm]{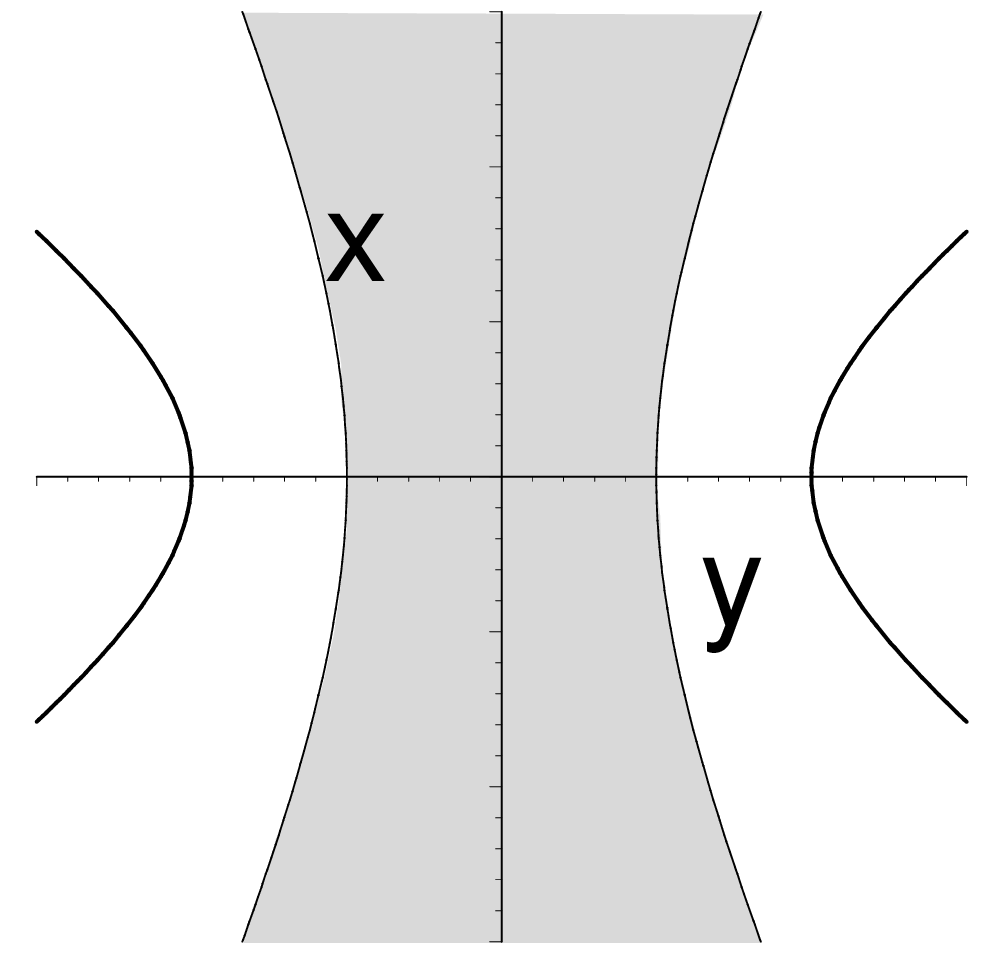}
\includegraphics[angle=0,height=2cm]{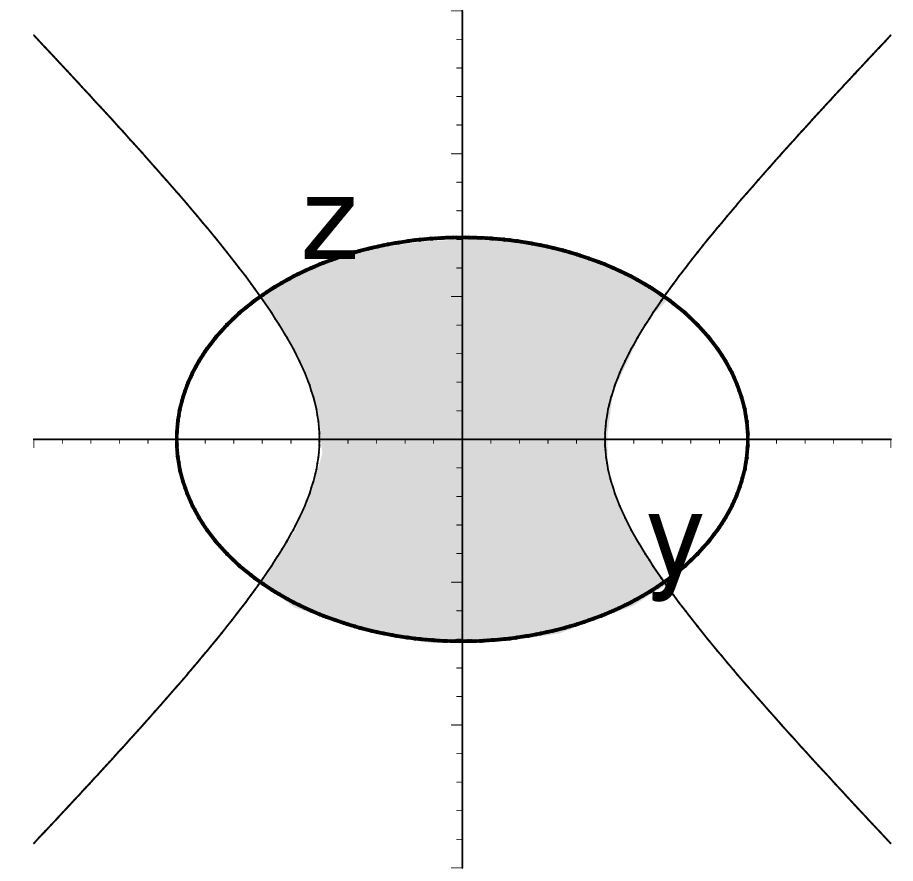}
}
\centerline{
\raisebox{2cm}{BB$_3$}
\includegraphics[angle=0,height=2cm]{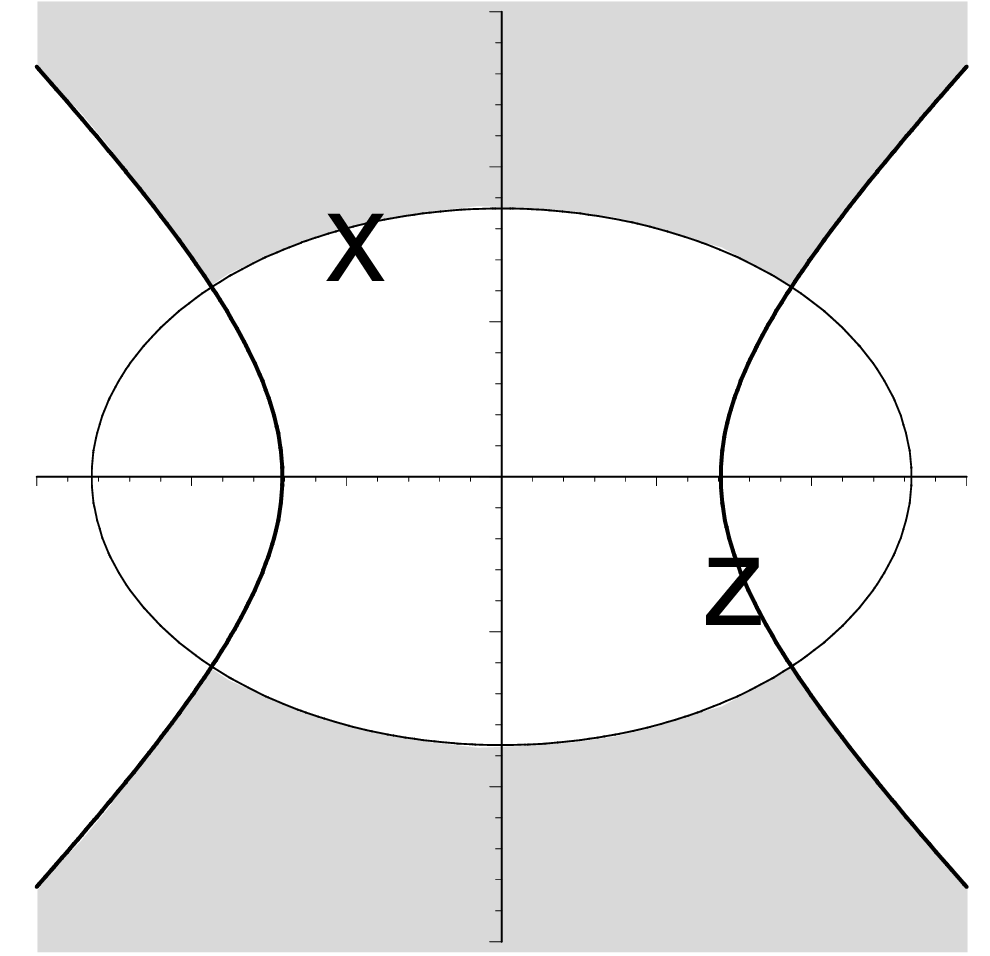}
\includegraphics[angle=0,height=2cm]{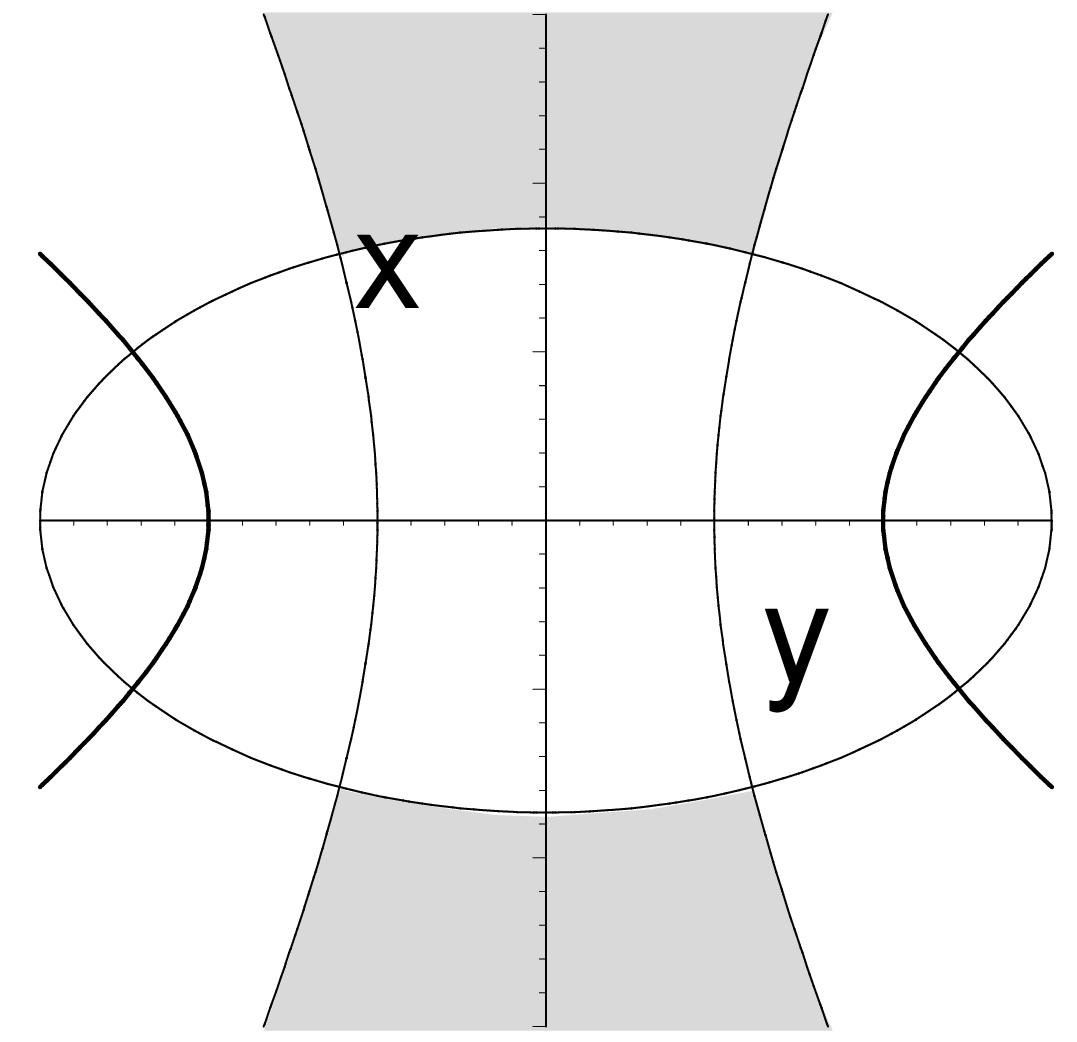}
\includegraphics[angle=0,height=2cm]{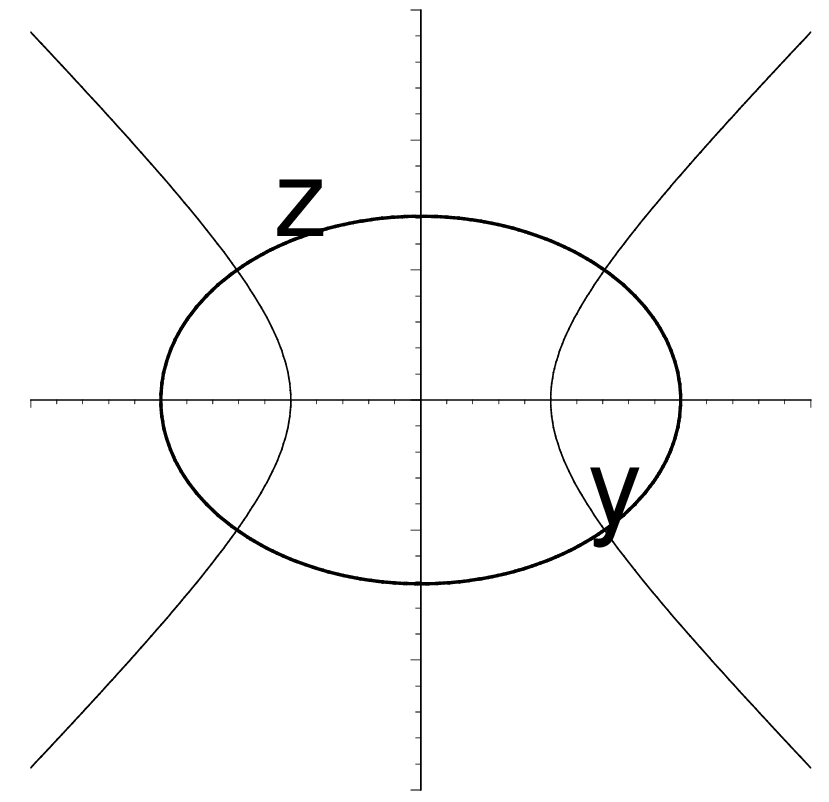}
}
\centerline{
\raisebox{2cm}{WG$_1$}
\includegraphics[angle=0,height=2cm]{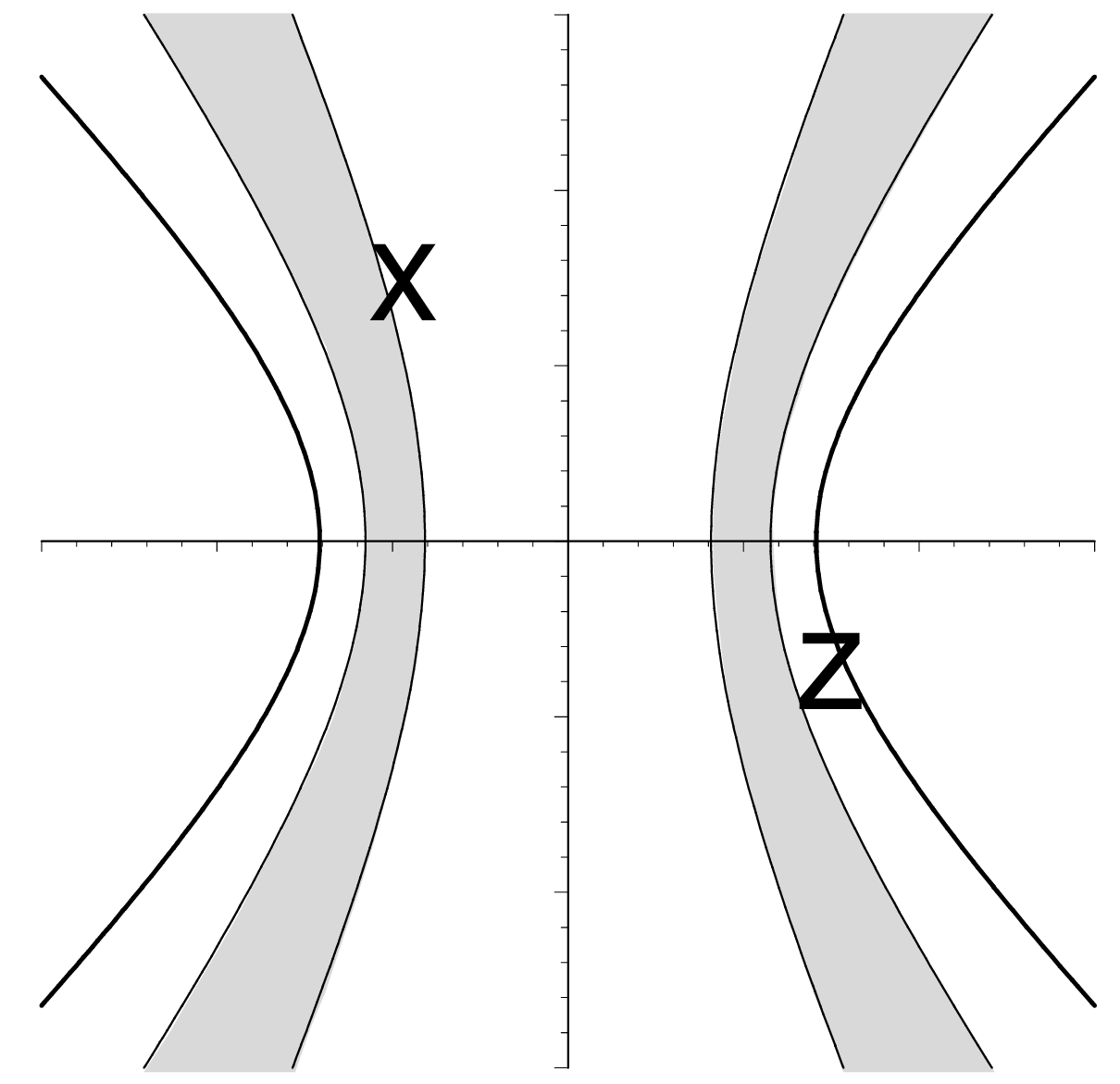}
\includegraphics[angle=0,height=2cm]{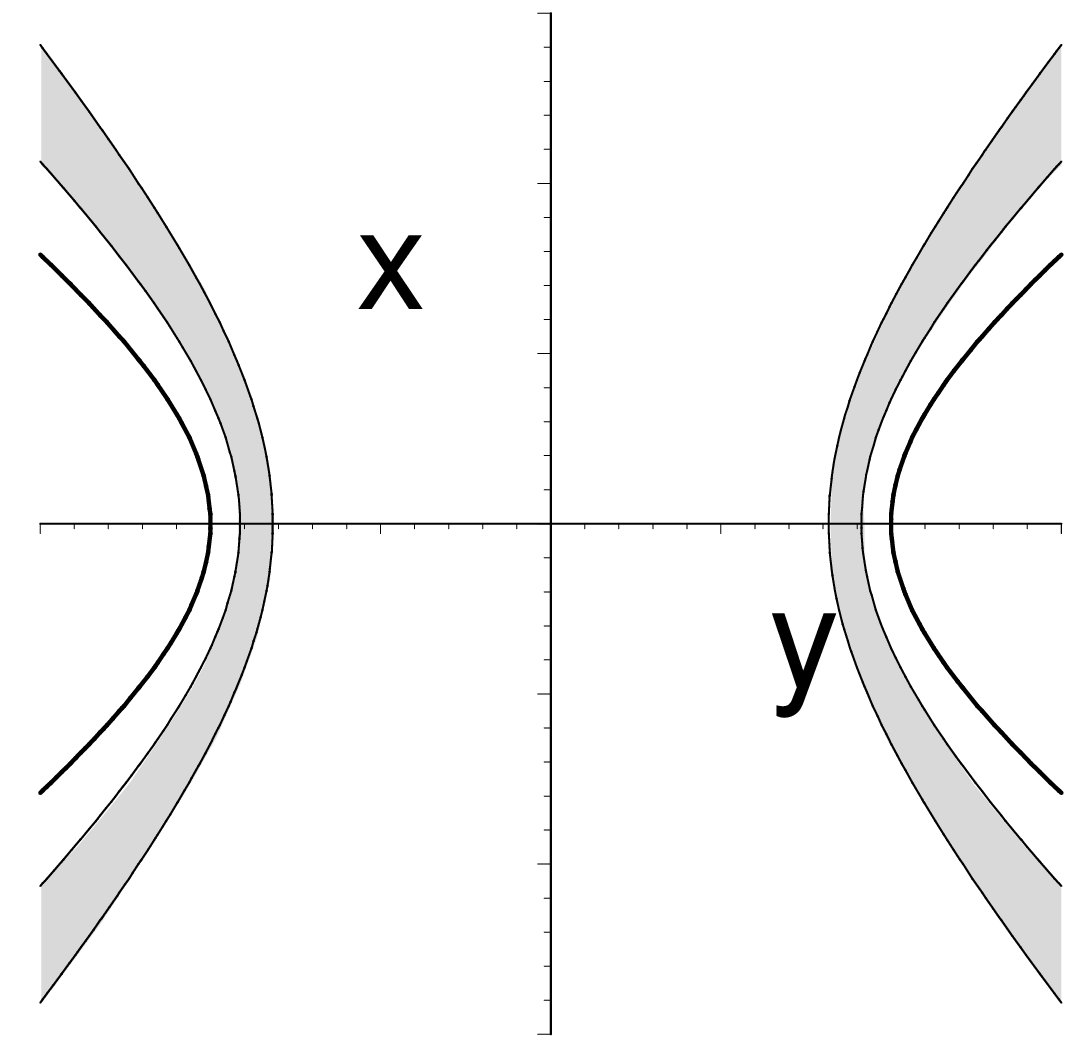}
\includegraphics[angle=0,height=2cm]{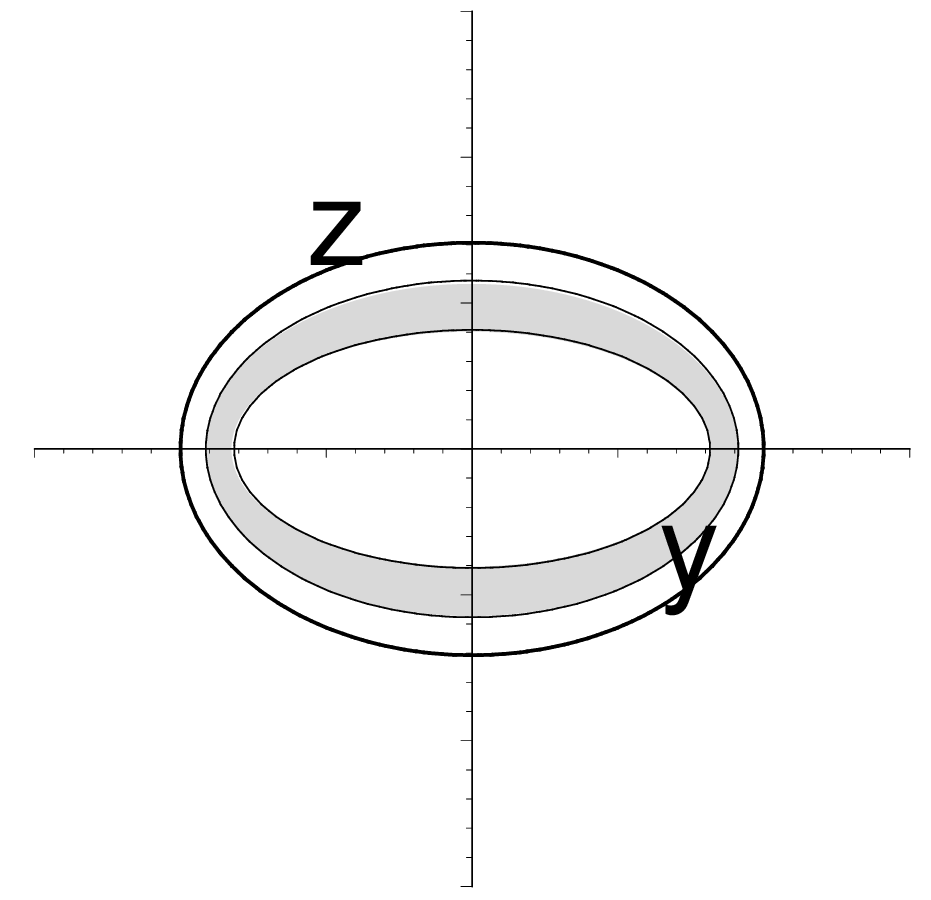}
}
\centerline{
\raisebox{2cm}{WG$_2$}
\includegraphics[angle=0,height=2cm]{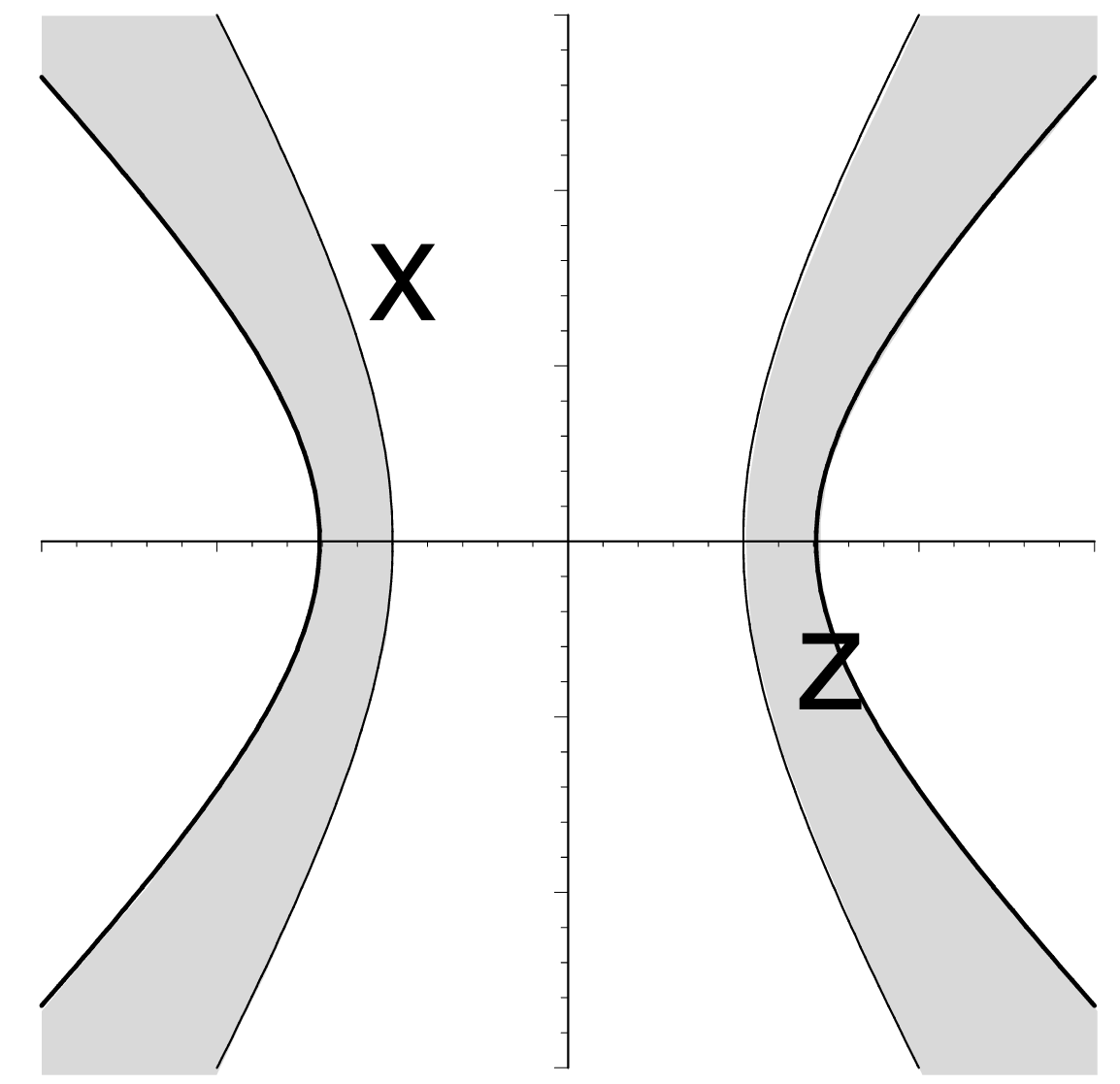}
\includegraphics[angle=0,height=2cm]{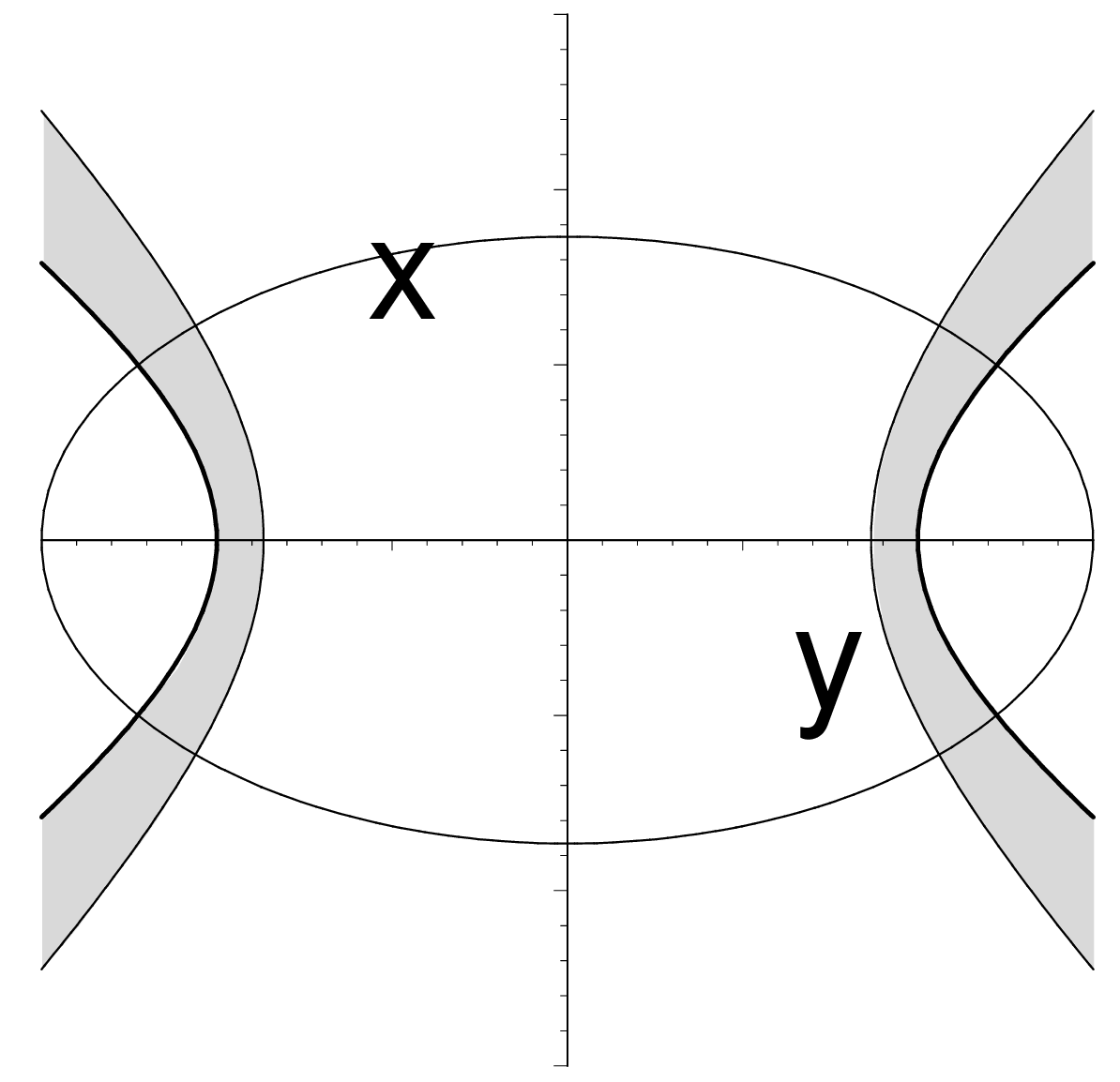}
\includegraphics[angle=0,height=2cm]{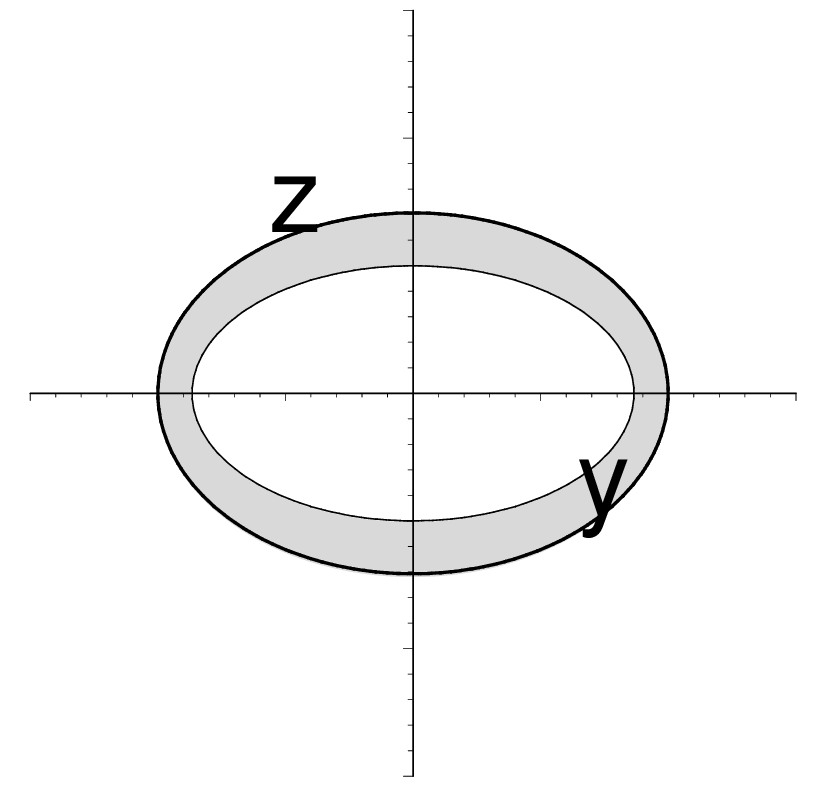}
}
\centerline{
\raisebox{2cm}{WG$_3$}
\includegraphics[angle=0,height=2cm]{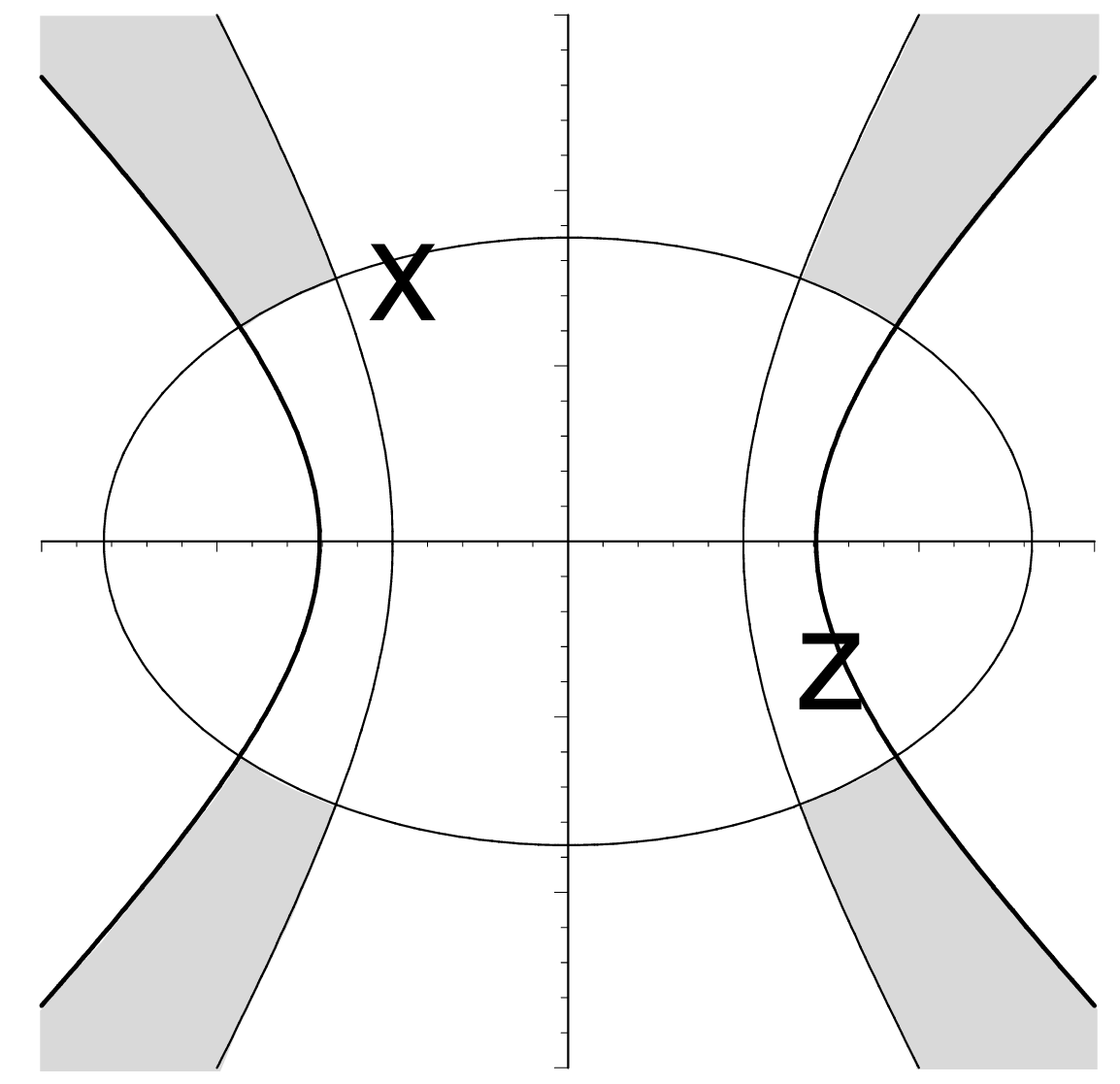}
\includegraphics[angle=0,height=2cm]{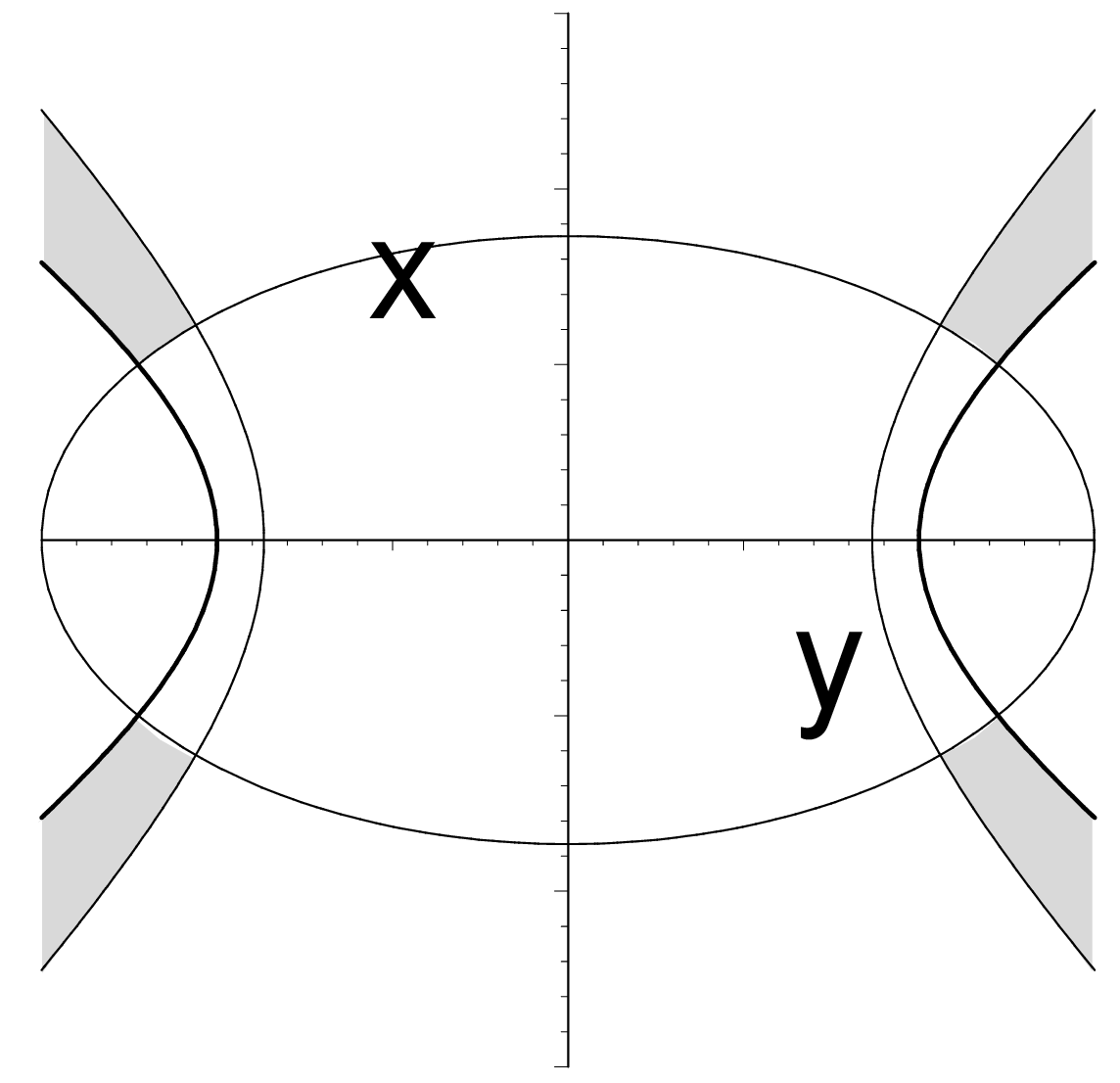}
\includegraphics[angle=0,height=2cm]{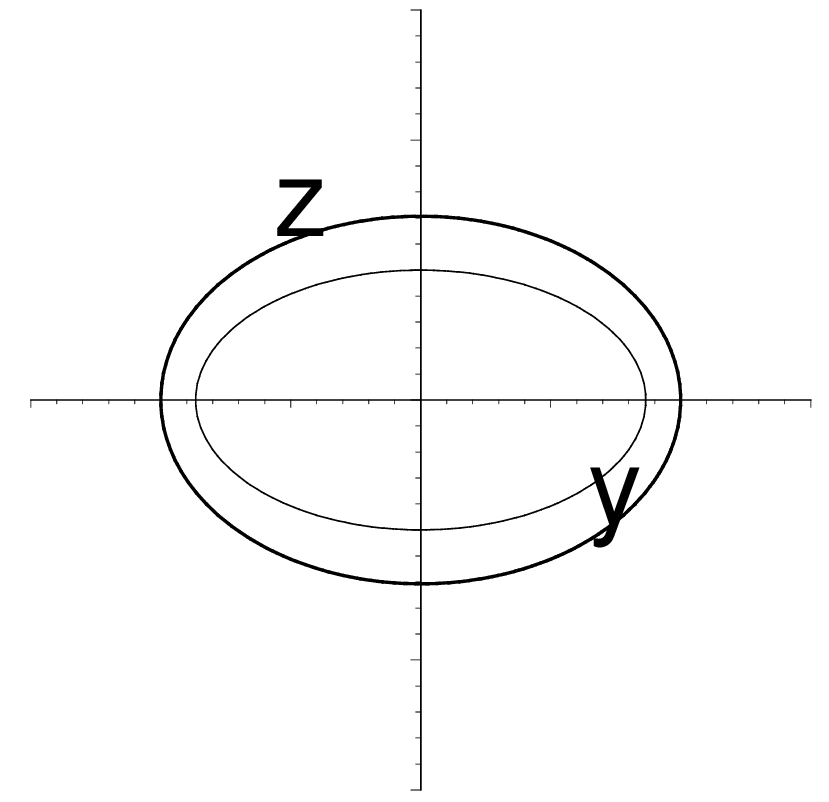}
}
\caption{
\new{\label{fig:caustics}
Configuration space projections of the invariant cylinders corresponding to the motions
 BB$_1$, BB$_2$, BB$_3$, WG$_1$, WG$_2$ and WG$_3$ as their intersections (shaded regions) with the Cartesian coordinate plane. 
The bold lines mark the intersections of the boundary hyperboloid (\ref{eq:hyper}). The ranges for $x$, $y$ and $z$ are $[-3/2,3/2]$. ($(a^2,c^2)=(3/2,1/2)$.)
}
}
\end{figure}

In order to describe the motions on the different families of cylinders it is again useful to illustrate the corresponding effective energies and potential, and phase portraits and also the intersections of these cylinders with the various Cartesian coordinate planes. This is shown in Figures~\ref{fig:pots_xfig_c} and \ref{fig:caustics}, respectively. To simplify the discussion we will consider, like in the 2D case, the specular reflections at the billiard boundary to be smooth. In the 3D case this implies that we identify $p_\mu$ and $-p_\mu$ when $\mu=\mu_{\text{B}}$ or $\mu=2K(q')-\mu_{\text{B}}$ (see Fig.~\ref{fig:pots_xfig_c}(b)).

For a fixed energy $E>0$ a pair $(s^2_1,s^2_2)$ 
(or the corresponding pair $(k,l)$)  in BB$_1$ or BB$_2$ has as its level set a  toroidal cylinder  $\R \times \T^2$ which we illustrate in
terms of  its projection to configuration space in Fig.~\ref{fig:caustics}.
It is unbound in the direction of $\lambda$ and the motion is oscillatory in the transverse degrees of freedom $\nu$ and $\mu$.
In  BB$_2$ the motion oscillates with reflections at the boundary hyperboloid. 
The intersection of the cylinders of type BB$_2$ with the $y-z$ plane
is bounded by the  two branches of the hyperbolas $\eta=s_1$, similar to the 
``bouncing ball modes'' which one 
finds in the billiard in a planar ellipse \cite{WWD97}. 

In contrast to that the motion in  BB$_1$, though oscillatory in $\eta$ and $\zeta$, does not touch the boundary
hyperboloid, i.e.  the corresponding toroidal cylinders are foliated by straight lines of free motions without reflections. 
A pair $(s^2_1,s^2_2)$ in   BB$_3$ represents motion which does not cross the $y-z$ plane.
The corresponding level sets consist of two toroidal cylinders $\R \times \T^2$ which are bounded away from the $y-z$ plane by the ellipsoid $\xi=s_2$.

Pairs $(s^2_1,s^2_2)$ in  WG$_1$ or WG$_2$ involve motions 
which are rotational in $\zeta$ (or, equivalently, in $\nu$). They represent two toroidal cylinders $\R \times \T^2$
which differ by the sense of rotation (see the corresponding panels in Fig~\ref{fig:caustics}).
In the elliptical cross-section in the $x-y$ plane the motion  WG$_2$ is bounded by the ellipse $\eta=s_1$,  similar to the ``whispering gallery modes'' which one finds in planar elliptic billiards. 
As in the case of BB$_1$, motions in  WG$_1$ do not touch the hyperboloidal boundary. The corresponding toroidal cylinders are again foliated by lines of free 
motion without reflections.
For $(s^2_1,s^2_2)$ in WG$_3$ the rotational motions are again bound away from the $y-z$ plane by the ellipsoid $\xi=s_2$.
The corresponding level set consists of four toroidal cylinders which have $x>0$ or $x<0$ combined with different senses of rotation.

The smooth families of cylinders bifurcate along the  boundaries of the $s_1^2-s_2^2$ parameterization intervals in Fig.~\ref{fig:bifdiag_a_b}.
Along $s_2^2=1$ we have the (minor) bifurcation from cylinders consisting reactive trajectories which have reflections at the boundary hyperboloid to cylinders with trajectories having no reflections.   

Along $s_1^2=c^2$ the motions bifurcate from bouncing ball to whispering gallery type.
Note that the distinction between BB$_1$ and WG$_1$ is only `artificial' . They both consists of free motions (without reflections). For such motions, there are more constants of motion than degrees of freedom  (the free motion can be separated in several coordinate systems). In such so called superintegrable systems (a multidimensional harmonic oscillator is a simple example) the foliation by invariant cylinders (or equivalently invariant tori in the case of a compact system like a harmonic oscillator), is therefore not uniquely defined \cite{Evans90}.  

\begin{figure}
\centerline{
\includegraphics[angle=0,width=8.6cm]{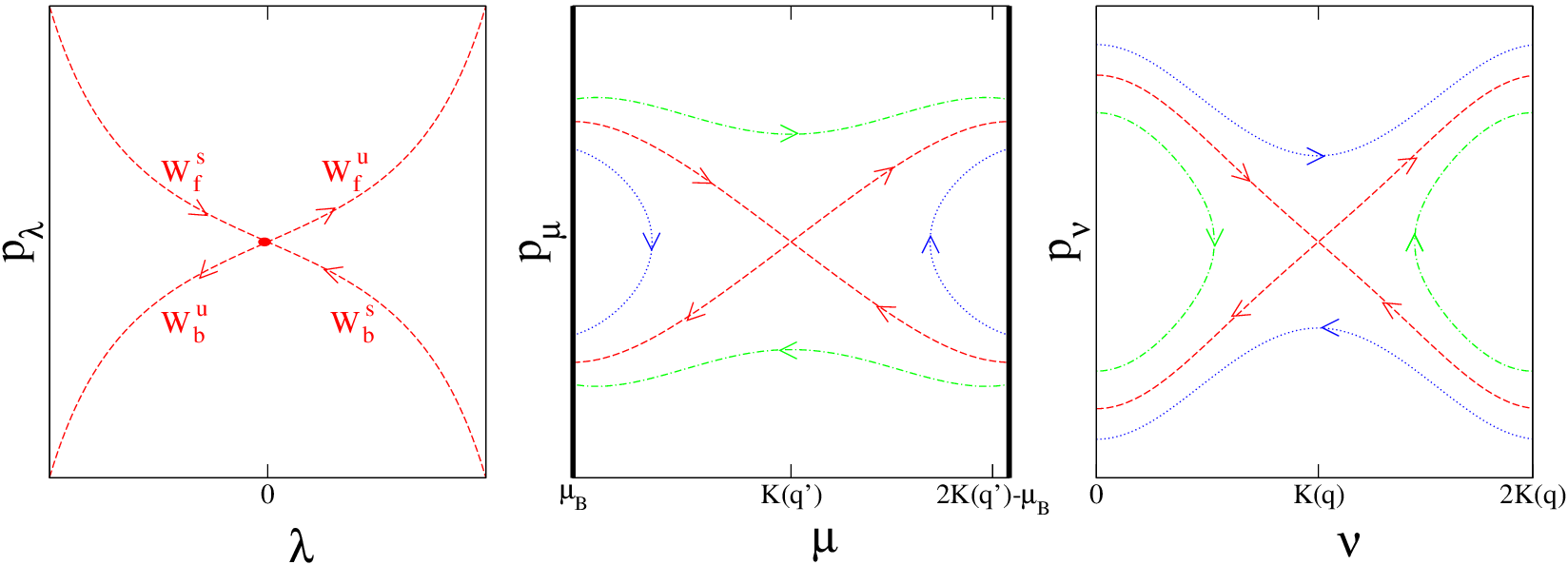}
}
\caption{\label{fig:invmanifold}
Foliation of the  level set $E=\text{const.}>0$, $s_2^2=a^2$ which consists of the invariant billiard in the bottleneck ellipse that forms the transition state of the 3D system, and its stable and unstable manifolds with branches $W^{s}_{f/b}$ and  $W^{u}_{f/b}$, respectively.  
In the $\lambda-p_\lambda$ phase plane the invariant billiard appears as the dot at the origin.
The corresponding phase curves in the $\mu-p_\mu$ plane and $\nu-p_\nu$ plane are parametrized by $s_1^2\in[0,1]$.
For $0<s_1^2<c^2$,  the phase curves are of bouncing ball type (green dashed curves);
for $c^2<s_1^2<1$,  the phase curves are of whispering gallery  type (blue dashed curves). The separatrix between bouncing ball and whispering gallery motions has $s_1^2=c^2$ (red dashed curves in the right panels).
Note that the two pieces of the phase space curve corresponding to the bouncing ball motion are mapped onto each other by the involution \eqref{eq:involution_S}.
}
\end{figure}

Most importantly for the reaction dynamics is the bifurcation from nonreactive motions to reactive motions along $s_2^2=a^2$.
In fact, the joint level set of the \emph{two} constants of motion $E$ and $s^2_2$ for a fixed $E>0$ and $s_2^2=a^2$ consists of the
energy surface of the
 unstable invariant two-degree-of-freedom subsystem which consists of the billiard in the bottleneck ellipse which has $x=0$ and $p_x=0$ and its stable and unstable manifolds.
We illustrate the foliation of this level set in Fig.~\ref{fig:invmanifold}.
The two-degree-of-freedom billiard in the bottleneck ellipse can be viewed to form the transition state for the 3D system. The transition state at energy $E$ is then given by
\begin{equation}\label{eq:TS3D}
\begin{split}
\text{TS} &=  \{ (x,y,z,p_x,p_y,p_z) \in \R^6\,:\, x=0, \, p_x=0,\,\\
&\phantom{=} y^2 + \frac{z^2}{\tilde{c}^2}  \le 1  , \, p_y^2 + p_z^2=2mE  \} \,.
\end{split}
\end{equation} 
The billiard in an ellipse is foliated by two different smooth families of twodimensional tori $\T^2$ which in this case are parameterized by $s_1^2$. For $0<s_1^2<c^2$, these tori are of bouncing ball type, and for $c^2<s_1^2<1$, the tori are of 
whispering gallery type. The  bifurcation between theses two families at  $s_1^2=c^2$ 
involves the unstable periodic orbit along the  major axis of the bottleneck ellipse. At $s_1^2=0$ the bouncing ball tori degenerate to the stable periodic orbit  along the minor axis of the ellipse. At $s_1^2=1$ the whispering gallery motions 
degenerate to the two periodic orbits sliding along the perimeter of the ellipse in opposite directions  
(see \cite{WWD97} for a detailed discussion). 
Regarding the specular reflections to be smooth, the energy surface of this invariant subsystem forms a three-dimensional sphere, $\sphere^3$. In the full original 3D system 
this sphere is unstable with respect to the transverse directions parametrized by $\lambda$ and $p_\lambda$  and therefore has stable and unstable manifolds $W^s$ and $W^u$ which are also contained in the level set (see the lines in the $\lambda-p_\lambda$ plane in the left panel of Fig.~\ref{fig:invmanifold}). The topology of $W^s$ and $W^u$ can be inferred  from taking the Cartesian product of the 3-sphere of the invariant subsystem with the lines in the $\lambda-p_\lambda$ plane in the left panel of Fig.~\ref{fig:invmanifold}, i.e. the stable and unstable manifolds have topology $\R\times \sphere^3$.
Like in the case of the 2D system discussed in Sec.~\ref{subsection:hyperbola_classical} these stable and unstable manifolds are again of codimension 1 in the energy surface. This way they again have sufficient dimensionality to act as separatrices, and in fact they again separate the reactive trajectories from the nonreactive trajectories. 
Similar to the case of the 2D system described in Sec.~\ref{subsection:hyperbola_classical} the manifolds $W^s$ and $W^u$  again each
have two branches. We again denote the branch of $W^s$ which has $p_\lambda>0$ (resp. $p_\lambda<0$) the \emph{forward} (resp. \emph{backward}) branch, $W_f^s$ (resp. $W_b^s$), of the stable manifold. Similarly we again denote the branch of $W^u$ which has $p_\lambda>0$ (resp. $p_\lambda<0$) the \emph{forward} (resp. \emph{backward}) branch, $W_f^u$ (resp. $W_b^u$) of the unstable manifold (see Fig.~\ref{fig:invmanifold}). Also, we again call the union of the forward banches,
\begin{equation}
W_f   := W^s_f \cup W^u_f\,,
\end{equation}
the \emph{forward reactive cylinder}, and the union of the backward branches, 
\begin{equation}
W_b  := W^s_b \cup W^u_b\,,
\end{equation}
the \emph{backward reactive cylinder}.  These forward and backward reactive cylinders then again enclose the forward and reactive trajectories, respectively, and separate them from the nonreactive trajectories in the energy surface under consideration.

Moreover, we can define a dividing surface DS by setting $x=0$ on the energy surface which gives
\begin{equation} \label{eq:def_DS_3D}
\begin{split}
\text{DS} &= \{ (x,y,z,p_x,p_y,p_z) \in \R^6\,:\, x=0, \, y^2 + \frac{z^2}{\tilde{c}^2}  \le 1  , \, \\
&\phantom{=} p_x^2+p_y^2 + p_z^2=2mE  \} 
\end{split}
\end{equation}
With our convention to consider the specular reflections to be smooth the dividing surface DS has the topology of a four-dimensional sphere, $\sphere^4$.
Similar to the situation in the 2D system the three-dimensional sphere associated with the transition state TS in \eqref{eq:TS3D}
can again be viewed to form the equator of the  DS 4-sphere. 
In fact the transition state TS divides the dividing surface into two hemispheres, the forward dividing surface
\begin{equation}\label{eq:forward_div_surf}
\begin{split}
\text{DS}_{\text{f}} &= \{(x,y,z,p_x,p_y,p_z)\in \R^3\times\R^3  \,: \,   x=0, \, y^2+\frac{z^2}{\tilde{c}^2}\le 1,  \,\\
&\phantom{=} p_x^2+p_y^2+p_z^2=2mE \,,p_x>0 \}\,.
\end{split}
\end{equation}
and the backward dividing surface
\begin{equation}
\begin{split}
\text{DS}_{\text{b}}  &= \{(x,y,z,p_x,p_y,p_z)\in \R^3\times\R^3  \,: \,   x=0, \, y^2+\frac{z^2}{\tilde{c}^2}\le 1,  \, \\
&\phantom{=} p_x^2+p_y^2+p_z^2= 2m E \,,p_x<0 \}\,.
\end{split}
\end{equation}
Each forward reactive trajectory has a single intersection with the forward hemisphere; each backward reactive trajectory has a single intersection with the backward hemisphere. Nonreactive trajectories do not intersect the dividing surface DS at all. Like in the 2D case the dividing surface is everywhere transverse to the Hamiltonian flow apart from its equator which is an invariant manifold.

\subsubsection{Action Integrals}

As we have seen in the previous subsection 
the phase space of  the 3D system is foliated by six different families of invariant toroidal  cylincers, $\R \times \T^2$. For the toroidal base, which is associated with the degrees of freedom $\zeta$ and $\eta$ (or equivalently $\nu$ and $\mu$), we can again define action-angle variables. The actions in this case are given by
\begin{equation} \label{eq:action_ints}
\begin{split}
 I_s=\frac{1}{2\pi}\oint p_s \ud s =&  \sqrt{2mE}  \frac{m_s}{2\pi}\int_{s_{-}}^{s_{+}} \sqrt{ \frac{s^4-2ks^2+l}{(s^2-a^2)(s^2-c^2)}}  \ud s\  \\=& 
 \sqrt{2mE}   \frac{m_s}{2\pi}\int_{s_{-}}^{s_{+}}     \sqrt{  \frac{(s^2-s_1^2)(s^2-s_2^2)}{(s^2-a^2)(s^2-c^2)}  }  \ud s\,, 
\end{split} 
\end{equation}
with $s\in \{\zeta,\eta\}$ and the $p_s$ being taken from  (\ref{eq:sep_s}). The integers $m_s$ and the integration boundaries $s_{-}$ and $s_{+}$ can be found in Tab.~\ref{int_bound_3D}.

 For the actions of the symmetry reduced system,  which we again denote by $\tilde{I}_s$, we always have $m_s=2$, $s\in\{\zeta,\eta\}$, i.e.
\begin{equation}
\label{eq:action_ints_red}
\tilde{I}_s= \sqrt{2mE} \frac{1}{\pi}\int_{s_{-}}^{s_{+}} p_s \ud s\,.
\end{equation}

\begin{table}
\begin{center}
\begin{tabular}{|c|c|c|c|c|c|c|}
\hline
type & $m_{\zeta}$ & $m_{\eta}$ & $\zeta_-$ & $\zeta_+$ & $\eta_-$ & $\eta_+$  \\
\hline\hline
BB$_1$  & 4 & 4 & $0$ & $s_1$ & $c$ & $s_2$  \\
\hline
BB$_2$  & 4 & 4 & $0$ & $s_1$ & $c$ & $1$  \\
\hline
BB$_3$  & 4 & 4 & $0$ & $s_1$ & $c$ & $1$ \\
\hline
WG$_1$  & 4 & 2 & $0$ & $c$ & $s_1$ & $s_2$  \\
\hline
WG$_2$  & 4 & 2 & $0$ & $c$ & $s_1$ & $1$ \\
\hline
WG$_3$  & 4 & 2 & $0$ & $c$ & $s_1$ & $1$  \\
\hline
\end{tabular}
\end{center}
\caption{\label{int_bound_3D} Integration boundaries, $s_-$ and $s_+$, and multipliers $m_s$ in (\ref{eq:action_ints}) for the six types of 3D motion BB$_1$, BB$_2$, BB$_3$, WG$_1$, WG$_2$ and WG$_3$.}
\end{table}

\begin{figure}
\centerline{
\includegraphics[angle=0,width=12cm]{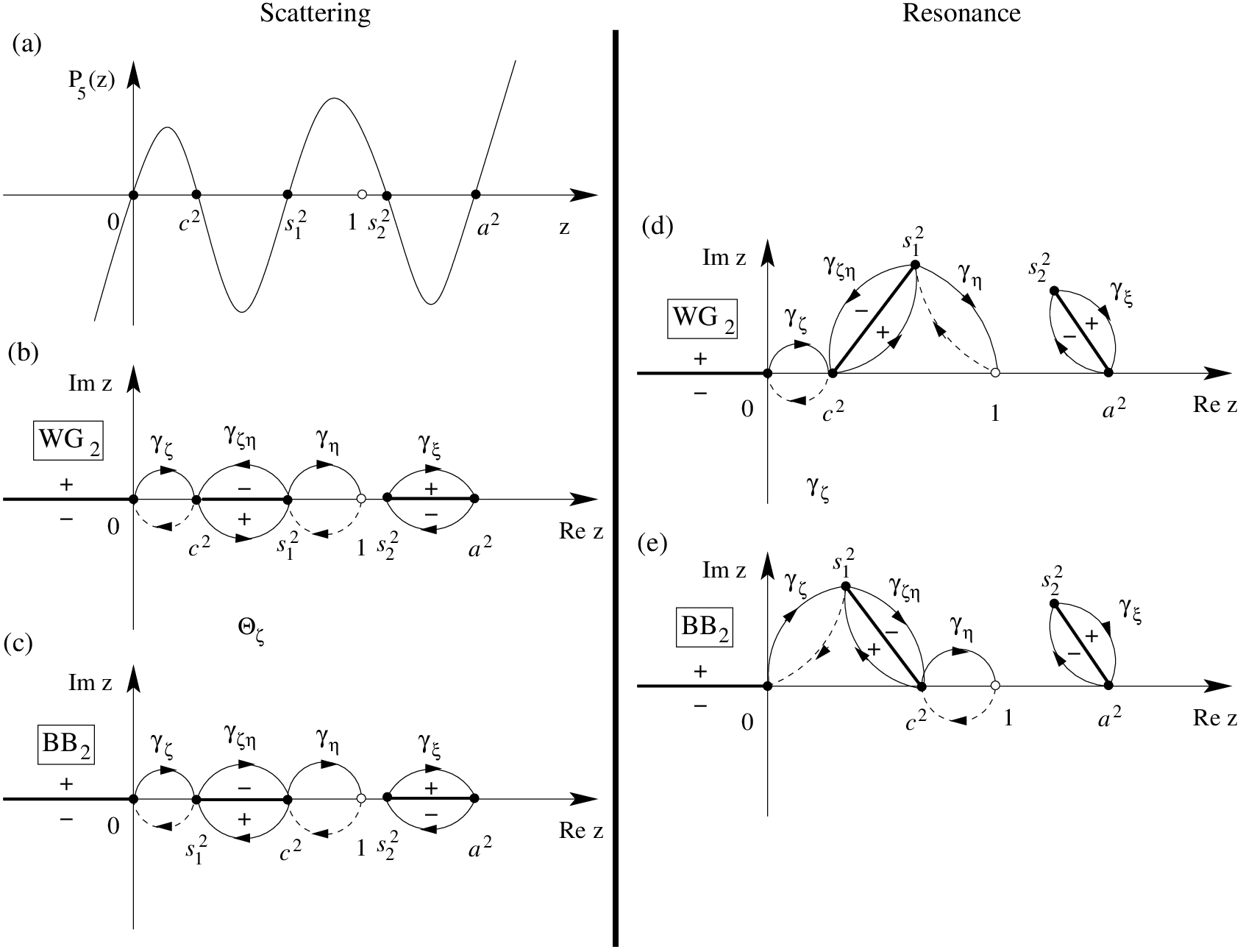}
}
\caption{\label{fig:ellcurve_res}
(a) The graph of the polynomial $P_{5}(z)$ defined in (\ref{eq:hyperelliptic_curve}) for real separation constants $s_1^2$ and $s_2^2$ satisfying $c^2<s_1^2<1<s_2^2<a^2$ (motion type WG$_2$). (b), (c), (d), (e) Complex planes with definitions of the integration paths $\gamma_\zeta$, $\gamma_\eta$, $\gamma_\xi$ and $\gamma_{\zeta\eta}$,  along which $I_\zeta$, $I_\eta$ , $I_\xi$ and 
$I_{\zeta\eta}$ in \eqref{eq:I_s_on_Gamma_3D},  \eqref{eq:def_I_xi_3D},  and   \eqref{eq:def_Izetaeta}  are computed; (b), (c) are for real $s_1^2$ and $s_2^2$ (scattering states of type WG$_2$ and BB$_2$, respectively), and (d), (e) are the corresponding continuations of (b) and (c) when $s_1^2$ and $s_2^2$ leave the real axis (resonance states).  
Similarly to  Fig.~\ref{fig:ellcurve_res_2D},  
the complex planes can be viewed as one half  of the hyperelliptic curve $\Gamma_w$ (the Riemann sheet of one `sign' of the square root $w$). To make the square root $w$ well defined in the complex plane three branch cuts connecting consecutive branch points of $w$ are introduced (bold lines). The left cut connects the branch points 0 and $\infty$; the middle cut connects $c^2$ and $s_1^2$; the right one connects $s_2^2$ and $a^2$. The integration paths $\gamma_{\zeta\eta}$ and $\gamma_\xi$ are closed loops which encircle the middle and right branch cuts, respectively. The integration path $\gamma_\zeta$ is also a closed loop of $\Gamma_w$. In the picture shown it consists of two parts. It starts at either $z=c^2$ (WG) or $z=s_1^2$ (BB) on the other Riemann sheet of $w$ (which is a copy of the one shown and which joins the copy shown at the branch cuts) and intersects the branch cut between 0 and $\infty$. This part is marked by a dashed line. The integration path then continues on the shown Riemann sheet where it ends at either  $z=c^2$ (WG) or $z=s_1^2$ (BB).   The integration path $\gamma_\eta$ also has two parts. It starts at $z=1$ on the other Riemann sheet of $w$ and intersects the branch cut between $c^2$ and $s_1^2$. This part is marked by a dashed line. The integration path continues on the  shown Riemann sheet where it ends at the point 1. Like in Fig.~\ref{fig:ellcurve_res_2D}  the signs $+$ or $-$ indicate whether $w=+\ui |w|$ or $w=-\ui |w|$ `just above' or `just below' the respective branch cut. 
}
\end{figure}

Substituting $z=s^2$ in Eq.~(\ref{eq:action_ints}) shows  that the action integrals $I_\zeta$ and $I_\eta$ are both of the form
\begin{equation}
\label{eq:riemann_ints}
I_s= \sqrt{2mE} \frac{m_s }{4\pi}\int_{z_-}^{z_+}\frac{(z-s_1^2)(z-s_2^2)}{w(z)} \ud z\,,
\end{equation}
where
\begin{equation}
\label{eq:riemann_poly}
w^2(z)=P_5(z):=\prod_{i=1}^{5}(z-z_i)\,,
\end{equation}
and  $z_-$ and $z_+$ are consecutive elements of the set $\{z_1=0,z_2=s_1^2,z_3=b^2,z_4=s_2^2,z_5=a^2,z_{\text{b}}= 1\}$. 
Again $z_{\text{b}}=1$  corresponds to the boundary hyperboloid. The differential $\ud z/w(z)$ has the six critical points $\{z_1,z_2,z_3,z_4,z_5,\infty\}$ which means that the integrals (\ref{eq:riemann_ints}) are \textit{hyperelliptic}. There do not exist tabulated standard forms for these integrals like for the elliptic integrals \eqref{eq:riemann_ints_2D} in the case of the 2D system. 
However we can again view them as Abelian integrals, which in this case are defined on the 
hyperelliptic curve
\begin{equation}
\label{eq:hyperelliptic_curve}
\Gamma_w = \{ (z,w)\in \overline{\C}^2 \, :  \, w^2  =P_{5}(z) \}\,,
\end{equation}
which is an algebraic  curve of genus 2.
The integrals \eqref{eq:riemann_ints} then become
\begin{equation} \label{eq:I_s_on_Gamma_3D}
I_s =  \sqrt{2mE} \frac{m_s}{8\pi}  \int_{\gamma_s}  (z-s_1^2)(z-s_2^2) \frac{\ud z}{w} \,,
\end{equation}
where $s\in\{\zeta,\eta\}$ and the integration paths $\gamma_\zeta$ and $\gamma_\eta$ for motions of type WG$_2$ and BB$_2$ are  shown in Fig.~\ref{fig:ellcurve_res}. For motions of type WG$_3$ and BB$_3$, the order of $s_2^2$ and $a^2$ along the real axis in Fig.~\ref{fig:ellcurve_res} is reversed which does not affect the definitions of $\gamma_\zeta$ and $\gamma_\eta$.
The  integration path $\gamma_\zeta$ is  a closed path on $\Gamma_w$, and hence the integral $I_\zeta$ is a complete hyperelliptic integral. Due to the billiard boundary the integration path $\gamma_\eta$ is not closed; the integral $I_\eta$ is an incomplete hyperelliptic integral.

Similarly to \eqref{eq:def_I_xi_2D}, we can also define a closed hyperelliptic integral associated with $\xi$,
\begin{equation} \label{eq:def_I_xi_3D}
I_\xi =   \ui  \sqrt{2mE} \frac{1}{2\pi}  \int_{\gamma_\xi}  (z-s_1^2)(z-s_2^2) \frac{\ud z}{w}\,,
\end{equation}
where $\gamma_\xi$ for motions of type WG$_2$ and BB$_2$ is defined in Fig.~\ref{fig:ellcurve_res} and leads to a real positive 
$I_\xi$ for these motions. For motions of type WG$_3$ and BB$_3$ the order of $s_2^2$ and $a^2$ along the real axis in Fig.~\ref{fig:ellcurve_res} is reversed, and $I_\xi$ becomes real negative.  

Moreover we define the integral
\begin{equation}\label{eq:def_Izetaeta}
I_{\zeta\eta} =  \ui  \sqrt{2mE} \frac{1}{\pi}  \int_{\gamma_{\zeta\eta}}  (z-z_1)(z-z_2) \frac{\ud z}{w}\,,
\end{equation}
where $\gamma_{\zeta\eta}$ for WG$_2$ and BB$_2$ is also defined in Fig.~\ref{fig:ellcurve_res}. 
The change of the order of $s_2^2$ and $a^2$ again does not affect the definition of $\gamma_{\zeta\eta}$. This way we get a real positive $I_{\zeta\eta} $ for WG$_{2/3}$ and a real negative $I_{\zeta\eta} $ for BB$_{2/3}$.
 
The integrals $I_\xi$ and $I_{\zeta\eta}$ will play an important role in 
Sections~\ref{sec:cum_reac_prob} and \ref{sec:resonances}.

\todo{Change notation $\gamma_{\Theta_\zeta}$ to $\gamma_{\zeta\eta}$ in Fig.~\ref{fig:ellcurve_res}.}

%% file: sec5_transtate.tex
\section{Computation of the classical and quantum transmission from transition state theory}
\label{sec:cum_reac_prob}

In this section we compute the  transmission probabilities for the classical and quantum transport from the regions $x\ll -1$ (the `reactants' region) to the region $x\gg 1$ (the `products' region) in the geometries \eqref{eq:2Dregion} (2D) and \eqref{eq:3Dregion} (3D). In the quantum case we are interested in the cumulative reaction probability which is defined as 
\begin{equation}\label{eq:def_N_trace}
N(E) = \text{Tr} \, \hat{T}(E)\hat{T}^\dagger(E)\,,
\end{equation}
where $\hat{T}(E)$ is the transmission block of the scattering matrix at energy $E$ (for references in the chemistry literature see, e.g., \cite{Miller98}; in the context of  ballistic electron transport problems \eqref{eq:def_N_trace} is known as the
Landauer-B{\"u}ttiker formula \cite{Landauer57,Landauer70,Buettiker92}).

According to its definition,  $N(E)$ can be computed from the scattering matrix. However, this is a very inefficient (and for the systems with many degrees of freedom even infeasible) procedure since one has to determine all the state-to-state reactivities while $N(E)$ is merely a  sum over these reactivities and hence no longer contains information about the individual state-to-state reactivities. Much effort has been and still is put into finding a computationally cheap method to compute $N(E)$.   
In the chemistry literature (see, e.g., \cite{Yamamoto60,MillerSchwartzTromp83,Miller98}) a method has been developed to compute $N(E)$ on the basis of transition state theory where the classical transmission probability is computed from the flux through a dividing surface which for a given energy $E$ separates the energy surface into a reactants and a  products region. 
For such a dividing surface DS the flux from reactants to products can be computed as
\begin{equation}\label{eq:Miller_flux}
f(E) = \int_{\R^f}  \int_{\R^f}  \delta(E-H(q,p)) F(q,p) P_r(q,p)\ud^f q \ud^f p\,.
\end{equation}
Describing the dividing surface by a zero level set of a function $s$ on phase space, or more precisely a function which is negative on the reactants side of the dividing surface and positive on the products side of the dividing surface, $F$ in \eqref{eq:Miller_flux} is defined as the following composition of functions
\begin{equation}
F(q,p)= \frac{\ud }{\ud t}    \left. \Theta \circ s \circ  \Phi_H^t(q,p)   \right|_{t=0} = \delta(s(q,p))\{s,H\}(q,p)\,.
\end{equation}
Here $\Theta$ denotes the Heaviside step function, $\Phi_H^t$ is the Hamiltonian flow generated by $H$ acting for the time $t$, and $\{\cdot,\cdot\}$ denotes the Poisson bracket.  The function $P_r$ in \eqref{eq:Miller_flux} is defined as
\begin{equation}\label{eq:def_Pr_classical}
P_r(q,p) = \lim_{t\to \infty} \Theta \circ s \circ  \Phi_H^t  (q,p) \,,
\end{equation}
which acts as a characteristic function on the dividing surface. In fact, by construction $P_r(q,p)=1$  if the trajectory through the point $(q,p)$ proceeds for $t\to \infty$ to products which is the region where the function $s$ is positive and  $P_r(q,p)=0$ otherwise (see \cite{WaalkensSchubertWiggins08} for a detailed discussion). 

The quantum analogue of \eqref{eq:Miller_flux} is given by 
\begin{equation} \label{eq:Miller_quantum}
N(E) = 2\pi \hbar \text{Tr }\, \delta(E-\hat{H})\hat{F} \hat{P}_r\,,
\end{equation}
where 
\begin{equation}
\hat{F} = -\frac{\ui}{\hbar} [ \widehat{\Theta \circ s},\hat{H}]
\end{equation}
and 
\begin{equation} \label{eq:def_Pr_quantum}
  \hat{P}_r = \lim_{t \to \infty} \ue^{\frac{\ui}{\hbar} \hat{H}t } \widehat{\Theta\circ s} \, \ue^{-\frac{\ui}{\hbar} \hat{H}t }\,.
\end{equation}
Here  $\widehat{\Theta\circ s} $ denotes a quantization of the classical function $\Theta \circ s $ \cite{WaalkensSchubertWiggins08}.

Like its classical analogue \eqref{eq:Miller_flux} 
the evaluation of \eqref{eq:Miller_quantum} 
involves a computationally expensive time integration which is manifested in 
\eqref{eq:def_Pr_classical} and \eqref{eq:def_Pr_quantum}, respectively.
The computational advantage of 
the transition state theoretical formulation of $N(E)$
over the original definition in \eqref{eq:def_N_trace} is therefore not obvious. 
In practice one can carry out the time integration only to a finite time. This time has to be large enough so that it can be decided 
that after this time the resulting trajectory (classical) or wavefunction (quantum) will stay in the products region.
In order to minimize this integration time one has to choose a ``good'' dividing surface.  In fact, for a dividing surface 
that, classically, is crossed exactly once by all reactive trajectories and not crossed at all by nonreactive trajectories (see our discussion
 in Sec.~\ref{sec:classical}) no time integration is required at all. The characterestic function $P_r$ in 
 \eqref{eq:def_Pr_classical} can then be replaced by a function  which at a point $(q,p)$ on the dividing surface is one if the Hamiltonian vector field at this point pierces the dividing surface in the forward direction and zero if the Hamiltonian vector field at that point  pierces the dividing surface in the backward direction. In other words this means that we can omit the function $P_r$ in  \eqref{eq:Miller_quantum} and restrict  the integral  \eqref{eq:Miller_quantum}  to the foward hemisphere of the dividing surfaces that we constructed in Sec.~\ref{sec:classical}. 
The choice of a good dividing surface  is thus crucial to benefit from the transition state theoretical approach to compute classical and quantum transmission probabilities.

In Sec.~\ref{sec:classical} we used the separability of the transmission problem discussed in this paper to construct the dividing surface which has the desired properties.  As a consequence of this separability we similarly get in the quantum mechanical case that
the transmission subblock of the scattering matrix in \eqref{eq:def_N_trace} is diagonal. Using
\begin{equation}
t_{nm}(E) =  t_n(E) \delta_{nm}\,,
\end{equation}
where $n$ and $m$ label the scattering states at the energy $E$, and $\delta_{nm}$ is the Kronecker symbol we have
\begin{equation}\label{eq:cum_reac_prob_sum}
N(E) =  \sum_{n} T_n(E)\,,
\end{equation}
with the state-to-state transmission probabilities defined as $T_n(E):= |t_{nn}(E)|^2$.
In the following we present the computation of these transmission probabilites and the comparison of the resulting cumulative reaction probabilities with the classical flux.


\subsection{The 2D system}
\label{sec:cond_2D}

\subsubsection{The quantum transmission}
\label{sec:cond_2D_qm}

To compute the transmission probabilities $T_n$ in the 2D case 
we look for solutions of the form
\begin{equation}
\label{eq:exactsol_2D_top}
\psi_{\zeta;n}(\zeta) \psi_{\xi;n}(\xi) + r_n \psi_{\zeta;n}(\zeta)   \psi^*_{\xi;n}(\xi)
\end{equation}
at the bottom ($x\ll -1$) and
\begin{equation}
\label{eq:exactsol_2D_bot}
t_n \psi_{\zeta;n}(\zeta)   \psi^*_{\xi;n}(\xi)
\end{equation}
at the top ($x\gg 1$).
Such solutions can be computed from first solving
 the transversal component 
$\zeta$ of
the wave equations 
(\ref{eq:sephelmholtz_2D}) for  the corresponding boundary conditions with $E$ as a parameter. 
As discussed in Sec.~\ref{sec:separation} 
the boundary conditions for $\psi_\zeta$ at $\zeta=0$ is determined by the paritiy $\pi_y$. For 
$\pi_y=+$, we have $\psi'_\zeta(0)=0$ (and choose $\psi_\zeta(0)=1$), and for $\pi_y=-$, we have $\psi_\zeta(0)=0$ (and choose $\psi'_\zeta(0)=1$).
The Dirichlet boundary condition at $\zeta=1$ requires $\psi_\zeta(1)=0$.
This way we obtain  modes which we label by the Dirac ``kets'' 
$\vert n_\zeta;\pi_y \rangle$ where $n_\zeta$ is a non-negative quantum number which
gives the number of nodes of $\psi_\zeta$ in the open 
interval $0<\zeta<1$. 
The modes of energy $E$ determine the separation constants 
$s^2_{2\, (n_\zeta;\pi_y)}(E)$.
This separation constant can then be used in the equation for the $\xi$ component of the separated wave equations (\ref{eq:sephelmholtz_2D}) to find solutions of the form \eqref{eq:exactsol_2D_top} and \eqref{eq:exactsol_2D_bot}.
This however is not completely straightforward and does not give much insight into the structure of the solutions.  We therefore resort to a semiclassical computation which will also lead to the semiclassical computation of resonances as we will discuss in 
Sec.~\ref{sec:resonances}.
The semiclassical approximation is obtained from using $s^2_{2\, (n_\zeta;\pi_y)}(E)$ 
to compute the transmission probability  $T_{(n_\zeta;\pi_y)}(E) $ as 
\begin{equation}\label{eq:transmission_prob_2D}(a)
T_{(n_\zeta;\pi_y)}(E) = 
\frac{1}{1+\exp (\theta_{(n_\zeta;\pi_y)}(E)/\hbar)}\,,
\end{equation}
where  $\theta_{(n_\zeta;\pi_y)}$ is a \emph{tunnel integral}. This tunnel integral describes the quantum mechanical 
tunneling  through the dynamical barrier which in terms of the $\lambda$ coordinate occurs as the barrier in the associated 
effective potential
$V_{\lambda,\textrm{eff}}$
for motions of type T$_3$ (see Fig.~\ref{fig:pots_xfig_c_2D}).
This tunnel integral is given by
\begin{eqnarray}
\label{eq:tunnelintegr_2D}
\theta_{(n_\zeta;\pi_y)}(E) =
- 2 \ui \int_{\lambda_-}^{\lambda_+} p_\lambda \,\mbox{d}\lambda 
= - 4 \ui \sqrt{2m E} \int_a^{s_2}
\sqrt{\frac{\xi^2-s^2_{2}(E)}{\xi^2-a^2}}
\,\mbox{d}\xi\,,
\end{eqnarray}
where $s^2_2=s^2_{2\, (n_\zeta;\pi_y)}(E)$ and $\lambda_+$ and $\lambda_-=-\lambda_+$ are the corresponding turning points using the phase space coordinates $(\lambda,p_\lambda)$ (see, e.g., \cite{BM72} for a derivation of the expression \eqref{eq:transmission_prob_2D}). 
For $s^2_2>a^2$ or equivalently $E_{\lambda, \textrm{eff}}<V_{\lambda, \textrm{eff}}(0)$ which corresponds to classical reflection of type T$_3$,
$p_\lambda$  is imaginary along the integration interval which is bounded by the real classical turning points $\lambda_-$ and $\lambda_+=-\lambda_-$. 
This integral can be identified with two-times the integral of $p_\xi$ from $a$ to the corresponding turning point $s_2$ which gives the second equality in (\ref{eq:tunnelintegr_2D}). 
For  $s^2_2<a^2$ or equivalently $E_{\lambda, \textrm{eff}}>V_{\lambda, \textrm{eff}}(0)$ which corresponds to classical transmission of types T$_2$ and T$_3$, the classical turning points $\lambda_\pm$ become imaginary (with $\lambda_-$ being complex
conjugate to $\lambda_+$) whereas $p_\lambda$ is real on the imaginary axis between $\lambda_\pm$. 
The branches of the square root in (\ref{eq:tunnelintegr_2D}) are
chosen such that the tunnel integral is positive if $a^2<s^2_2$ and negative if $s^2_2<a^2$. 
This choice of the branches can be  described more precisely from relating $\theta_{(n_\zeta;\pi_y)}(E)$ to the integral $I_\xi$ that we defined in \eqref{eq:def_I_xi_2D}. In fact we have,
\begin{equation}\label{xi_Theta_relation_2D}
I_\xi = -\frac{1}{2\pi} \theta_{(n_\zeta;\pi_y)}(E)\,.
\end{equation}

The boundary value problem for $\psi_\zeta$ can be solved numerically using a shooting method which relates the solution of the boundary value problem to a Newton procedure (see \cite{Press88}, and \cite{WWD97,WWD99} for similar applications).  Since $\psi_\zeta$ is an oscillatory function which leads to multiple zeroes in the resulting Newton procedure the shooting method requires good starting values $s^2_2$.  These are obtained from a semiclassical approximation also of the boundary value problem for 
$\psi_\zeta$. 
To this end we note that the phase portraits of the motions of type T$_2$ and T$_3$ between which the classical motion switches from transmission to reflection are identical in the $\nu-p_\nu$ plane (see Fig.~\ref{fig:pots_xfig_c_2D})(b).
For these types of motions we can thus use the EBK quantization condition for the action $I_\zeta$,
\begin{equation}\label{eq:EBKscatt_2D}
I_\zeta
=\hbar(n_\nu+1)\,,\quad n_\nu\in\N_0\,,
\end{equation}
which is the same as the EBK quantization of a one-dimensional square well problem.
We can also rewrite this quantization condition in terms of the action $\tilde{I}_\zeta=I_\zeta/2$ 
of the symmetry reduced system which gives
\begin{equation} \label{eq:EBK_tilde_I_zeta}
\tilde{I}_\zeta = \hbar( n_\zeta+\frac{1}{4}(3-\pi_y))\,,\quad n_\zeta \in \N_0\,.
\end{equation} 
This decomposes the semiclassical modes in terms of the parity $\pi_y$. The quantum numbers $n_\zeta$ and $n_\nu$ are related by
\begin{equation}
n_\nu = 2n_\eta+\frac12(1-\pi_y)\,.
\end{equation} 

We note that for the type T$_1$ the motions involve a smooth rather than a hard wall reflection in the $\nu$ degree of freedom. 
As a result the EBK quantization for T$_1$ would be different from the EBK quantization for T$_2$ and T$_3$,  and hence, in order to describe the transission from  T$_2$ to T$_1$  a unifrom semiclassical quantization scheme would be desirable. However this transition plays no role for the transition from transmission  to reflection (see below) and we therefore do not consider this aspect in more detail.  The quantization condition \eqref{eq:EBKscatt_2D} can be solved by a standard Newton procedure.
The solutions for $E$ and $s_2^2$ for a given quantum number $n_\zeta$ and parity $\pi_y$ are then used as the starting value for the shooting method described above.

\begin{figure}
\centerline{
\begin{tabular}{c}
\tiny
\includegraphics[angle=0,width=12cm]{fig14}\\
\begin{tabular}{|c||c|c|c|c|c|c|c|}
\hline
$\kappa$ axis tick & $\vert 1\rangle$ & $\vert 2\rangle$ & $\vert 3\rangle$ & $\vert 4\rangle$ & $\vert 5\rangle$ & $\vert 6\rangle$ & $\vert 7\rangle$  \\ \hline
$\vert n_\zeta, \pi_y \rangle$ & $\vert 0, +\rangle$ & $\vert 0, -\rangle$ & $\vert 1, +\rangle$ & $\vert 1, -\rangle$ & $\vert 2, +\rangle$ & $\vert 2, -\rangle$ & $\vert 3, +\rangle$  \\ \hline
\end{tabular}
\end{tabular}
}
\caption{
\label{fig:conductance_a_2D} 
\textit{Top Panel:} Cumulative reaction probability $N(E)$ as a function of the wavenumber $\kappa=\sqrt{2mE}/\hbar$  for the shape parameter $a^2=5$, and, for comparison, $a^2=\infty$, which corresponds to the transmission through a rectangular strip. The ticks on the wavenumber axis mark the energies at which, for $a^2=5$, the modes $\vert n_\zeta;\pi_y\rangle$ ``open''  as transmission channels (see text) (the key to the tick labels is given in the table). 
The smooth dot-dashed blue curve and the solid blue curve show the Weyl approximations of $N(E)$ defined in \eqref{eq:NWeyl2D} and \eqref{eq:NWeyltilde2D}, respectively.
\textit{Bottom Panel:} Resonances in the complex wavenumber ($\kappa$) plane, for $a^2=5$. Semiclassical resonances are marked by pluses ($+$) and exact resonances by diamonds ($\Diamond$). Note that we use the symbol $\kappa$ for wavenumber to distinguish it from the separation constant $k$ of the 3D system. ($\hbar=1$, $m=1$.)
}
\end{figure}

The  cumulative reaction probability  $N(E)$ is then the sum over all the $T_{(n_\zeta;\pi_y)}(E)$ in
\eqref{eq:transmission_prob_2D} for all quantum numbers $n_\zeta$  and parities $\pi_y$. For the numerical computation of $N(E)$ we need only consider the finite number of modes which, at a value $E>0$, have a nonnegleglibile transmission probability. A graph of $N(E)$ is shown in 
Fig.~\ref{fig:conductance_a_2D}.  
We note that on the scale of the picture one can notice no difference between the exact and the semiclassically computed $N(E)$.
Depending on the shape parameter $a^2$ for the boundary hyperbola the cumulative reaction probability  shows more or less pronounced steps with unit step size.
A detailed analysis of the graphs of $N(E)$ can be obtained from relating the modes $\vert n_\zeta;\pi_y\rangle$ to the classical motions. For a given energy $E$,
this relationship is established via the separation constant $s^2_{2(n_\zeta;\pi_y)}(E)$ 
which determines the classical invariant cylinder the mode is associated with. As can be seen 
from the projections of the cylinders in Fig.~\ref{fig:caustics_2D} these projections become increasingly confined  in the order 
T$_3\rightarrow$ T$_2\rightarrow$ T$_1$.
Since high confinement in configuration space implies high kinetic energy via the Heisenberg uncertainty principle, 
the modes, which classically correspond to the type of motion T$_1$ have highest energy.
In fact, for low energies all modes have $s^2_{2(n_\zeta;\pi_y)}(E)$ in the classically
reflecting type of motion T$_3$. Upon increasing the energy the $s^2_{2(n_\zeta;\pi_y)}(E)$  wander
towards the transmitting mode T$_2$, and for even higher energy to T$_1$, see Fig.~\ref{fig:bifdiag_a_b}(a). 
Concerning the classical mechanics, the border between reflection and transmission is given by 
$s^2_2=a^2$.
This border is crossed for the modes $\vert n_\zeta;\pi_y\rangle$ for different energies.
Upon crossing the border the tunnel integral \eqref{eq:tunnelintegr_2D} changes sign 
and the transmission probability \eqref{eq:transmission_prob_2D}   changes from 0 to 1. 
The energy for which the tunnel integral of a given mode $\vert n_\zeta;\pi_y\rangle$ is zero, and hence gives $T_{(n_\zeta;\pi_y)}(E)=1/2$, can be defined as the energy at which the mode ``opens'' as a transmission channel.
Marking these energies on the energy axis in Fig.~\ref{fig:conductance_a_2D} 
we see in which order the transmission channels open and this way contribute a step of $N(E)$. Semiclassically these ``opening'' energies are identical to the eigenenergies of a square well.


\subsubsection{The classical transmission}

The classical transmission probability can be computed from the 
\emph{directional}  flux through  the dividing surface DS of energy $E$ defined in \eqref{eq:def_DS_2D}, or following our discussion at the beginning of this section by an integral over the forward hemsiphere DS$_{\text{f}}$ of this dividing surface.
In a more modern notation which also reveals the symplectic nature of the flux (see \cite{MacKay1,MacKay2} and also \cite{WaalkensWiggins04}) the flux is given by
\begin{equation}
f(E) = \int_{\text{DS}_{\text{f}} }    \omega\,,
\end{equation}
where $\omega$ is the symplectic  2-form
\begin{equation}
\omega = \text{d}x\wedge \text{d}p_x +  \text{d}y\wedge \text{d}p_y\,.
\end{equation}
Since $\omega=\ud \phi$ where $\phi$ is the Liouville 1-form 
\begin{equation}
\phi = p_x\, \text{d}x + p_y\, \text{d}y
\end{equation}
we can utilize Stokes' theorem to compute $f(E)$ from integrating $\phi$ over the boundary of the forward hemisphere DS$_{\text{f}}$. Using the fact that the boundary  of DS$_{\text{f}} $ is given by the transition state TS consisting of the periodic orbit along the $y$ axis at energy $E$ (see 
Sec.~\ref{subsection:hyperbola_classical}) we find that the flux is given by the Liouville action of the periodic orbit,
\begin{equation}
f(E) = \int_{\text{TS}} 
    \phi  = 4 \sqrt{2mE} \,.
\end{equation}

In order to make the comparison to the cumulative reaction probability $N(E)$ we consider the dimensionless quantity
\begin{equation}  \label{eq:NWeyl2D}
N_{\text{Weyl}} (E) = \frac{1}{2\pi\hbar} f(E)\,,
\end{equation}
which is shown together with $N(E)$ in Fig.~\ref{fig:conductance_a_2D}. We see that  $N_{\text{Weyl}} (E)$ gives an approximate smooth local average of
$N(E)$ which however overestimates the local average of $N(E)$ as the graph of $N_{\text{Weyl}} (E)$ intersects the graph of $N(E)$ at the top of its steps. 
In fact disregarding the tunneling, $N(E)$ simply gives the integrated density of states of the transition state or activated complex (the one-dimensional square well along the $y$ axis) to energy $E$. The term  $N_{\text{Weyl}} (E)$ is the Weyl approximation of this quantity. 
As also shown in Fig.~\ref{fig:conductance_a_2D}  one can obtain a better local average by modifying $N_{\text{Weyl}} (E)$ to
\begin{equation} \label{eq:NWeyltilde2D}
\tilde{N}_{\text{Weyl}} (E) = \frac{1}{2\pi\hbar} f(E)-\frac12 \,,
\end{equation}
which can be formally derived from counting the mean number of states to energy $E$ in a one-dimensional square well potential.


\subsection{The 3D system}
\label{sec:cond_3D}

\subsubsection{The quantum transmission}
\label{sec:quantum_trans_3D}

To compute the transmission probabilities $T_n$ in the 3D case 
we look for solutions of the form
\begin{equation}
\label{eq:exactsol_top}
\psi_{\zeta;n}(\zeta) \psi_{\eta;n}(\eta)     \psi_{\xi;n}(\xi) + r_n \psi_{\zeta;n}(\zeta)    \psi_{\eta;n}(\eta)   \psi^*_{\xi;n}(\xi)
\end{equation}
at the bottom ($x\ll -1$) and
\begin{equation}
\label{eq:exactsol_bot}
t_n \psi_{\zeta;n}(\zeta)  \psi_{\eta;n}(\eta)     \psi^*_{\xi;n}(\xi)
\end{equation}
at the top ($x\gg 1$).
In this case we first solve the components of the 
separated wave equations 
\eqref{eq:sephelmholtz} and the corresponding boundary conditions 
which belong to the transversal coordinates  $\zeta$ and $\eta$ with the energy $E$ as a parameter. 
The boundary conditions for $\psi_\eta$ are given by the parity
$\pi_z$ which yields the index of $\psi_\eta$ at $\eta=c$ and the Dirichlet boundary condition $\psi_\eta(1)=0$. 
The boundary conditions for $\psi_\zeta$ are determined by the parities $\pi_z$ and $\pi_y$: $\pi_z$ determines the index of $\psi_\zeta$ at 
$\zeta=c$ and $\pi_y$ determines whether $\psi'_\zeta(0)=0$, $\psi_\zeta(0)=1$ ($\pi_y=+$) or $\psi_\zeta(0)=0$, $\psi_\zeta'(0)=1$ ($\pi_y=-$). 
This way we obtain modes that are parametrized by $E$ and which we label by the Dirac kets
$\vert n_\zeta,n_\eta;\pi_y,\pi_z\rangle$, where $n_\zeta$ and $n_\eta$ are non-negative quantum numbers which
give the number of nodes of $\psi_\eta$ and $\psi_\zeta$ in the open 
intervals $c<\eta<1$ and $0<\zeta<c$, 
respectively. The modes for energy $E$ determine the separation constants 
$(k_{(n_\zeta,n_\eta;\pi_y,\pi_z)}(E),l_{(n_\zeta,n_\eta;\pi_y,\pi_z)}(E))$.
These can then be used in the $\xi$ component of the equations (\ref{eq:sephelmholtz}) to find solutions of the form (\ref{eq:exactsol_top}) and (\ref{eq:exactsol_bot}). Like in the 2D case we resort to a semiclassical computation of the transmission probabilities instead. Analogously to (\ref{eq:transmission_prob_2D})
we obtain
\begin{equation}\label{eq:transmission_prob}
T_{(n_\zeta,n_\eta;\pi_y,\pi_z)}(E) = 
\frac{1}{1+\exp (\theta_{(n_\zeta,n_\eta;\pi_y,\pi_z)}(E)/\hbar)}\,,
\end{equation}
where $\theta_{(n_\zeta,n_\eta;\pi_y,\pi_z)}(E) $ is the tunnel integral
\begin{eqnarray}
\label{eq:tunnelintegr}
&\theta&_{(n_\zeta,n_\eta;\pi_y,\pi_z)}(E) = -2\ui  \int_{\lambda_-}^{\lambda_+} p_\lambda \,\mbox{d}\lambda \\\nonumber
&=& - 4 \ui  \sqrt{2mE} \int_a^{s_2}
\sqrt{\frac{\xi^4-2k_{(n_\zeta,n_\eta;\pi_y,\pi_z)}(E)\xi^2+l_{(n_\zeta,n_\eta;\pi_y,\pi_z)}(E)}{(\xi^2-a^2)(\xi^2-c^2)}}
\,\mbox{d}\xi\,,
\end{eqnarray}
which describes the tunneling through the potential barrier of the effective potential $V_{\lambda, \textrm{eff}}$ for types of motion WG$_{2/3}$ and BB$_{2/3}$ (see the corresponding phase portraits in Fig.~\ref{fig:pots_xfig_c}).
The branches of the square root in (\ref{eq:tunnelintegr}) are again chosen in such a way 
that the tunnel integral is positive if $a^2<s^2_2$ (corresponding to classical reflection of type WG$_{3}$ or BB$_{3}$) and negative if $s^2_2<a^2$ (corresponding to classical transmission of type WG$_{2}$ or BB$_{2}$). 
We can again make this more precise by relating $\theta_{(n_\zeta,n_\eta;\pi_y,\pi_z)}(E) $ to the integral $I_\xi$ that we defined in \eqref{eq:def_I_xi_3D}. This gives
\begin{equation}\label{xi_Theta_relation_3D}
I_\xi = -\frac{1}{2\pi} \theta_{(n_\zeta,n_\eta;\pi_y,\pi_z)}(E) \,.
\end{equation}
Like in the case of the 2D system we solve the boundary value problems for $\psi_\zeta$ and $\psi_\eta$ by a shooting method. To this end we again need good starting values for the separation constants $l_{(n_\zeta,n_\eta;\pi_y,\pi_z)}(E)$ and $k_{(n_\zeta,n_\eta;\pi_y,\pi_z)}(E)$ which we obtain from a semiclassical computation. In contrast to the 2D system we here face the problem that the 
types of motion  WG$_{2/3}$ and BB$_{2/3}$ differ with respect to their degrees of freedom $\zeta$ and $\eta$, or equivalently $\nu$ and $\mu$ (see the corresponding phase portraits in Fig.~\ref{fig:pots_xfig_c}). 
Accordingly, the EBK quantizations of the actions are different. The conditions are
\begin{equation}\label{eq:BB_EBK_full}
I_\nu = \big( n_\nu  + \frac{2}{4}\big ) \hbar  \,,\quad I_\mu =  \big( n_\mu +  \frac{4}{4}\big) \hbar
\end{equation}
for the  bouncing ball motions  of type BB$_{2/3}$, and 
\begin{equation}\label{eq:WG_EBK_full}
I_\nu = \big(n_\nu  + \frac{0}{4}\big) \hbar  \,,\quad 
I_\mu =  \big( n_\mu + \frac{3}{4}\big) \hbar
\end{equation}
for the whispering gallery motions  of type WG$_{2/3}$.
Writing these EBK quantization conditions in terms of the actions of the symmetry reduced system one finds
\begin{equation}\label{eq:BB_EBK}
\tilde{I}_\zeta = \big( n_\zeta  + \frac{1}{4} (2- \pi_y) \big ) \hbar  \,,\quad \tilde{I}_\eta =  \big( n_\eta +  \frac{1}{4} (3- \pi_z)  \big) \hbar
\end{equation}
for the  bouncing ball motions  of type BB$_{2/3}$, and 
\begin{equation}\label{eq:WG_EBK}
\tilde{I}_\zeta = \big(n_\zeta  + \frac{1}{4}(2-\pi_y-\pi_z)  \big) \hbar  \,,\quad 
\tilde{I}_\eta =  \big( n_\eta + \frac{1}{4}( 3 ) \big) \hbar
\end{equation}
for the whispering gallery motions  of type WG$_{2/3}$.
The quantum numbers $(n_\nu,n_\mu)$ of the full system and the quantum numbers $(n_\zeta,n_\eta)$ of the symmetry reduced system are related by
\begin{equation}\label{eq:full_reduced_bb}
n_\nu = 2n_\zeta+ \frac12 (2-\pi_y) \,,\quad 
n_\mu = 2n_\eta+\frac12 (1-\pi_z)\,,
\end{equation}
for the bouncing ball motions and
\begin{equation}\label{eq:full_reduced_wg}
n_\nu = 2n_\zeta+ \frac12 (2-\pi_y-\pi_z) \,,\quad
n_\mu = n_\eta\,,
\end{equation}
for the whispering gallery motions. We can overcome this problem of differing quantization conditions by introducing a {\em uniform} quantization of the actions $I_\zeta$ and $I_\eta$ which interpolates the EBK quantizations  in the regions  BB$_{2/3}$ and WG$_{2/3}$ in a smooth way. Using ideas similar to \cite{WWD99}
one finds that \eqref{eq:BB_EBK} and \eqref{eq:WG_EBK} can be written in the uniform way
\begin{equation}\label{eq:EBKzeta_eta_scatt}
\tilde{I}_\zeta=\big( n_\zeta + \frac{\alpha_\zeta}{4} \big) \hbar\,,\quad
\tilde{I}_\eta = \big( n_\eta + \frac{\alpha_\eta}{4} \big) \hbar\,,
\end{equation}
where again $n_\zeta,n_\eta \in \N_0$, and $\alpha_\zeta$ and $\alpha_\eta$ are ``effective Maslov indices'' given by
\begin{eqnarray}\label{eq:Maslov_zeta}
\alpha_\zeta &=& \pi_z\frac{2}{\pi}\arctan e^{\theta_\zeta/\hbar}+2-\pi_y-\pi_z\,, \\\label{eq:Maslov_eta}
\alpha_\eta &=& \pi_z\frac{2}{\pi}\arctan e^{\theta_\eta/\hbar}+3-\pi_z\,.
\end{eqnarray}
Here 
$\theta_\zeta$ and $\theta_\eta$
are again tunnel integrals which in this case describe the tunneling through the barriers of the effective potentials $V_{\nu, \textrm{eff}}$ and $V_{\mu, \textrm{eff}}$ in Fig.~\ref{fig:pots_xfig_c} that one needs to overcome to change
 between whispering gallery and bouncing 
ball type motions (see  \cite{WWD97} for more details).  The tunnel integrals $\theta_\zeta$ and $\theta_\eta$ are again best described
as integrals on the hyperelliptic curve $\Gamma_w$ in \eqref{eq:hyperelliptic_curve}. Interestingly  $\theta_\zeta=-\theta_\eta$ and
\begin{equation} \label{eq:Theta_zeta}
\theta_\zeta=-\theta_\eta=  -\frac{1}{2\pi} I_{\zeta\eta} \,,
\end{equation}
where $I_{\zeta\eta}$ is the integral we defined in \eqref{eq:def_Izetaeta}.
For $\theta_\zeta=-\theta_\eta \gg \hbar$, and using (\ref{eq:full_reduced_bb}), one recovers the quantization conditions for the bouncing ball type motions in \eqref{eq:BB_EBK_full}; for  $-\theta_\zeta=\theta_\eta \gg \hbar$, and using (\ref{eq:full_reduced_wg}), one recovers the quantization conditions for the whispering gallery type motions in \eqref{eq:WG_EBK_full}.

In contrast to the other motions, we note that the types of motion BB$_1$ and WG$_1$ involve a smooth rather than a hard wall reflection in the $\mu$ degree of freedom which leads to yet another set of EBK quantization conditions. However, as we will see below, for the energies under considerations BB$_1$ and WG$_1$ play no role for the transmission problem (see the discussion for the analogous effect in the 2D in Sec.~\ref{sec:cond_2D_qm}).   The uniform quantization conditions \eqref{eq:EBKzeta_eta_scatt} can be solved by 
a standard Newton procedure.
The resulting values for $s_1^2$ and $s_2^2$ for given quantum numbers $n_\zeta$ and $n_\eta$, and parities 
$\pi_y$ and $\pi_z$, are then used as the starting values for the shooting method to solve the $\zeta$ and $\eta$ components of the wave equations as described above, and hence  to compute the transmission probability in \eqref{eq:transmission_prob}. The cumulative reaction probability $N(E)$ is the sum over all these transmission probabilities. As in the 2D case, to numerically compute $N(E)$ we need only consider the finite number of modes which, at a value $E>0$, have a nonnegleglibile transmission probability. A graph of $N(E)$ is shown in Fig.~\ref{fig:conductance_a}. Depending on the shape parameters $(a^2,c^2)$ for the boundary hyperboloid the cumulative reaction probability  shows more or 
less pronounced steps which in contrast to the 2D case (see Fig.~\ref{fig:conductance_a_2D}) are of size 1 or 2. 
\begin{figure}
\centerline{
\begin{tabular}{c}          
\includegraphics[angle=0,width=12cm]{fig15}\\
\tiny
\begin{tabular}{|c|c|c|c|c|c|c|c|}
\hline
$\kappa$ axis tick & $\vert 1\rangle$ & $\vert 2\rangle$ & $\vert 3\rangle$ & $\vert 4\rangle$ & $\vert 5\rangle$ & $\vert 6\rangle$ & $\vert 7\rangle$ \\ \hline
$\vert n_\zeta, n_\eta; \pi_y, \pi_z \rangle$ & 
$\vert 0,0;+,+\rangle$ & 
$\vert 0,0;-,+\rangle$ & 
$\vert 0,0;+,-\rangle$ & 
$\vert 1,0;+,+\rangle$ & 
$\vert 0,0;-,-\rangle$ & 
$\vert 0,1;+,+\rangle$ & 
$\vert 1,0;-,+\rangle$  \\ \hline\hline
$\vert 8\rangle$ & $\vert 9\rangle$ & $\vert 10\rangle$ & $\vert 11\rangle$ & $\vert 12\rangle$ & $\vert 13\rangle$ & $\vert 14\rangle$ & $\vert 15\rangle$  \\ \hline
 $\vert 1,0;+,-\rangle$ &
$\vert 0,1;-,+\rangle$ & 
$\vert 0,1;+,-\rangle$& 
$\vert 2,0;+,+\rangle$ &
 $\vert 1,0;-,-\rangle$ & 
 $\vert 1,1;+,+\rangle$ & 
 $\vert 0,1;-,-\rangle$ & 
 $\vert 2,0;-,+\rangle$  \\ \hline\hline
$\vert 16\rangle$ & $\vert 17\rangle$ & $\vert 18\rangle$ & $\vert 19\rangle$ & $\vert 20\rangle$ & $\vert 21\rangle$ & $\vert 22\rangle$ & \phantom{1} \\ \hline
 $\vert 2,0;+,-\rangle$ &
$\vert 0,2;+,+\rangle$ & 
$\vert 1,1,-,+\rangle$ &
 $\vert 1,1;+,-\rangle$ & 
 $\vert 3,0;+,+\rangle$ & 
 $\vert 2,0;-,-\rangle$ & 
 $\vert 0,2;-,+\rangle$ & 
\phantom{1} \\ \hline\hline
\end{tabular}
\end{tabular}
}
\caption{
\label{fig:conductance_a} 
\textit{Top Panel:} Cumulative reaction probability $N(E)$ as a function of the wavenumber $\kappa=\sqrt{2mE}/\hbar$ for shape parameters $(a^2,c^2)=(5,0.2)$, and, for comparison, $(a^2,c^2)=(\infty,0.2)$, which corresponds to the transmission through a cylinder with elliptical cross-section. The ticks on the wavenumber axis mark the energies at which, for $(a^2,c^2)=(5,0.2)$, the transmission channels  $\vert n_\zeta,n_\eta;\pi_y,\pi_z\rangle$ ``open'' (see text) (the key to the tick labels is given in the table). 
For pairs of near degenerate states the one corresponding to the higher wavenumber is marked above the $\kappa$ axis.
The smooth dot-dashed  blue curve and the solid  blue curve are the Weyl approximations of $N(E)$ defined in \eqref{eq:Weyl3D} and \eqref{eq:Weyl3D_correction}, respectively.
\textit{Bottom Panel:} Resonances in the complex wavenumber ($\kappa$) plane, for $a^2=5$. Semiclassical resonances are marked by pluses ($+$) and exact resonances by diamonds ($\Diamond$). Note that we use the symbol $\kappa$ for wavenumber to distinguish it from separation constant $k$. ($\hbar=1$, $m=1$.)
}
\end{figure}
\begin{figure}
\centerline{
\includegraphics[angle=0,width=7.5cm]{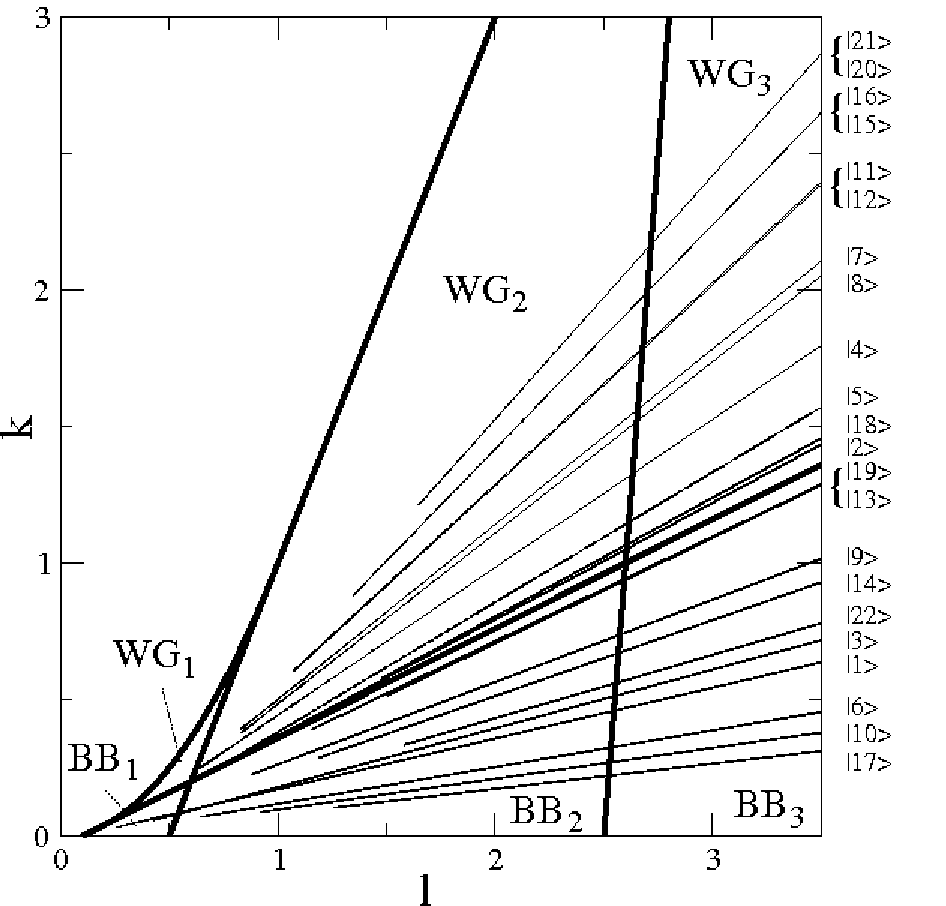}
}
\caption{\label{fig:conductance_b}
The $(k,l)$-spectra of the modes  leading to the jumps of the cumulative reaction probability in Fig.~\ref{fig:conductance_a} for $(a^2,c^2)=(5,0.2)$. 
For each shown mode 
$\vert n_\zeta,n_\eta;\pi_y,\pi_z\rangle$ (see the table in Fig.~\ref{fig:conductance_a}) the energy $E$ is varied from 2 (for which $(k_{(n_\zeta,n_\eta;\pi_y,\pi_z)}(E),l_{(n_\zeta,n_\eta;\pi_y,\pi_z)}(E))$ are in the reflective 
types of motion BB$_3$ or WG$_3$ beyond the right border of the shown region) to $E=100$ for which
$(k_{(n_\zeta,n_\eta;\pi_y,\pi_z)}(E),l_{(n_\zeta,n_\eta;\pi_y,\pi_z)}(E))$ are in one of the classically transmitting  types of motion BB$_1$,
BB$_2$, WG$_1$ or WG$_2$. The bold lines mark the classical bifurcation diagram.
}
\end{figure}
This can be understood in more detail if we relate the modes  $\vert n_\zeta,n_\eta;\pi_y,\pi_z\rangle$ to the classical motions. 
For a given energy $E$,
this relationship is established via the separation constants $(k_{(n_\zeta,n_\eta;\pi_y,\pi_z)}(E),l_{(n_\zeta,n_\eta;\pi_y,\pi_z)}(E))$ 
which determine the corresponding toroidal cylinders. The wave functions of the modes $\vert n_\zeta,n_\eta;\pi_y,\pi_z\rangle$ 
are mainly ``concentrated'' on the projections of the corresponding toroidal cylinders to configuration space. 
As can be seen in Fig.~\ref{fig:caustics}, for the whispering gallery types of motion,
these projections become increasingly confined  in the order 
WG$_3\rightarrow$ WG$_2\rightarrow$ WG$_1$. For the bouncing ball types of motion the confinement increases 
in the order BB$_3\rightarrow$ BB$_2\rightarrow$ BB$_1$.
Since high confinement in configuration space implies high kinetic energy via the Heisenberg uncertainty principle, 
the modes, which classically correspond to the types of motion WG$_1$ or BB$_1$, have highest energy.
In fact, for low energies all modes have $(k_{(n_\zeta,n_\eta;\pi_y,\pi_z)}(E),l_{(n_\zeta,n_\eta;\pi_y,\pi_z)}(E))$ in the classically
reflective  types of motion
WG$_3$ or BB$_3$. Upon increasing the energy  the $(k_{(n_\zeta,n_\eta;\pi_y,\pi_z)}(E),l_{(n_\zeta,n_\eta;\pi_y,\pi_z)}(E))$  wander
towards the transmitting  modes WG$_2$ or BB$_2$, and for even higher energy to WG$_1$ or BB$_1$, see Fig.~\ref{fig:bifdiag_a_b}. 
Concerning the classical mechanics, the border between reflection and transmission is given by 
$s^2_2=a^2$ or $l = k^2 - (a^2-k)^2$.
This border is crossed for the modes $\vert n_\zeta,n_\eta;\pi_y,\pi_z\rangle$ for different energies.
Upon crossing the border the tunnel integral changes sign 
and the transmission probability changes from 0 to 1.
The energy for which the tunnel integral of a given mode $\vert n_\zeta,n_\eta;\pi_y,\pi_z\rangle$  is zero, and hence gives $T_{(n_\zeta,n_\eta;\pi_y,\pi_z)}(E) =1/2$, can be defined can be defined as the energy at which the mode opens as a transmission channel (see the analogous definition for the 2D case). These energies are marked on the energy axis in Fig.~\ref{fig:conductance_a}.
Semiclassically these opening energies are identical to the eigenenergies of the ellipse billiard.

Classically, the border $s^2_2=a^2$ corresponds to the unstable invariant motion in the $y-z$ plane. 
This is the planar billiard in the bottleneck ellipse which is an invariant subsystem 
with one degree of freedom less than the full three-dimensional billiard. 

\begin{figure*}
\centerline{
\includegraphics[angle=0,width=12cm]{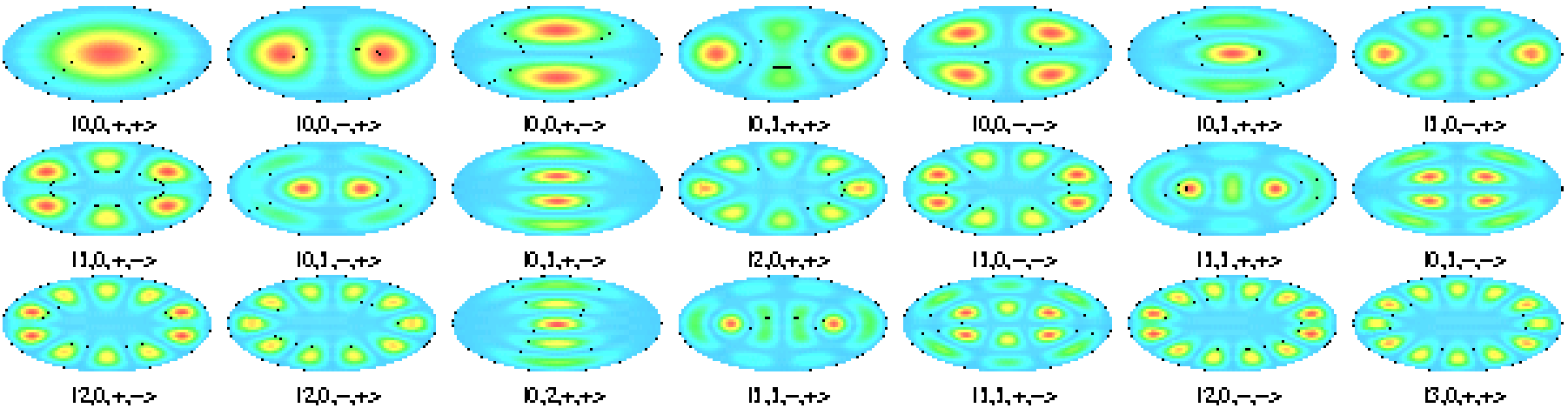}
}
\caption{\label{fig:waves}
Probability contours in the section $x=0$ of the wave functions   of the modes 
$\vert n_\zeta,n_\eta;\pi_y,\pi_z\rangle$ at the moment when they ``open'' as transition channels (see text). 
The wave functions are displayed in the order they contribute to the cumulative reaction probability  in Fig.~\ref{fig:conductance_a}.
Light blue corresponds to low probability, red corresponds to high probability.
The black lines mark the envelopes  of the corresponding classical motion. Ellipses indicate whispering gallery modes; hyperbolas indicate bouncing ball modes.
}
\end{figure*}

Due to the dynamical barrier
the wave functions of the modes deep in the reflective types of motion WG$_3$ and BB$_3$
have negligible amplitudes in the $y-z$ plane. As the energy increases 
the increase of the amplitudes is indicated by the switching of the corresponding transmission probability 
$T_{(n_\zeta,n_\eta;\pi_y,\pi_z)}$ from 0 to 1, i.e. 
the ``opening'' of a new transmission channel. 
The wave functions  of the transmission channels which lead to the step in Fig.~\ref{fig:conductance_a} 
are shown  in Fig.~\ref{fig:waves} as their intersection with the $y-z$ plane.

The quantum mechanical manifestation of the two senses of rotation in the whispering gallery types of motion is the energetic quasi-degeneracy of the 
corresponding modes $\vert n_\zeta,n_\eta;\pi_y,\pi_z\rangle$. 
The further the separation constants $(k_{(n_\zeta,n_\eta;\pi_y,\pi_z)}(E)$, $l_{(n_\zeta,n_\eta;\pi_y,\pi_z)}(E))$ in the whispering gallery types of motion
lie away from the border 
$s^2_1=c^2$ to bouncing ball motions, the higher  the effective energy 
$E_{\nu, \textrm{eff}}$ lies above the effective potential $V_{\nu, \textrm{eff}}$.
In this limit the role of the potential becomes negligible and the energy is essentially determined by the total number of nodes of $\psi_\nu$ along
a complete $\nu$-loop which is an ellipse in Fig.~\ref{fig:waves}.
The relation (\ref{eq:full_reduced_wg}) between the quantum numbers of the full system and the quantum numbers of the symmetry reduced system leads to the energetic degeneracy of the two pairs of modes 
\begin{eqnarray}
\label{eq:deg1}
&\vert n_\zeta+1,n_\eta,+,+\rangle &\leftrightarrow \quad \vert n_\zeta,n_\eta,-,-\rangle\,,\\
\label{eq:deg2}
&\vert n_\zeta,n_\eta,+,-\rangle   &\leftrightarrow \quad \vert n_\zeta,n_\eta,-,+\rangle\,.
\end{eqnarray}
In Fig.~\ref{fig:conductance_a}   this effect is seen for the pairs of modes
$\vert1,0,+,+\rangle$ and $\vert0,0,-,-\rangle$, 
$\vert1,0,-,+\rangle$ and $\vert1,0,+,-\rangle$, 
$\vert2,0,+,+\rangle$ and $\vert1,0,-,-\rangle$, 
$\vert2,0,+,-\rangle$ and $\vert2,0,-,+\rangle$,
$\vert2,0,-,-\rangle$ and $\vert3,0,+,+\rangle$
which,  on the energy axis, become more and more indistinguishable as energy increases, and this way effectively lead to steps  of 
size 2 (see also Fig.~\ref{fig:conductance_b} and the plot of the wavefunctions in Fig.~\ref{fig:waves}).
There is no analogous degeneracy for the bouncing ball type states, as can be deduced from the relation (\ref{eq:full_reduced_bb}).


\subsubsection{The classical transmission}

In order to compute the directional flux through the 
fourdimensional dividing surface \eqref{eq:def_DS_3D} of the 3D system 
we consider the symplectic 2-form
\begin{equation}
\omega =  \text{d}p_x\wedge \text{d}x +  \text{d}p_y\wedge \text{d}y + \text{d}p_z\wedge \text{d}z\,,
\end{equation}
from which we can define the 4-form
\begin{equation}
\Omega' = \frac12 \omega^2 =  \text{d}p_x\wedge \text{d}x\wedge \text{d}p_y\wedge \text{d}y + \text{d}p_x\wedge \text{d}x\wedge \text{d}p_z\wedge \text{d}z + \text{d}p_y\wedge \text{d}y\wedge \text{d}p_z\wedge \text{d}z \,.
\end{equation}
The directional flux through the dividing surface \eqref{eq:def_DS_3D} is then obtained from integrating $\Omega'$ over the forward hemisphere DS$_{\text{f}}$ defined in \eqref{eq:forward_div_surf}, i.e.
\begin{equation}\label{eq:3D_flux}
    f (E) = \int_{\text{DS}_{\text{f}} }    \Omega' \,.
\end{equation}
Noting that
\begin{equation}
\Omega' = \text{d} \phi \,,
\end{equation}
where $\phi$ is the 3-form
\begin{equation} 
\phi = \frac12 (p_x \text{d}x + p_y \text{d}y + p_z \text{d}z  )\wedge \omega 
\end{equation}
we can again use Stokes' theorem to compute the flux from the integral over the boundary of DS$_{\text{f}}$ which is the transition state
TS consisting of 
the invariant billiard in the bottleneck ellipse at energy $E$. Hence, 
\begin{equation} \label{eq:3Dflux_Stokes}
f(E) =  \int_{\text{TS}} \phi \,.
\end{equation}
Using either \eqref{eq:3D_flux} or  \eqref{eq:3Dflux_Stokes} we get 
\begin{equation}
    f (E) =  2m  \pi^2 \sqrt{1-c^2}  E\,,
\end{equation}
which is the product of the area, $A=\pi \sqrt{1-c^2}$, of the bottleneck ellipse and the area of the circular disk of radius $\sqrt{2mE}$ in the two-dimensional momentum space $(p_y,p_z)\in\R^2$.

In order to relate the flux to the cumulative reaction probability we introduce the dimensionless quantity
\begin{equation}\label{eq:Weyl3D}
N_{\text{Weyl}} (E) = \frac{1}{(2\pi \hbar)^2} f(E) = \frac{A}{4\pi}  \frac{2mE}{\hbar^2}\,.
\end{equation}
Comparing the graphs of  $N_{\text{Weyl}} (E) $ and $N(E)$ in Fig.~\ref{fig:conductance_a} we see that $N_{\text{Weyl}} (E) $ overestimates the local average of $N(E)$. This is an indication that quantum effects are quite severe in this system. 
Using the fact that, neglecting quantum mechanical tunneling through the dynamic barrier, $N(E)$ is essentially the number of states of the billiard in the bottleneck ellipse to energy $E$  we can 
introduce correction terms to $N_{\text{Weyl}} (E) $ of which the first is proportional to $\sqrt{E}$ and depends on the length, $L$, of the perimeter of the boundary ellipse and the second is a constant term resulting from integrating the Gauss curvature along the perimeter of the bottleneck ellipse \cite{BH76}. This way we get
\begin{equation}\label{eq:Weyl3D_correction}
\tilde{N}_{\text{Weyl}} (E) = \frac{A}{(2\pi \hbar)^2}  \frac{2mE}{\hbar^2}   - \frac{L}{4\pi} \frac{\sqrt{2mE}}{\hbar} + \frac{1}{6}   \,,
\end{equation}
where $L=4E(c)$ with  $E(c)$ denoting Legendre's complete elliptic integral of the second kind with modulus $c$. The graph of $\tilde{N}_{\text{Weyl}} (E)$ 
is also shown in Fig.~\ref{fig:conductance_a} and in fact gives a very good local average of $N(E)$.

%% file: sec6_resonances.tex
\section{Quantum resonances}
\label{sec:resonances}

In Sec.~\ref{sec:classical} we have seen that the classical systems possess invariant subsystems  of one degree of freedom less than the full system contained in the respective phase space bottlenecks. These systems which form the transition states were given by a one-dimensional billiard in a square well in the 2D case and the invariant elliptic billiard in the 3D case.  The Heisenberg uncertainty relation rules out the existence of analogous invariant subsystems in the corresponding quantum mechanical problems. In fact, a wavepacket initialised on such an invariant subsystem will decay exponentially fast in time. This exponential decay is described by the resonances \cite{WaalkensSchubertWiggins08}.

The resonances can be formally defined as the poles of the meromorphic continuation of the transmission probabilities to the lower half of the complex energy plane.  A semiclassical approximation of the resonances can thus be obtained from  
the poles  of the  expressions of the transmission probabilities  we have given in  \eqref{eq:transmission_prob_2D} (for the 2D case) and \eqref{eq:transmission_prob} (for the 3D case). This leads to the complex EBK type quantization condition for the tunnel integrals $\theta$ defined  in \eqref{eq:tunnelintegr_2D} or \eqref{eq:tunnelintegr} given by 
\begin{equation} \label{eq:res_cond}
 \theta =  \ui  \pi  \hbar (2 n_\lambda + 1)\,, n_\lambda \in \N_0\,,
\end{equation}
or equivalently
\begin{equation} \label{eq:res_cond_I_xi}
I_\xi = -\ui \hbar(n_\lambda + \frac12 ) \,, n_\lambda \in \N_0\,,
\end{equation}
(see Equations~\eqref{xi_Theta_relation_2D} and \eqref{xi_Theta_relation_3D}.)
In the following we study this semiclassical approach to the computation of resonances and compare it with a numerical computation of the exact resonances based on the complex scaling method.


\subsection{The 2D system}

 \subsubsection{Semiclassical computation of resonances}
 \label{sec:res_2D_semi}

We at first  compute the resonances semiclassically. To this end we have to simultaneously solve the (standard) EBK quantization condition \eqref{eq:EBK_tilde_I_zeta}, i.e. 
\begin{equation} \label{eq:zeta_EBK_reson_2D}
  \tilde{I}_\zeta = \hbar( n_\zeta+\frac{1}{4}(3-\pi_y))\,,\quad n_\zeta \in \N_0\,,
\end{equation}
in combination with the 
complex EBK quantization condition
\eqref{eq:res_cond_I_xi} which rewritten in terms of the action $\tilde{I}_\xi$ of the symmetry reduced system defined in \eqref{eq:def_I_xi_reduced_2D} gives
\begin{equation} \label{eq:I_xi_red_EBK_2D}
   \tilde{I}_\xi = -\ui \hbar (n_\xi + \frac{1}{4}(2-\pi_x) \,,\quad n_\xi \in \N_0 \,.
\end{equation}
As we will see below when discussing the numerically exact  resonances this quantization condition decomposes the resonance states with respect to their parity $\pi_x$. Note that similar to the semiclassical computation of the cumulative reaction probability in Sec.~\ref{sec:cond_2D_qm} we assume that the type of motion T$_1$ also plays no role for the computation of the resonances. Otherwise \eqref{eq:zeta_EBK_reson_2D} would have to be replaced by a uniform quantization condition which interpolates between T$_1$ and T$_2$, i.e. across $s_2^2=1$ (see Sec.\ref{subsection:hyperbola_classical}). This assumption is justified by the fact that the resonances are associated with the activated complex consisting of the classically invariant billiard in the 
one-dimensional square well potential which has $s_2^2=a^2$, and the resonances can be expected to have values $s_2^2$ near $a^2$ and hence stay away from $s_2^2=1$. We will see that this assumption is indeed fulfilled.

The solutions  $E$ and $s^2_2$ of the quantization conditions \eqref{eq:zeta_EBK_reson_2D}  and 
\eqref{eq:I_xi_red_EBK_2D} are complex valued.  
The integration paths $\gamma_\zeta$ and $\gamma_\xi$ defining $I_\zeta$ and $I_\xi$ in \eqref{eq:I_s_on_Gamma} and \eqref{eq:def_I_xi_2D} therefore have to  be continued accordingly into the complex plane (see Fig.~\ref{fig:ellcurve_res_2D})(c).
They can be found numerically using a standard Newton procedure. To this end one has to decompose \eqref{eq:zeta_EBK_reson_2D}  and \eqref{eq:I_xi_red_EBK_2D} with respect to their real and imaginary parts which leads to a four-dimensional Newton procedure. We note that this procedure is less robust than in the real case in 
Sec.~\ref{sec:cond_2D_qm}. In particular, the procedure struggles when the energies $E$ are close to the imaginary axis.
For fixed quantum number $n_\zeta$ and parity $\pi_y$ we find the resonances by starting near to, but not at, the corresponding ``opening'' tick on the real energy axis of Fig.~\ref{fig:conductance_a_2D} and by smoothly moving the parameter $s_2^2$ into the complex plane. We then go through the grid $(n_\xi,\pi_x)\in \N_0\times\{-1,1\}$. We give a list of the resulting complex energies $E$ and separation constants $s_2^2$ in Tab.~\ref{tab:resonances_2D}.

\begin{table*}[htbp]
\begin{center}
\tiny
\begin{tabular}{|cc|cc||c|c||c|c||c|}
\hline
$n_\zeta$ & $n_\xi$ & $\pi_y$ & $\pi_x$ & $E_{\text{qm}}$ & $s^2_{2,\,\text{qm}}$ & $E_{sc}$ & $s^2_{2,\,\text{sc}}$  & $\Delta E$ \\ \hline\hline
0 & 0 & $+$ & $+$ & $1.1525-\ui0.3548$ & $4.6850+\ui1.4005$ & $1.1891-\ui0.3760$ & $4.7356+\ui1.3858$ & $3.42$ \\ \hline
0 & 0 & $+$ & $-$ & $0.8269-\ui1.0267$ & $2.8677+\ui3.3911$ & $0.8295-\ui1.0807$ & $2.9113+\ui3.3371$ & $3.35$ \\ \hline
0 & 0 & $-$ & $+$ & $4.8468-\ui0.7422$ & $4.9205+\ui0.7090$ & $4.8902-\ui0.7549$ & $4.9331+\ui0.7070$ & $0.91$ \\ \hline
0 & 0 & $-$ & $-$ & $4.5011-\ui2.2053$ & $4.4057+\ui2.0161$ & $4.5330-\ui2.2413$ & $4.4171+\ui2.0098$ & $0.89$ \\ \hline
1 & 0 & $+$ & $+$ & $11.0135-\ui1.1243$ & $4.9646+\ui0.4737$ & $11.0587-\ui1.1332$ & $4.9702+\ui0.4731$ & $0.41$ \\ \hline
1 & 0 & $+$ & $-$ & $10.6624-\ui3.3580$ & $4.7303+\ui1.3877$ & $10.7019-\ui3.3839$ & $4.7356+\ui1.3858$ & $0.41$ \\ \hline
1 & 0 & $-$ & $+$ & $19.6487-\ui1.5045$ & $4.9801+\ui0.3555$ & $19.6946-\ui1.5113$ & $4.9832+\ui0.3553$ & $0.23$ \\ \hline
1 & 0 & $-$ & $-$ & $19.2954-\ui4.5021$ & $4.8472+\ui1.0525$ & $19.3379-\ui4.5222$ & $4.8502+\ui1.0517$ & $0.23$ \\ \hline
2 & 0 & $+$ & $+$ & $30.7517-\ui1.8838$ & $4.9872+\ui0.2845$ & $30.7979-\ui1.8893$ & $4.9893+\ui0.2844$ & $0.15$ \\ \hline
2 & 0 & $+$ & $-$ & $30.3972-\ui5.6423$ & $4.9018+\ui0.8463$ & $30.4413-\ui5.6586$ & $4.9038+\ui0.8459$ & $0.15$ \\ \hline
2 & 0 & $-$ & $+$ & $44.3222-\ui2.2627$ & $4.9911+\ui0.2372$ & $44.3686-\ui2.2673$ & $4.9925+\ui0.2371$ & $0.11$ \\ \hline
2 & 0 & $-$ & $-$ & $43.9671-\ui6.7805$ & $4.9317+\ui0.7072$ & $44.0120-\ui6.7942$ & $4.9331+\ui0.7070$ & $0.1$ \\ \hline
\end{tabular}\end{center}
\caption{\label{tab:resonances_2D} The exact resonances $(E_{\text{qm}},s^2_{2,\,\text{qm}})$ and the semiclassical resonances $(E_{\text{sc}},s^2_{2,\,\text{sc}})$ of the 2D system for $\Re E<50$ and  $\{n_\xi,\pi_x\}=\{0,\pm\}$ (the first two families of resonances). The relative error $\Delta E=(\lvert E_{\text{sc}}\rvert - \lvert E_{\text{qm}}\rvert)/\lvert E_{\text{qm}}\rvert$ 
is given in percent.  ($\hbar=1$, $m=1$.)
}
\end{table*}

\begin{figure*}
\centerline{
\includegraphics[angle=0,width=9cm]{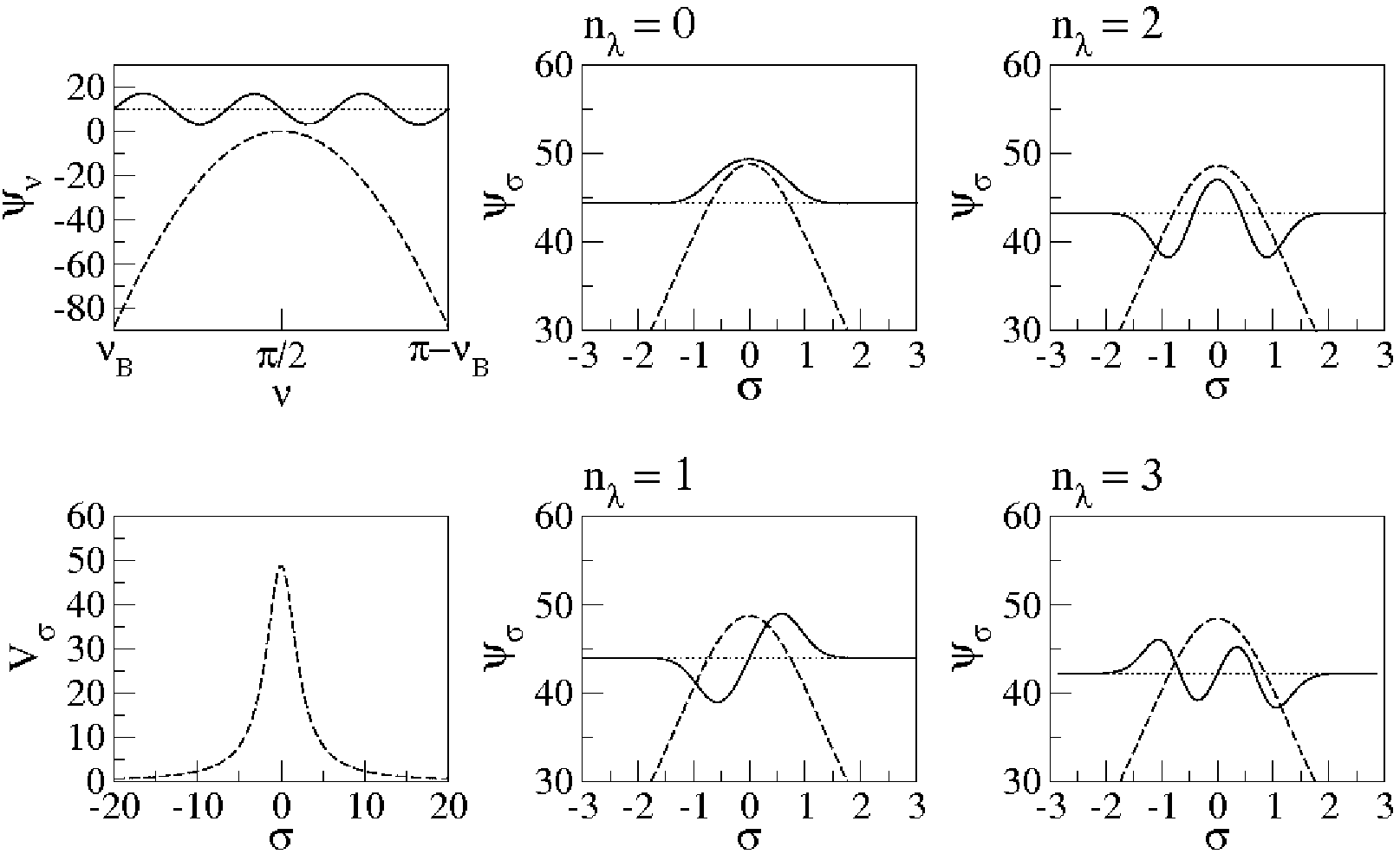}
}
\caption{\label{fig:wave_func_2D}
Real parts of the separated resonance wavefunctions (solid lines) of the four 
states $\vert n_\zeta,n_\xi;\pi_y,\pi_x\rangle$ with $n_\zeta= 2$, $n_\xi\in \{0,1\}$, $\pi_y=-$, and $\pi_x=\pm$. 
For fixed $\{n_\zeta,\pi_y\}$ (and hence fixed $n_\nu$), and different $n_\xi$ and $\pi_x$,  the wavefunctions $\psi_\nu$ are qualitatively the same 
(upper left panel).  For increasing $n_\lambda$, the four right hand panels show
 the increase in the number of nodes of the separated wavefunction in the reaction coordinate 
$\sigma$. The dashed lines are the real parts of the corresponding effective 
potentials $V_{\hat{s}, \textrm{eff}}(\hat{s})$, $\hat{s}\in\{ \nu, \lambda \}$. We add the real parts  of the effective 
energies $E_{\hat{s}, \textrm{eff}}$ (dotted lines)
to the amplitude of the  wavefunctions to visualize their energies relative the height of the potential barrier. 
The bottom left panel is 
the real part of a sample effective potentials $V_{\sigma, \textrm{eff}}$ over a larger $\sigma$ interval.
}
\end{figure*}


\subsubsection{Exact computation of resonances}
\label{sec:exact_res_2D}

We compute the (numerically) exact resonances from the complex scaling method \cite{Simon79,Moiseyev98}.
The main idea here is to turn the  wavefunctions which are associated with the resonances and exponentially divergent when the reaction coordinate goes to infinity into square integrable functions by a complex scaling of the reaction coordinate. In our case the reaction coordinate is given by $\xi$, or equivalently $\lambda$. 
In order to apply the complex scaling method we at first transform the $\lambda$ component of the separated Helmholtz equation (see \eqref{eq:helmreg_2D}),
\begin{equation}  \label{eq:lambda_wave_equation_2D}
    -\frac{\hbar^2}{2m} \frac{\ud^2\psi_\lambda }{\ud \lambda^2} = 
    E\, \big( \cosh\lambda^2 - s_2^2 \big)\, \psi_\lambda\,.
\end{equation}
The goal of the transformation is to get a wave equation which no longer involves an exponentially decreasing potential which would cause problems in the complex scaling method.
This can be achieved by rewriting \eqref{eq:lambda_wave_equation_2D} in terms of
\begin{equation}
\label{eq:restrans_2D}
\sigma(\lambda) =  a\sinh\lambda
\end{equation}
and the scaled wavefunction $\psi_\sigma$ defined by
\begin{equation}
\label{eq:restrans_wfunc_2D}  
\psi_\lambda(\lambda)  =  \frac{\psi_\sigma(\sigma)}{\sqrt{\ud\sigma/\ud\lambda}}= \frac{\psi_\sigma(\sigma)}{(\sigma^2+a^2)^{1/4}}  \,.
\end{equation}
This gives
\begin{equation}
\label{eq:res_wfunc_2D}
-\frac{\hbar^2}{2m }\frac{\ud^2 \psi_\sigma}{\ud \sigma^2} = \big( E_{\sigma, \text{eff}}-V_{\sigma, \text{eff}} \big)\psi_\sigma  \,,
\end{equation}
with the effective energy and potential given by
\begin{eqnarray}
E_{\sigma, \text{eff}}(\sigma) &=& E \,,\\
V_{\sigma, \text{eff}}(\sigma) &=&  \frac{Es_2^2+\frac{\hbar^2}{4m}}{\sigma^2+a^2}-  \frac{\hbar^2}{2m} \frac{3\sigma^2}{4(\sigma^2+a^2)^2}  \, ,\label{eq:effenerpot_res_2D}
\end{eqnarray}
respectively.
As opposed to the effective potential in \eqref{eq:lambda_wave_equation_2D},  the potential  $V_{\sigma, \text{eff}}$ goes to zero as $| \sigma |  \to \infty$, and accordingly, for real $E$ and $s_2^2$, the wave function $\psi_\sigma$ is a plane wave as $\sigma\to \infty$.
In fact, in terms of the original Cartesian coordinates   $(x,y)$ the solutions have to become plane waves for  $|x|\to \infty$.
From \eqref{eq:eqb3d_transxeztoxyz_a_2D} and noting that $\xi=a\cosh(\lambda)$ we see that  $\sigma=a \sinh(\lambda)$ is proportional to  $x$ and this is the motivation for the transformation  \eqref{eq:restrans_2D}.
The scaling of the wavefunction \eqref{eq:restrans_wfunc_2D} is performed in order to again obtain a system of type `kinetic-plus-potential'.

In order to compute the resonances we substitute  $\sigma$ in \eqref{eq:res_wfunc_2D} with $ \sigma \ue^{\ui\alpha}$. Upon this scaling the outgoing plane waves become asymptotically decreasing as $\sigma\to \infty $ provided that $\alpha>-\arg(E)$. \todo{Is this condition on $\alpha$ correct?}  

We implement the complex scaling method numerically using a shooting method. To this end we choose a suitably large but finite value  $\sigma_\infty$ at which we require the scaled wave function  $\psi_\sigma$ to vanish. 
The other boundary condition on $\psi_\sigma$ is given by  $\psi_\sigma(0)=0$, $\psi_\sigma'(0)=1$ if $\pi_x=-$, and $\psi_\sigma(0)=1$, $\psi_\sigma'(0)=0$ if $\pi_x=+$.
The boundary conditions for $\psi_\zeta$ are the same as for the scattering states in Sec.~\ref{sec:cond_2D_qm}.
In order to implement the complex scaling method we decompose the two equations into their real and imaginary parts. This then leads to a (real) four-dimensional Newton procedure on the complex two dimensional $E-s_2^2$ plane.  
As the starting values we use the semiclassical values for the resonances obtained as described above in Sec.~\ref{sec:res_2D_semi}.
We note that the complex scaling method is quite sensitive with respect to the choice of $\sigma_\infty$ and 
the scaling angle $\alpha$.
Like in the semiclassical computation this is particularly true for values of the energy near the imaginary axis. 
For the scaling used in these systems a ``suitable'' value for $\sigma_\infty$ ranges from $\approx1$ for resonances with $\Re (E)\approx500$ to $\approx5$ for resonances with $\Re (E)\approx3$.  An inappropriate value leads to apparent but false convergence.
In Tab.~\ref{tab:resonances_2D} the exact resonances computed this way  are compared to the corresponding semiclassical values. The relative error reaches its  maximum value of about 5 percent at the first state $\vert0,0;+,+\rangle$. The relative error shrinks rapidly for larger real parts of the resonance energies.

As shown in Fig.~\ref{fig:conductance_a_2D} the resonance energies (resp. the corresponding wavenumbers) form a grid in the complex energy (resp. wavenumber) plane. The grid sites can be labeled by the quantum numbers $n_\zeta$ and $n_\xi$ and the parities $\pi_x$ and $\pi_y$.  As also shown in Fig.~\ref{fig:conductance_a_2D} each step of the cumulative reaction probability is associated with one `string' of resonances of fixed $n_\zeta$ and $\pi_y$. We note that interestingly the values of $s_2^2$ are located along a smooth line in the complex $s_2^2$ plane. This can be understood semiclassically from taking the quotient of the EBK quantized actions $I_\xi$ and $I_\zeta$ whose integrals both scale with $\sqrt{E}$, and hence leads to the energy independent condition
\begin{equation}
- \frac{n_\xi+\frac{1}{4}(2-\pi_x)}{ n_\zeta +\frac{1}{4} (3-\pi_y)} = \frac{ \int_{\gamma_\xi} (z-s_2^2) \frac{\ud z}{w} }{ \int_{\gamma_\zeta} (z-s_2^2) \frac{\ud z}{w}  }
\end{equation} 
on $s_2^2$.

In Fig.~\ref{fig:wave_func_2D} we present the separated wavefunctions $\psi_\nu$ and $\psi_\sigma$ for a selection of resonances. 
We see that the scaled wavefunction obey the boundary conditions determined by the parities $\pi_x$ and $\pi_y$ and the exponential decay as $\sigma\to \infty$, and also have the expected number of nodes determined by the quantum numbers $n_\zeta$ and $n_\xi$.

\begin{figure*}
\centerline{
\includegraphics[angle=0,width=12cm]{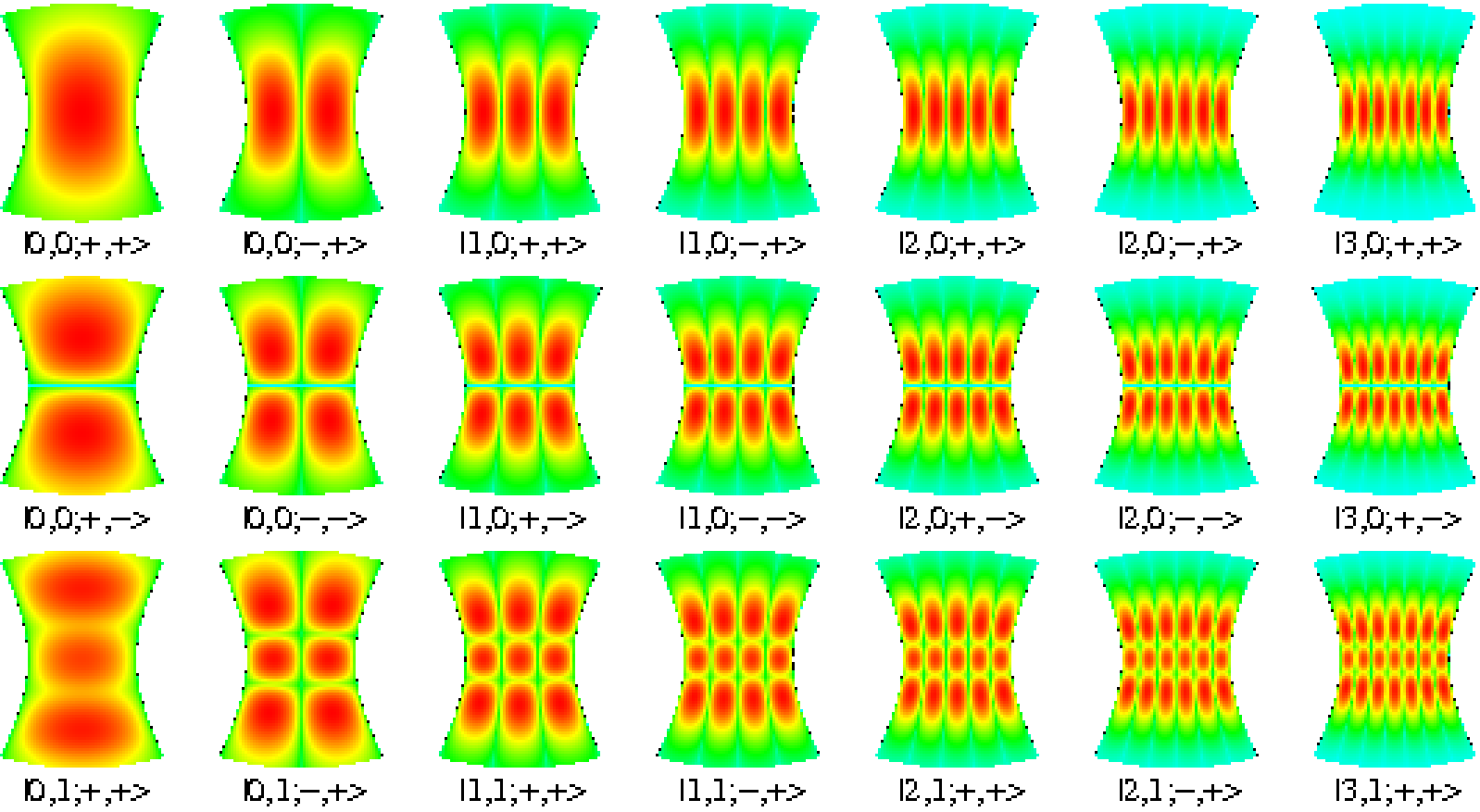}
}
\caption{\label{fig:isosurfaces_2D}
Two-dimensional contourplots of the densities of position in the $x-y$ plane  
for all the resonance wavefunctions  \eqref{eq:tot_wfunc_2D} of
 the 2D system, with $a^2=5$, for all the resonance states shown in 
 Fig.~\ref{fig:conductance_a_2D}. Light blue corresponds to low probability, red 
 corresponds to high probability. 
}
\end{figure*}

For a selection of resonances, 
the total density of position  given by
\begin{equation}
\label{eq:tot_wfunc_2D}
\vert\psi(\nu,\lambda)\vert^2=\vert\psi_\nu(\nu)  \psi_\lambda(\lambda)\vert^2
\end{equation}
is shown in Fig.~\ref{fig:isosurfaces_2D}.  
Note that 
there are no nodal lines other than the coordinate axes for states with negative parities. 
This is due to the complex valuedness of the energy $E$ and the separation constant $s_2^2$ for these states.
In particular 
sections of these plots along the bottleneck  $x=0$ (which would give the 2D analogue of Fig.~\ref{fig:waves_res} in the 3D 
case) would lead to densities that are greater than zero at every $y\in(-1,1)$ apart from a possible zero at $y=0$ if $\pi_y=-$.

\subsection{The 3D system}

\subsubsection{Semiclassical computation of resonances}

In the 3D case we combine the complex EBK quantization condition \eqref{eq:res_cond_I_xi} which we write in terms of the symmetry reduced action $\tilde{I}_\xi$ analogously to \eqref{eq:I_xi_red_EBK_2D} with the uniform quantization conditions \eqref{eq:EBKzeta_eta_scatt}.
The resulting set of equations
\begin{eqnarray} \label{eq:res_cond_zeta_3D}
\tilde{I}_\zeta=\big( n_\zeta + \frac{\alpha_\zeta}{4} \big) \hbar \,, \quad
 \tilde{I}_\eta=\big( n_\eta + \frac{\alpha_\eta}{4} \big) \hbar  \,, \quad
\tilde{I}_\xi = -\ui \big(n_\xi + \frac{1}{4}(2-\pi_x) \big)\hbar    \,,
\end{eqnarray}
where $n_\zeta ,n_\eta, n_\xi \in \N_0$, and $\alpha_\nu$ and $\alpha_\mu$ are defined in \eqref{eq:Maslov_zeta} and \eqref{eq:Maslov_eta}, can again be solved by  a Newton procedure on the (real) six-dimensional space of complex $s_1^2$, $s_2^2$ and $E$. To this end
the integration paths defining the integrals that enter the quantization conditions \eqref{eq:res_cond_zeta_3D} have to be continued into the complex plane as shown in Fig.~\ref{fig:ellcurve_res}(d,e).
We label the resulting resonance modes by the Dirac kets $\vert n_\zeta,n_\eta,n_\xi;\pi_y,\pi_z,\pi_x \rangle$. 
For fixed quantum numbers $n_\zeta,n_\eta$ and parities $\pi_y,\pi_z$ we find the resonances by starting near to, but not at, the corresponding ``opening'' tick on the real energy axis of Fig.~\ref{fig:conductance_a} and by smoothly moving the parameter $s_2^2$ into the complex plane. We then go through the grid $(n_\xi,\pi_x)\in \N_0\times\{-1,1\}$.
As with the 2D case, special care has to be taken of resonances which have energies close to the imaginary axis. We give a list of semiclassically computed resonances in Tab.~\ref{tab:resonances}. 

\begin{sidewaystable*}[htbp]
\begin{center}
\tiny
\begin{tabular}{|ccc|ccc||c|c|c||c|c|c||c|}
\hline
$n_\zeta$ & $n_\eta$ & $n_\xi$ & $\pi_y$ & $\pi_z$ & $\pi_x$ & $E_{qm}$ & $k_{qm}$ & $l_{qm}$ & $E_{sc}$ & $k_{sc}$ & $l_{sc}$ & $\Delta E$ \\ \hline\hline


0 & 0 & 0 & + & + & + & $3.1688-\ui0.5835$ & $2.4498+\ui0.4141$ & $0.4408+\ui0.0771$ & $2.9886-\ui0.5942$ & $2.5052+\ui0.4454$ & $0.5802+\ui0.1081$ & $5.4$ \\ \hline
0 & 0 & 0 & + & + & - & $2.8348-\ui1.7219$ & $2.0997+\ui1.1533$ & $0.3756+\ui0.2147$ & $2.6279-\ui1.7526$ & $2.0976+\ui1.2224$ & $0.4814+\ui0.2966$ & $4.8$ \\ \hline

0 & 0 & 0 & - & + & + & $7.7066-\ui0.9376$ & $2.5633+\ui0.2801$ & $1.0338+\ui0.1196$ & $7.5633-\ui0.9414$ & $2.5844+\ui0.2860$ & $1.0528+\ui0.1237$ & $1.8$\\ \hline
0 & 0 & 0 & - & + & - & $7.3521-\ui2.7939$ & $2.4011+\ui0.8131$ & $0.9646+\ui0.3473$ & $7.2031-\ui2.8049$ & $2.4139+\ui0.8284$ & $0.9790+\ui0.3584$ & $1.7$  \\ \hline

0 & 0 & 0 & + & - & + & $8.6233-\ui0.9976$ & $2.5145+\ui0.2627$ & $0.5052+\ui0.0566$ & $8.4030-\ui0.9971$ & $2.5386+\ui0.2689$ & $0.5767+\ui0.0661$ & $2.5$ \\ \hline
0 & 0 & 0 & + & - & - & $8.2655-\ui2.9751$ & $2.3693+\ui0.7649$ & $0.4739+\ui0.1647$ & $8.0398-\ui2.9732$ & $2.3853+\ui0.7815$ & $0.5391+\ui0.1921$ & $2.4$ \\ \hline

1 & 0 & 0 & + & + & + & $14.2036-\ui1.2881$ & $2.6094+\ui0.2104$ & $1.3141+\ui0.1140$ & $14.0702-\ui1.2859$ & $2.6202+\ui0.2116$ & $1.3142+\ui0.1147$ & $0.93$ \\ \hline
1 & 0 & 0 & + & + & - & $13.8402-\ui3.8499$ & $2.5176+\ui0.6197$ & $1.2643+\ui0.3359$ & $13.7074-\ui3.8435$ & $2.5266+\ui0.6231$ & $1.2634+\ui0.3376$ & $0.90$ \\ \hline

0 & 0 & 0 & - & - & + & $14.7294-\ui1.3158$ & $2.5926+\ui0.2060$ & $1.1386+\ui0.0981$ & $14.6744-\ui1.3161$ & $2.6009+\ui0.2065$ & $1.1176+\ui0.0963$ & $0.37$ \\ \hline
0 & 0 & 0 & - & - & - & $14.3639-\ui3.9332$ & $2.5040+\ui0.6072$ & $1.0964+\ui0.2892$ & $14.3106-\ui3.9344$ & $2.5112+\ui0.6083$ & $1.0757+\ui0.2837$ & $0.34$ \\ \hline

0 & 1 & 0 & + & + & + & $17.5729-\ui1.4364$ & $2.5143+\ui0.1856$ & $0.3219+\ui0.0250$ & $17.3973-\ui1.4300$ & $2.5255+\ui0.1865$ & $0.3476+\ui0.0270$ & $1.0$ \\ \hline
0 & 1 & 0 & + & + & - & $17.2100-\ui4.2966$ & $2.4407+\ui0.5486$ & $0.3120+\ui0.0738$ & $17.0367-\ui4.2775$ & $2.4507+\ui0.5510$ & $0.3368+\ui0.0799$ & $0.97$ \\ \hline

$\vdots$ & $\vdots$ & $\vdots$ & $\vdots$ & $\vdots$ & $\vdots$ & $\vdots$ & $\vdots$ & $\vdots$ & $\vdots$ & $\vdots$ & $\vdots$ & $\vdots$ \\ \hline




7 & 3 & 0 & + & + & + & $514.6176-\ui7.8748$ & $2.6479+\ui0.0357$ & $1.4856+\ui0.0219$ & $514.4407-\ui7.7908$ & $2.6483+\ui0.0353$ & $1.4858+\ui0.0217$ & $0.04$ \\ \hline
7 & 3 & 0 & + & + & - & $514.2416-\ui23.6217$ & $2.6453+\ui0.1069$ & $1.4840+\ui0.0656$ & $514.0764-\ui23.3700$ & $2.6457+\ui0.1058$ & $1.4842+\ui0.0650$ & $0.03$\\ \hline
6 & 3 & 0 & - & - & + & $514.6179-\ui7.8748$ & $2.6479+\ui0.0357$ & $1.4856+\ui0.0219$& $514.4407-\ui7.7908$ & $2.6483+\ui0.0353$ & $1.4858+\ui0.0217$ & $0.04$ \\ \hline
6 & 3 & 0 & - & - & - & $514.2419-\ui23.6217$ & $2.6453+\ui0.1069$ & $1.4840+\ui0.0656$ & $514.0764-\ui23.3701$ & $2.6457+\ui0.1058$ & $1.4842+\ui0.0650$ & $0.04$ \\ \hline
8 & 2 & 0 & - & + & + & $514.8034-\ui7.9062$ & $2.6919+\ui0.0360$ & $1.9256+\ui0.0287$ & $514.5142-\ui7.8196$ & $2.6923+\ui0.0356$ & $1.9263+\ui0.0284$ & $0.06$ \\ \hline
8 & 2 & 0 & - & + & - & $514.4227-\ui23.7160$ & $2.6893+\ui0.1080$ & $1.9234+\ui0.0859$ & $514.1450-\ui23.4562$ & $2.6897+\ui0.1069$ & $1.9242+\ui0.0851$ & $0.06$ \\ \hline
8 & 2 & 0 & + & - & + & $514.8034-\ui7.9062$ & $2.6919+\ui0.0360$ & $1.9256+\ui0.0287$ & $514.5142-\ui7.8196$ & $2.6923+\ui0.0356$ & $1.9263+\ui0.0284$ & $0.06$ \\ \hline
8 & 2 & 0 & + & - & - & $514.4227-\ui23.7160$ & $2.6893+\ui0.1080$ & $1.9234+\ui0.0859$& $514.1450-\ui23.4562$ & $2.6897+\ui0.1069$ & $1.9242+\ui0.0851$ & $0.06$ \\ \hline
2 & 7 & 0 & + & + & + & $515.1296-\ui7.8539$ & $2.5566+\ui0.0349$ & $0.5719+\ui0.0083$& $514.9221-\ui7.7729$ & $2.5570+\ui0.0346$ & $0.5730+\ui0.0082$ & $0.04$ \\ \hline
2 & 7 & 0 & + & + & - & $514.7588-\ui23.5592$ & $2.5540+\ui0.1047$ & $0.5713+\ui0.0249$& $514.5623-\ui23.3165$ & $2.5544+\ui0.1036$ & $0.5723+\ui0.0246$ & $0.04$\\ \hline
4 & 5 & 0 & - & + & + & $518.0938-\ui7.8509$ & $2.5998+\ui0.0351$ & $1.0038+\ui0.0143$& $518.0400-\ui7.7650$ & $2.6000+\ui0.0348$ & $1.0033+\ui0.0141$ & $0.01$ \\ \hline
4 & 5 & 0 & - & + & - & $517.7245-\ui23.5502$ & $2.5972+\ui0.1053$ & $1.0028+\ui0.0429$& $517.6817-\ui23.2929$ & $2.5974+\ui0.1042$ & $1.0022+\ui0.0423$ & $0.03$ \\ \hline
0 & 8 & 0 & + & - & + & $519.5687-\ui7.8903$ & $2.5062+\ui0.0344$ & $0.0680+\ui0.0010$ & $519.4248-\ui7.8110$ & $2.5066+\ui0.0340$ & $0.0688+\ui0.0010$ & $0.01$\\ \hline
0 & 8 & 0 & + & - & - & $519.1991-\ui23.6686$ & $2.5036+\ui0.1031$ & $0.0680+\ui0.0030$ & $519.0642-\ui23.4308$ & $2.5040+\ui0.1021$ & $0.0687+\ui0.0030$ & $0.03$ \\ \hline
\end{tabular}
\end{center}
\caption{\label{tab:resonances} The quantum mechanical complex eigenvalues $\{E_{qm},k_{qm},l_{qm}\}$ and the semiclassical complex eigenvalues $\{E_{sc},k_{sc},l_{sc}\}$ of the 3D asymmetric hyperboloidal billiard for the ranges $\mathrm{Re}(E)<20$ and $500<\mathrm{Re}(E)<520$ and for $\{n_\xi,\pi_x\}=\{0,\pm\}$ (the first two families of resonances). The relative error $\Delta E=(\lvert E_{qm}\rvert - \lvert E_{sc}\rvert)/\lvert E_{qm}\rvert$ 
is given in percent. ($\hbar=1$, $m=1$.)
}
\end{sidewaystable*}


\subsubsection{Exact computation of resonances}

For the computation of the exact quantum resonances using the complex scaling method we again transform
the $\lambda$ component of the separated wave equation (see \eqref{eq:helmreg})
\begin{equation}\label{eq:lambda_wave_equation}
    -\frac{\hbar^2}{2m} \frac{\ud^2\psi_\lambda }{\ud \lambda^2}  = 
    \frac{E}{a^2}\, \big( \xi^4(\lambda) -2k \xi^2(\lambda) +l \big)\, \psi_\lambda\,.
\end{equation}
We are now looking for a transformation of $\lambda$ and $\psi_\lambda$ analogous to  
\eqref{eq:restrans_2D} and \eqref{eq:restrans_wfunc_2D} in the 2D case
which yields a system of type `kinetic-plus-potential' with the potential going to zero at infinity. 
In the 3D case this can be achieved by setting
\begin{equation}
\label{eq:restrans}
\sigma(\lambda) = aq'\tn(\lambda,q)= aq' \frac{\sn(\lambda,q)}{\cn(\lambda,q)}
\end{equation}
and
\begin{equation}
\label{eq:restrans_wfunc}
  \psi_\lambda(\lambda)  =  \frac{\psi_\sigma(\sigma)}{\sqrt{\ud\sigma/\ud\lambda}}= \sqrt{\frac{a}{q'}}\frac{\psi_\sigma(\sigma)}{\big((\sigma^2+a^2)(q'^2\sigma^2+a^2)\big)^{1/4}}  \,,
\end{equation}
where the modulus of the elliptic functions is again given by $q=c/a$ and $q'=(1-q^2)^{1/2}$.
This leads to the new wave equation
\begin{equation}\label{eq:res_wfunc}
-\frac{\hbar^2}{2m }\frac{\ud^2 \psi_\sigma}{\ud \sigma^2} = \big(  E_{\sigma, \text{eff}}-V_{\sigma, \text{eff}} \big) \psi_\sigma  \,,
\end{equation}
with the effective energy and potential given by
\begin{eqnarray}
\label{eq:effenerpot_res}
E_{\sigma, \text{eff}} &=& E\,, \nonumber\\
V_{\sigma, \text{eff}}(\sigma) &=& \frac{E[(2k-a^2q^2)(q'^2\sigma^2+a^2)-l]/q'^2+\hbar^2(6q'^2\sigma^2+a^2+q'^2)/4m}{(\sigma^2+a^2)(q'^2\sigma^2+a^2)} \nonumber\\
&-&\frac{\hbar^2}{2m}\frac{3(2q'^2\sigma^3+
  (a^2+q'^2)\sigma)^2}{4(\sigma^2+a^2)^2(q'^2\sigma^2+a^2)^2}    \, ,
\end{eqnarray}
respectively. 

As opposed to the effective potential in \eqref{eq:lambda_wave_equation},  the potential  $V_{\sigma, \text{eff}}$ goes to zero as $| \sigma |  \to \infty$, and accordingly, for real $E$, $s_1^2$ and $s_2^2$, the wave function $\psi_\sigma$ is a plane wave as $\sigma\to \infty$.
In fact, in terms of the original Cartesian coordinates   $(x,y,z)$ the solutions have to become plane waves for  $|x|\to \infty$.
From \eqref{eq:eqb3d_transxeztoxyz_a} and noting that $\xi=a\dn(\lambda,q)/\cn(\lambda,q)$ we see that  $\sigma=aq' \tn(\lambda,q)$ is proportional to  $x$ and this is the motivation for the transformation  \eqref{eq:restrans}.
The scaling of the wavefunction \eqref{eq:restrans_wfunc} is performed in order to again obtain a system of type `kinetic-plus-potential'.

Similarly to the 2D case, to compute the resonances we substitute  $\sigma$ in \eqref{eq:res_wfunc} with $ \sigma \ue^{\ui\alpha}$. We have to solve the $\zeta$ and $\eta$ components of the wave equation and the corresponding boundary conditions described in Sec.~\ref{sec:quantum_trans_3D} in combination with the equation for $\psi_\sigma$ in  
\eqref{eq:res_wfunc} with the boundary conditions at zero determined by the parity $\pi_x$ and $\psi_\sigma(\sigma)\to 0$ as 
$|\sigma| \to \infty$ (see the analogous conditions for the 2D case in Sec.~\ref{sec:exact_res_2D}). 
In our numerical procedure, which  consists of a shooting method like in the 2D case, we again choose a sufficiently large $\sigma_\infty$  at which we require $\psi_\sigma (\sigma_\infty)=0$. Decomposing all equations with respect to their real and imaginary parts the shooting method results in a (real) six-dimensional Newton procedure acting on the complex three-dimensional plane $E-s_1^2-s_2^2$ (or equivalently $E-k-l$). Using the semiclassical values from the previous section as the starting values the  shooting method always converges to the expected resonance state.
Like in the 2D case, special care has to be taken of those resonance states of energy  close the imaginary axis. 
In Tab.~\ref{tab:resonances} the exact resonances are compared to the semiclassically computed resonances. 
The relative error of the semiclassical complex energy eigenvalues reaches its maximum value of about 5 percent for the first state $\vert0,0,0;+,+,+\rangle$. The relative error shrinks rapidly for larger resonance energies.
Plotting  the resonances in the  complex energy (resp. wavenumber) plane in Fig.~\ref{fig:conductance_a} we see that each string of resonances of fixed quantum numbers $n_\zeta$ and $n_\eta$, and parities $\pi_y$ and $\pi_z$ give rise to one step of the cumulative reaction probability. Since we have three quantum numbers for the 3D system (as opposed to the two quantum numbers in the 2D system) the resonances can be viewed to form the superposition of an infinite number of grids of the more regular type of grid found in the 2D system in Fig.~\ref{fig:conductance_a_2D}.

\begin{figure*}
\includegraphics[angle=0,width=13cm]{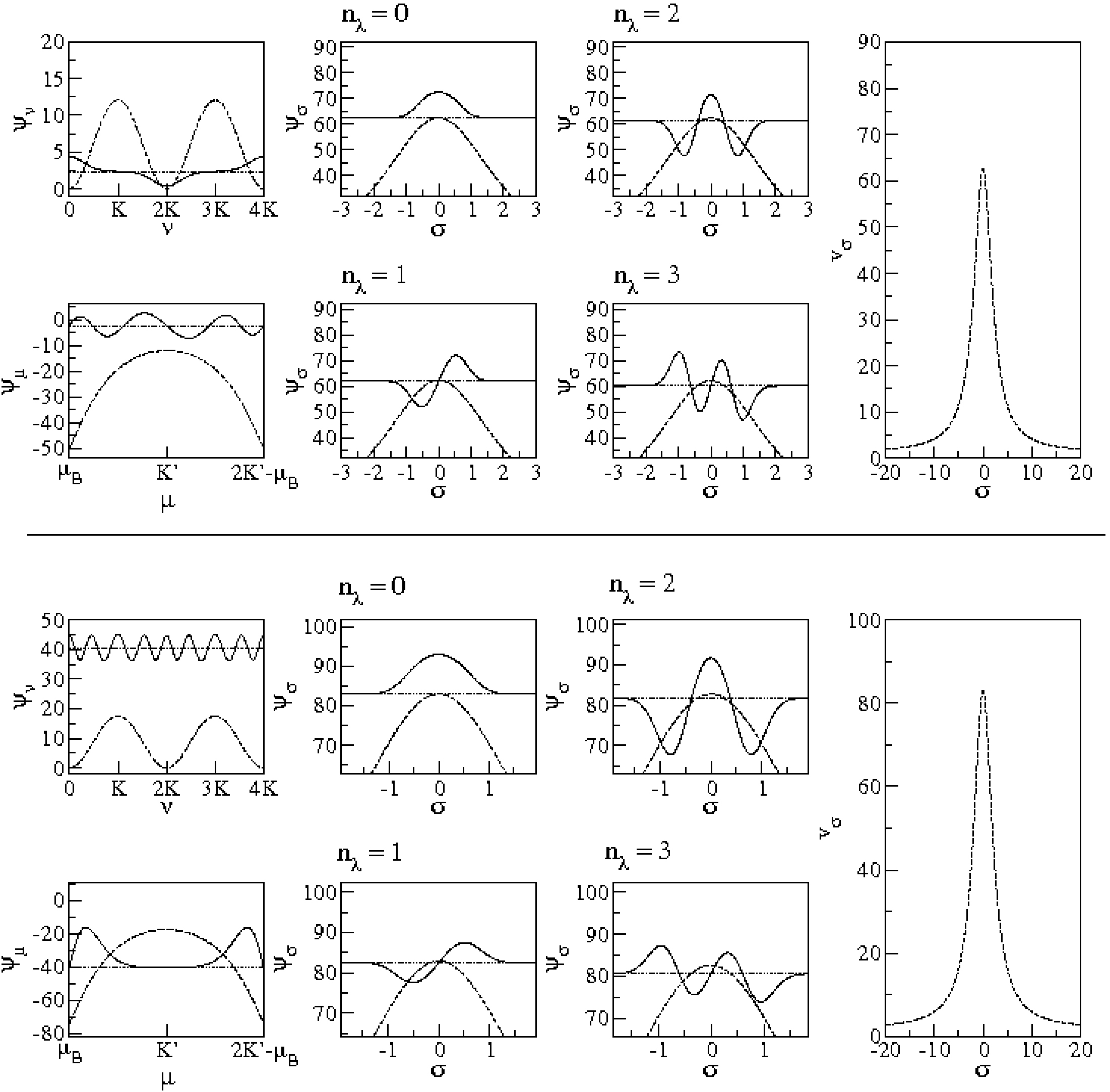}
\caption{\label{fig:wave_func}
Real parts of the separated resonance wavefunctions (solid lines) of the eight states $\vert n_\zeta,n_\eta,n_\xi;\pi_y,\pi_z,\pi_x\rangle=\vert 0,2,\{0,1\};+,-,\pm\rangle$ (classically BB, upper panels) and $\vert n_\zeta,n_\eta,n_\xi;\pi_y,\pi_z,\pi_x\rangle=\vert 4,0,\{0,1\};+,+,\pm\rangle$ (classically WG, lower panels). For fixed $\{n_\zeta,n_\eta,\pi_y,\pi_z\}$ (and hence fixed $\{n_\nu,n_\mu\}$), $\psi_\nu$ and $\psi_\mu$ are the same (far left panels) whilst for increasing $n_\lambda$ the center panels show the increase in nodes of the separated wavefunction in the reaction coordinate $\sigma$. The dashed lines are the real parts of the corresponding effective potentials $V_{\hat{s}, \textrm{eff}}(\hat{s})$. The wavefunctions are plotted at the real part of the effective energies $E_{\hat{s}, \textrm{eff}}$ (dotted lines). The two far right panels are the real parts of two sample effective potentials $V_{\sigma, \textrm{eff}}$ viewed at long range.
}
\end{figure*}

\begin{figure*}
\centerline{
\includegraphics[angle=0,width=12cm]{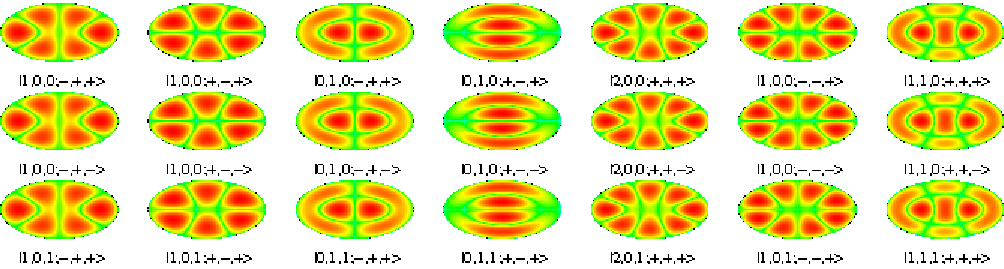}
}
\caption{\label{fig:waves_res}
Probability contours in the section $x=0$ of the wave functions of the resonance modes $\vert n_\zeta,n_\eta,n_\xi;\pi_y,\pi_z,\pi_x\rangle$. These correspond to the first three families of resonances associated with the steps labelled $\vert7\rangle$ to $\vert13\rangle$ in Fig.~\ref{fig:conductance_a}. For $\pi_x=-$ the ellipses shown here are actually nodal surfaces of the total wavefunction. The separated wavefunction $\psi_\xi$ is therefore ignored for the purposes of this figure. Light blue corresponds to low probability, red corresponds to high probability. Only axis lines (i.e. the symmetry lines) are nodal.
}
\end{figure*}

From studying the separated transverse wavefunctions $\psi_\nu$ and $\psi_\mu$ one can see that their real and imaginary parts are similar, but not equal. Also for increasing quantum number $n_\lambda$ these wavefunctions vary little (see Fig.~\ref{fig:wave_func}.)
However, the real and imaginary parts of $\psi_\nu$ and $\psi_\mu$ have their zeros at slightly different values of the respective coordinates $\nu$ and $\mu$. Due to the complex valuedness of $s_1^2$ and $s_2^2$  the total wavefunctions do not have any nodal surfaces apart from 
the ones along the symmetry planes if the corresponding parity is negative. This can also be seen in Fig.~\ref{fig:waves_res}, where we present the contours  of the resonance wavefunctions as their intersections with the $y-z$ plane (ignoring the component $\psi_\lambda$ of the total wave function which would lead to zero valued total wavefunctions when $\pi_x=-$). 
Note that these contours look quite different from the analogous contours for the scattering states in Fig.~\ref{fig:waves}. In addition to the absence of nodal lines there is also (as to be expected) no clear distinction between whispering gallery and bouncing ball modes. 

\begin{figure*}
\centerline{
\includegraphics[angle=0,width=14cm]{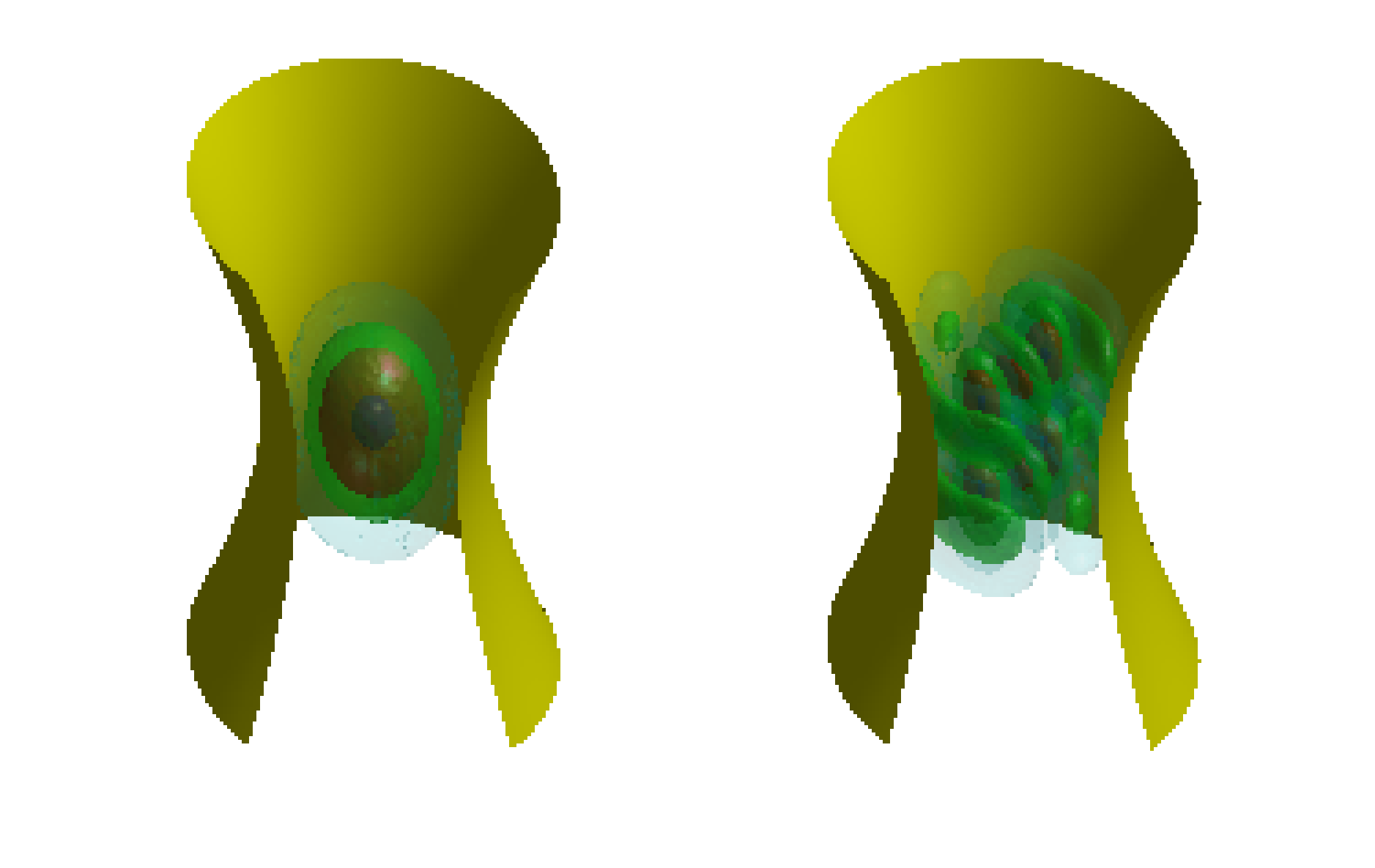}
}
\caption{\label{fig:isosurfaces}
Three-dimensional contourplots of the probabilty densities of the wavefunctions of the 3D system, with parameters $c^2=0.2$ and $a^2=5$, for the two resonance states $\vert n_\zeta,n_\eta,n_\xi;\pi_y,\pi_z,\pi_x\rangle=\vert 0,0,0;+,+,+\rangle$ (left) and $\vert n_\zeta,n_\eta,n_\xi;\pi_y,\pi_z,\pi_x\rangle=\vert 1,1,0;+,+,-\rangle$ (right).  The isosurfaces are $\vert\psi(\nu,\mu,\lambda)\vert^2=0.1$ (cyan), $\vert\psi(\nu,\mu,\lambda)\vert^2=0.3$ (green), $\vert\psi(\nu,\mu,\lambda)\vert^2=0.5$ (red) and $\vert\psi(\nu,\mu,\lambda)\vert^2=0.9$ (blue). Note that there are no nodal surfaces apart from the $y-z$ plane due to a negative parity $\pi_x$ for the second state.}
\end{figure*}

The total probability density is given by
\begin{equation}
\label{eq:tot_wfunc}
\vert\psi(\nu,\mu,\lambda)\vert^2=\vert\psi_\nu(\nu)   \psi_\mu(\mu)   \psi_\lambda(\lambda)\vert^2\,.
\end{equation}
In order to compute the separated wavefunction $\psi_\lambda(\lambda)$ we use again the transformation (\ref{eq:restrans_wfunc}) and the inverse of (\ref{eq:restrans}) which is given by the elliptic integral
\begin{equation}
\label{eq:restrans_inv}
\lambda(\sigma)=\tn^{-1}(\sigma/aq',q)=\int_{0}^{\sigma}\frac{a}{\sqrt{(a^2+x^2)(a^2q'^2+x^2)}}\ud x\,.
\end{equation}
In Fig.~\ref{fig:isosurfaces} we show the isosurfaces of (\ref{eq:tot_wfunc}) inside the hyperboloidal boundary for two examples of resonance states which further illustrates the absence of nodal surfaces (in addition to the Cartesian coordinate planes for negative parities).

%% file: sec7_conclusions.tex
\section{Conclusions and Outlook}
\label{sec:conclusions}

In this paper we demonstrated how ideas from transition state theory can be used to compute the classical and quantum mechanical transmission probabilities for transport  through entropic barriers. 
For barriers associated with saddle points of the potential  (and more generally saddle type equilibria of the Hamilton function) it has recently be shown that the transport through the phase space bottlenecks induced by such saddle points are controlled by a set of phase space structures. In the present paper we identified the analogous phase space structures  for potentialless  barriers where the phase space bottlenecks are induced by hard wall constrictions.  
We focused on the special case of hyperboloidal constrictions in two and three dimensions for which the classical and quantum transmission problems are separable, and hence facilitate a very detailed, and to a large extent also analytical, study. 
For these systems,
we showed that like in the case of smooth systems one  can construct a dividing surface which has the property that it is crossed exactly once by reactive trajectories, and not crossed at all by nonreactive trajectories. This, like in the smooth case, leads to a rigorous realization of Wigner's transition state theory, i.e. to an exact computation of the classical transmission probability from the (directional) flux through the dividing surface. Similarly the quantum cumulative reaction probability can be computed exactly from the quantum mechanical flux through the dividing surface. 
We showed that like in the  case of smooth systems the dividing surface is linked to an unstable invariant subsystem with one degree of freedom less than the full system. In the context of chemical reactions such an invariant subsystem located between reactants and products
is referred to as the \emph{transition state} or the \emph{activated complex}.
For the 2D and 3D systems studied in this paper 
the transition states consist of the billiard in a one-dimensional square well  and the billiard in an ellipse, respectively.
Like in the smooth case the transition states in the hard wall constrictions have stable and unstable manifolds which have sufficient dimensionality to form separatrices which separate reactive trajectories from nonreactive trajectories and thus play a key role for the classical transmission problems. 

Quantum mechanically the transition states manifest themselves on the one hand through a `quantization' of the cumulative reaction probability and quantum resonances. The quantization of the cumulative reaction probability refers to the stepwise  increase of the cumulative reaction probability as a function of energy  each time a new state fits into the respective transition state. In fact, as discussed in some detail, the cumulative reaction probability is approximately given by the integrated density of states of the invariant subsystems associated with the transition states.  
For the quantum systems, the Heisenberg uncertainty relation excludes the existence of invariant subsystems analogous to the classical case. Instead a wavepacket initialised on the classically invariant subsystems will decay (exponentially fast) in time with the lifetimes being described by the resonances. 
We computed such resonances semiclassically from the poles of a meromorphic continuation of the semiclassical expression for the transmission probability to the lower half of the complex energy plane. We showed that this leads to a very good agreement with the results obtained from the numerical computation of the exact resonances from the complex scaling method.
The separability of the systems yields an assignment of the resonances by  quantum numbers. We showed that each string of resonances corresponding to fixed quantum numbers associated with the transverse degrees of freedoms gives rise to a step of unit size of the cumulative reaction probability. 
Despite of their separability, the systems studied in this paper display quite a rich variety of dynamics. In particular the transition state of the 3D system involves two different types of modes that we referred to as whispering gallery and bouncing ball modes. We showed that the energetic quasidegeneracy of the whispering gallery modes leads to steps in the cumulative 
reaction probability of effective stepsize 2. Similarly the corresponding resonances are quasidegenerate with respect to their complex energies.

For many aspects of our study we heavily used the separability of the systems. 
We in particular presented a detailed study of the analytic nature of the integrals which enter  the  EBK and uniform semiclassical quantization schemes as action and tunnel integrals. We showed that all these integrals can be interpreted as Abelian integrals on an elliptic (2D system) or hyperelliptic (3D system) curve with branch points determined by the separation constants and the geometry parameters of the hyperboloidal constrictions. This interpretation led to a natural extension  of the semiclassical quantization conditions of scattering states to quantization conditions for resonances states by complexifying the separation constants and integration paths.

Though the systems discussed in this paper are special due their separability many of the phase space structures found are expected to exist in systems with more general constrictions that lead to nonseparable dynamics. 
In particular, it is to be expected that the phase space structures associated with the transition states persist under (small) deformations  of the hyperboloidal constrictions  which (generically) would destroy the integrability of the systems. For smooth systems, this has been studied already in quite great detail. The phase space structures for such systems can, e.g.,  be practically determined from a normal form expansion about the saddle equilibrium points \cite{ujpyw,WaalkensSchubertWiggins08}. For systems where the barriers are induced by hard wall constrictions the determination of the phase space structures associated with the transition state is more challenging in the generic nonseparable case. The best approach here seems to be the study of these structures in terms of a billiard map, i.e. a map resulting from taking snapshots of the billiard dynamics from one specular reflection to the next. For a 2D system the resulting  map is a  symplectic map from $\R^2$ to $\R^2$, with the periodic orbit that forms the transition state in this case appearing as a hyperbolic fixed point of the map. There are well established methods  for determining the fixed point and also (using methods based again on a normal form) for computing the stable and unstable manifolds of the fixed point. However,
the situation is much more involved in the 3D case, where the resulting billiard map is a symplectic map from 
 $\R^4$ to $\R^4$. The transition state then forms a two-dimensional manifold in the domain and image of this map. Its stable and unstable manifolds are three-dimensional. Computing these manifolds in practice is quite difficult. This gives an interesting field for future studies.